# Extended study of crystal structures, optical properties and vibrational spectra of polar 2-aminopyrimidinium hydrogen phosphite and three centrosymmetric salts - bis(2-aminopyrimidinium) sulfate monohydrate and two 2-aminopyrimidinium hydrogen sulfate polymorphs

Irena Matulková[a], Ladislav Bohatý[b], Petra Becker[b], Ivana Císařová[a], Róbert Gyepes[c], Michaela Fridrichová[a], Jan Kroupa[d], Petr Němec[e] and Ivan Němec[a]*

[a] *Charles University, Faculty of Science, Department of Inorganic Chemistry, Hlavova 8, 128 40 Prague 2, Czech Republic.*
[b] *University of Cologne, Institute of Geology and Mineralogy, Section Crystallography, Zülpicher Str. 49b, 50674 Köln, Germany.*
[c] *Czech Academy of Sciences, J. Heyrovsky Institute of Physical Chemistry, Department of Molecular Electrochemistry and Catalysis, Dolejškova 2155, 182 00 Prague 8, Czech Republic.*
[d] *Czech Academy of Sciences, Institute of Physics, Na Slovance 2, 182 21 Prague 8, Czech Republic.*
[e] *Charles University, Faculty of Mathematics and Physics, Department of Chemical Physics and Optics, Ke Karlovu 3, 121 16 Prague 2, Czech Republic.*

* Corresponding author. E-mail address: ivan.nemec@natur.cuni.cz

## Abstract

This study aimed primarily at completing and extending the characterization of the crystallographic, spectroscopic and optical properties of polar, biaxial, optically negative 2-aminopyrimidinium(1+) hydrogen phosphite. Besides the redetermination of the low-temperature crystal structure (space group $P2_1$), high-quality single crystals of this salt were grown from an aqueous solution, and their optical properties were studied. The determination of the refractive indices in the wavelength range of 435–1083 nm showed anomalous dispersion of the refractive indices, resulting in a point of uniaxiality. The crystal allows phase matching for collinear second harmonic generation (SHG) processes of both type I and type II in a broad wavelength range. SHG properties were studied for powdered size-fractioned samples and oriented single-crystal cuts. The optical damage threshold experiments confirmed excellent optical resistance - at least 220 $TWm^{-2}$ and 70 $TWm^{-2}$ for 800 and 1000 nm irradiation, respectively. The low-temperature crystallographic study was also extended for three monoclinic salts of 2-aminopyrimidine and sulfuric acid - i.e. bis(2-aminopyrimidinium(1+) sulfate monohydrate (space group $P2_1/n$) and two polymorphs of 2-aminopyrimidinium(1+) hydrogen sulfate (both with space group $P2_1/c$). The vibrational spectra of all title compounds were assigned using single-molecule quantum chemical computations (including Potential

Energy Distribution analysis) in combination with the nuclear site group analysis. Spectroscopic results concerning sulfates of 2-aminopyrimidine provided valuable “reference” materials for the vibrational spectroscopic study and also addressed the question of their polymorphism. An optimal computational approach employing solid-state DFT calculations has also been sought to model the vibrational spectra of 2-aminopyrimidinium (1+) hydrogen phosphite crystals.

## 1. Introduction

In Materials Science, part of the research focuses on identifying and designing new crystalline materials for nonlinear optics (NLO) using organic molecules and their salts and cocrystals [1-4]. Thanks to their $\chi^{(2)}$ and $\chi^{(3)}$ NLO effects (e.g., second (SHG) and third (THG) harmonic generation and cascaded self-frequency doubling and tripling), these materials find many technical applications, especially in new laser frequency generators and signal processing units, in addition to optical communication, all-optical switching, optical power limiting and image manipulation devices [1, 5]. Designing a new molecular NLO material under controlled conditions requires selecting (synthesis) a promising polarizable molecule (i.e., molecular engineering) and promoting its incorporation into a suitable crystal structure (i.e., crystal engineering) [6]. The resulting molecular crystals contain organic molecules (acting as carriers of NLO properties), which ideally can form non-centrosymmetric phases that meet symmetry conditions for $\chi^{(2)}$ NLO effects (e.g., SHG). Appropriate crystal packing results from supramolecular self-assembly controlled by intermolecular interactions (primarily hydrogen bonding) between these molecules or with co-crystallisation partners or by salt formation. The salts and cocrystals of these organic molecules combine favourable physicochemical properties, such as appropriate hyperpolarizability, high optical transparency, sufficient thermal stability, and an excellent optical damage threshold.

Some of the most intensively studied materials in this family are compounds based on heteroaromatic molecules, particularly nitrogen-containing heterocyclic bases derived from triazoles [7-9], triazines [10-14], pyrazines [15] and pyrimidines [16, 17].

Among these compounds, 2-aminopyrimidine (**2-Amp**) and the 2-aminopyrimidinium(1+) cation (**2-Amp**(1+) stand out as promising 2D moieties for new NLO materials. In previous studies, we have predicted and experimentally confirmed the potential of **2-Amp** and **2-Amp**(1+) for $\chi^{(2)}$ NLO processes through quantum-chemical computations and hyper-Rayleigh scattering measurements [18], highlighting their protonation-dependent NLO properties. The uncharged **2-Amp** molecule has approximately 1.5 times higher overall hyperpolarizability ($\beta_{tot}$) than the protonated **2-Amp**(1+) cation. In addition, protonation of the **2-Amp** heterocycle also strongly affects linear optical properties, such as birefringence [17]. Based on these findings, we have previously prepared **2-Amp** cocrystals with a weak inorganic acid, i.e., boric acid, based on the "pKa rule for acid-base complexes" [19] and studied their structural, spectroscopic and optical properties [20].

In turn, salts with stronger inorganic acids, containing a protonated **2-Amp**(1+) cation, may also be promising candidates for new NLO materials [21-23]. Given their high variability of symmetries and different donor-acceptor potentials for hydrogen bonding, inorganic anions can yield phases with appropriate crystal symmetry [1-4]. Unfortunately, in the group of **2-Amp** salts and cocrystals with inorganic acids (including structures of pure **2-Amp**) characterized so far, which are listed in Table S1 (Supplementary material), the centrosymmetric arrangement prevails, excluding the symmetry conditions required for $\chi^{(2)}$ NLO effects. Crystallographic data is the only information available for most of these compounds, and only in refs. [20, 24, 25], which concern 2-aminopyrimidinium(1+) dihydrogen phosphate monohydrate **2-AmpH$_2$PO$_4$H$_2$O** and the cocrystals 2-aminopyrimidine–boric acid (3/2) **(2-Amp)$_3$(H$_3$BO$_3$)$_2$** and 2-aminopyrimidine–boric acid (1/2) **2-Amp(H$_3$BO$_3$)$_2$**, the study was also extended to vibrational spectroscopic methods.

As the most exciting representative of 2-aminopyrimidine salts, the polar crystal of 2-aminopyrimidinium(1+) hydrogen phosphite **2-AmpH$_2$PO$_3$** was first characterized in our study on inorganic salts of aminopyrimidines [21] and subsequently further investigated with respect to the growth of bulk single crystals as well as spectroscopic and non-linear optical properties [22]. Obtained crystallographic data were deposited with the Cambridge Crystallographic Data Centre as a supplementary publication CCDC 1503404 in 2017. More recently, Zhang et al. [23] reported their characterization of **2-AmpH$_2$PO$_3$** as a promising NLO material while we were drafting the present manuscript.

In this context, the present study primarily aims to extend and refine the experimental analysis of the structural, vibrational spectroscopic and optical (linear and nonlinear) properties of polar 2-aminopyrimidinium(1+) hydrogen phosphite **2-AmpH$_2$PO$_3$** crystals. The accompanying structural and spectroscopic analysis of **2-Amp** sulfates provides not only valuable “reference” materials (necessary for the vibrational spectroscopic study) but also new experimental data on bis(2-aminopyrimidinium(1+)) sulfate monohydrate (**2-Amp)$_2$SO$_4$H$_2$O** and two polymorphs of 2-aminopyrimidinium(1+) hydrogen sulfate (**2-AmpHSO$_4$ (I)** and **(II)**) in a comprehensive study of inorganic salts of 2-aminopyrimidine.

An extensive vibrational spectroscopic study of four related title salts enhances our understanding of vibrational manifestations of the **2-Amp**(1+) cation as a carrier of NLO properties in the solid state. These findings are highly useful for IR and Raman spectroscopy applications, especially for phase analysis, including a study of polymorphism and phase transformations. Moreover, our detailed assignment of spectra of novel NLO materials promotes the understanding of vibrational contributions to hyperpolarizability [26] and processes associated with Stimulated Raman Scattering (SRS) [27, 28].

## 2. Experimental

### *2.1. Materials and methods*

The compound **2-AmpH$_2$PO$_3$** was synthesized from an aqueous solution of 2-aminopyrimidine (97%, Fluka) and 2 mol.L$^{-1}$ phosphorous acid (purum, p.a., Fluka), mixed in the molar ratio 1:1. Small crystals were obtained by evaporation of the solution at room temperature. The aqueous solutions of 2-aminopyrimidine (97%, Fluka) and 2 mol.L$^{-1}$ sulphuric acid (96%, p.a., Lachema), mixed in the molar ratios 1:1 and 2:1, by slow evaporation at room temperature provide **2-AmpHSO$_4$ (I)** and (**2-Amp)$_2$SO$_4$H$_2$O**, respectively. Only a few crystals of **2-AmpHSO$_4$ (II)** phase were isolated from the ethanol-water (ratio 1:2) solution of 2-aminopyrimidine (97%, Fluka) to which 2 mol.L$^{-1}$ solution of sulphuric acid (96%, p.a., Lachema) was added dropwise to achieve a molar ratio of 1:1.4 (base to acid). The solution was intensively stirred for 20 min and left to slowly evaporate at room temperature until the formation of colourless intergrown crystals of both polymorphs of **2-AmpHSO$_4$**, which were separated under a microscope. A limited amount of obtained **2-AmpHSO$_4$ (II)** crystals enabled only single-crystal X-ray diffraction and micro-spectroscopic characterization of this phase. The reference compound for the vibrational spectroscopic study – crystalline 2-aminopyrimidinium(1+) chloride hemihydrate (**2-AmpCl½H$_2$O**) – was isolated (at room temperature) from an aqueous solution of 2-aminopyrimidine (97%, Fluka) and 2 mol.L$^{-1}$ hydrochloric acid (35%, p.a, Lach-ner), mixed in the molar ratio 1:1.

FTIR spectra were recorded on a Thermo Fisher Scientific Nicolet Magna 6700 FTIR spectrometer. Micro-FTIR spectra of **2-AmpHSO$_4$** polymorphs were recorded using the ATR technique on a Thermo Fisher Scientific Nicolet iN10 FTIR microscope. FT Raman spectra of the powdered samples were recorded on a Thermo Fisher Scientific Nicolet 6700 FTIR spectrometer equipped with the Nicolet Nexus FT Raman module. Raman spectra of microcrystalline samples and aqueous **2-Amp** and **2-Amp**(1+) solutions were collected on a Thermo Scientific DXR Raman Microscope interfaced to an Olympus microscope. Raman spectra of microcrystalline samples were also collected on a dispersive confocal Raman microscope MonoVista CRS+ (Spectroscopy & Imaging GmbH, Germany) interfaced to an Olympus microscope. For the experimental details of all vibrational spectroscopic measurements, see the Supplementary material.

The UV-Vis-NIR absorption spectrum of a polished thin single crystal plate (0.7 mm thickness, no defined crystallographic orientation) of **2-AmpH$_2$PO$_3$** was recorded using non-polarized light with a UNICAM UV-300 spectrometer in the 190-1100 nm region with 1 nm spectral resolution.

The phase purity of prepared polycrystalline products was controlled by the powder X-ray diffraction using a Philips X'pert PRO MPD diffractometer (Bragg-Brentano geometry, ultrafast X'Celerator detector and Cu Kα radiation, λ = 1.5418 Å). The data were analysed using the FullProf software [29], and the results are presented in Tables S2-S5, Supplementary material. The theoretical diffraction patterns used to confirm the sample composition were calculated from the single-crystal data using the PLATON software [30].

Melting points of the polycrystalline samples (with the exception of **2-AmpHSO$_4$ (II)** phase) were determined using a melting-point apparatus Büchi B-540 (visual detection, heating rate 5 °C/min).

The DSC measurements (with the exception of **2-AmpHSO$_4$ (II)** phase) were performed on a Perkin Elmer DSC 8500 instrument with a Perkin Elmer CLN2 liquid nitrogen cooler. The measurement cycles (heating rate 10 °C/min, helium atmosphere, finely ground samples in sealed aluminium pans) were carried out in temperature regions ranging from -160°C to temperatures slightly below the melting points of studied salts.

*2.2. Crystal structure determination*

The collection of low-temperature X-ray diffraction data for **2-AmpH$_2$PO$_3$**, **2-AmpHSO$_4$ (I)** and (**2-Amp)$_2$SO$_4$H$_2$O** single crystals was performed on a Nonius Kappa CCD diffractometer (MoKα radiation, graphite monochromator). Data for **2-AmpHSO$_4$ (I)** and (**2-Amp)$_2$SO$_4$H$_2$O** were corrected for absorption by the methods incorporated in the diffractometer software (multi-scan routine [31]). The diffraction data for a selected single crystal of **2-AmpHSO$_4$ (II)** were collected on a Bruker D8 VENTURE Kappa Duo diffractometer (MoKα radiation, graphite monochromator) equipped with a PHOTON100 detector. The full-set data (±*h*, ±*k*, ±*l*, 2θ ≤ 55°) were reduced by the diffractometer software SAINT [32]. The diffractometer software (SADABS [33]), utilizing a multi-scan method, was used to correct the data for absorption. An Oxford Cryosystems liquid nitrogen Cryostream Cooler controlled the temperature of all studied crystals.

The direct methods (SHELXS97 [34]) were used for solving the phase problem, and the structures were refined using a full-matrix least-squares routine based on $F^2$ (SHELXL97 [34]). The non-hydrogen atoms were refined with anisotropic displacement parameters. The hydrogen atoms bonded to carbon were included in their calculated positions and refined as riding atoms. The other hydrogen atoms were localized on the difference electron density maps and were fixed during refinement using a rigid body approximation with the assigned displacement parameters equal to 1.2 $U_{iso}$ (pivot atom). The recent PLATON software [30] was used for the geometric analysis and creation of crystallographic figures. Presented graph-set descriptors were generated by MERCURY software [35].

Tables 1(a) and 1(b) summarize the basic crystallographic data, measurement and refinement details. The crystallographic data of **2-AmpH$_2$PO$_3$**, **2-AmpHSO$_4$ (I), 2-AmpHSO$_4$ (II)** and (**2-Amp)$_2$SO$_4$H$_2$O** have been deposited at the Cambridge Crystallographic Data Centre as supplementary publications CCDC 1503404, CCDC 1586910, CCDC 2320713 and CCDC 1586911, respectively. The selected geometric parameters (bond lengths, angles and hydrogen bonds) are listed in Tables S6-S9, Supplementary material.

The crystallographic data of **2-AmpCl½H$_2$O** (i.e. reference compound for the vibrational spectroscopic study) have been deposited at the Cambridge Crystallographic Data Centre as supplementary publication CCDC 2427845. Tables S10, S11 and Fig. S1 (Supplementary material) summarize basic crystallographic data, measurement and refinement details, selected geometric parameters, and atom numbering, respectively.

*2.3. Quantum Chemical Computations*

Gaussian 09W software [36] was used for quantum chemical computations concerning geometry optimization, followed by vibrational frequency calculations of **2-Amp**(1+) cation. Raman intensities were calculated by the RAINT programme [37], and the assignment of the computed vibrational modes (see Table S12, Supplementary material) is based on the visualization of the vibrations using the GaussView programme [38] and performed PED analysis by the VEDA4 programme [39]. Table S13, Supplementary material, contains the comparison of recorded vibrational spectra of reference 2-aminopyrimidinium(1+) chloride hemihydrate (**2-AmpCl½H2O**) with computed normal modes (scaled by dual scaling [40] or wavenumber-linear scale "WLS" [41] procedures).

Solid-state DFT computational studies of **2-AmpH$_2$PO$_3$** focused on vibrational spectra and optical properties were carried out using the CRYSTAL17 program [42]. Three approaches named "B3LYP", "B3LYP-advanced", and "PBESOL0", differing in functional and basis sets, were selected.

For details of all quantum chemical computations performed, see the Supplementary material.

## *2.4. Crystal growth of* ***2-AmpH$_2$PO$_3$***

The solubility of **2-AmpH$_2$PO$_3$** crystals in water is highly temperature-dependent and increases from 76.9 g/L (293 K), 93.6 g/L (298 K), 178.2 g/L (303 K), 237.2 g/L (308 K), and 311.5 g/L (313 K) to 373.1 g/L (318 K). Large single crystals of **2-AmpH$_2$PO$_3$** were grown from aqueous solutions using both methods, controlled solvent evaporation at constant temperature or slow temperature lowering within the 45-35°C range. The best results were obtained by controlled solvent evaporation at 38°C. During a 12 - 14 weeks growth period, crystal with dimensions of up to 4 x 3 x 1 cm$^3$ with optically clear volume of approx. 1 cm$^3$ resulted. The quality of the obtained crystals strongly depends on the purity of the starting chemicals and several steps of purification by recrystallization were therefore necessary. In Fig. 1 an example of a grown single crystal of **2-AmpH$_2$PO$_3$** is given.

## *2.5. Linear optical properties*

Refractive indices and their wavelength dispersion of the monoclinic crystals of **2-AmpH$_2$PO$_3$** were measured using the prism method and a high-precision goniometer-spectrometer (Möller-Wedel; for instrumental details see, e.g., [43]). For monoclinic crystals, three principal refractive indices $n_1^0$, $n_2^0$ and $n_3^0$, and the orientation angle $\psi$ of the principal axes of the optical indicatrix { $\boldsymbol{e}_i^0$ } with respect to the Cartesian reference system { $\boldsymbol{e}_i$ } ("crystal physical axes") and the crystallographic axes ***a***, ***b***, ***c*** (see Fig. 2) have to be determined. The used crystal physical axes { $\boldsymbol{e}_i$ } are related to the crystallographic axes in the standard way for monoclinic crystals, i.e., $\boldsymbol{e}_3 = \frac{1}{|\boldsymbol{c}|}\boldsymbol{c}$, $\boldsymbol{e}_2 = \frac{1}{|\boldsymbol{b}|}\boldsymbol{b}$ and $\boldsymbol{e}_1 = \boldsymbol{e}_2 \times \boldsymbol{e}_3$, see Fig. 2. The principal refractive indices $n_1^0$ and $n_3^0$ can be measured by normal incidence on a prism with incidence face (010), while the refractive index $n_2^0$, together with a mixed index $n'$ is obtained by normal incidence on a prism with incidence face ($h0l$). For the measurement of refractive indices of **2-AmpH$_2$PO$_3$**, besides a single crystal prism with incidence face (010), a second prism with incidence face (001) was prepared and used. Refractive index data were collected at 9 discrete wavelengths between 435.8 nm and 1083.0 nm. The measured data $n_{meas.}$ were corrected with respect to the refractive index of air according to $n_{corr.} = n_{meas.} \cdot n_{air}$ using data for standard air from [44]. All data in the following are corrected refractive index data. A modified Sellmeier equation was fitted to the refractive index data, which characterizes the wavelength dispersion of the refractive indices of **2-AmpH$_2$PO$_3$**:

$$n^2(\lambda) = P_1 + \frac{P_2}{(\lambda^2 - P_3)} - P_4 \cdot \lambda^2 \tag{1}$$

From the mixed index $n'$ the orientation angle $\psi$ was calculated according to $\psi = \varphi + \beta - 90°$ (see also Fig. 2), with $\beta$ = monoclinic angle $\angle(\boldsymbol{a}, \boldsymbol{c})$ and

$$\sin\varphi = \frac{n_3^0}{n'}\sqrt{\frac{(n_1^{0^2} - n'^2)}{(n_1^{0^2} - n_3^{0^2})}}\,. \tag{2}$$

The orientation angle $\psi$ was also checked and verified by direct observation of the extinction angle of a (010) crystal plate with respect to the ***a*** axis with a polarization microscope between crossed polarizers.

### *2.6. SHG measurements*

Initial SHG measurements on polar **2-AmpH$_2$PO$_3$** were performed by the modified Kurtz-Perry powder method [45] using 800 nm pulses of Ti: sapphire laser (MaiTai, Spectra-Physics). The first experiments were performed on powdered samples (100-150 μm particle size). The phase matching measurements were then carried out on the size fractioned samples (particle size in the 25-150 μm range). Lastly, the optical damage threshold experiments were performed with 800 and 1000 nm laser pulses.

The standard Maker fringe method [46] was used for the determination of the individual components of the second order nonlinear optical tensor $[d_{ijk}^{SHG}]$ (contracted Voigt notation for tensor components $d_{mn}$ is used in the following text) of **2-AmpH$_2$PO$_3$** single crystal samples. SHG measurements with plane parallel samples were performed using a Q-switched Nd-YAG laser (1064 nm).

Experimental details concerning all performed SHG measurements are available in Supplementary material. The details of the measurement strategy used for monoclinic crystals of point group 2 have been described, for example, in Ref. [47].

## 3. Results and discussion

### *3.1. Crystal structures*

Selected bond lengths and angles, including those of hydrogen bonds, are for **2-AmpH$_2$PO$_3$**, **2-AmpHSO$_4$ (I)**, **2-AmpHSO$_4$ (II)** and **(2-Amp)$_2$SO$_4$H$_2$O** presented in Tables S6-S9, Supplementary material. Atom numbering is shown in Figs. S2-S5, Supplementary material.

The crystals of **2-AmpH$_2$PO$_3$** belong to the monoclinic system with the space group $P2_1$. The crystal structure is based on chains of hydrogen-bonded (O-H…O) hydrogen phosphite anions, oriented along the ***c***-axis (graph-set motif $C_1^1(4)$ and O1…O2[b] distance 2.572(2) Å), which are interconnected by 2-aminopyrimidium(1+) cations *via* N-H…O and C-H…O hydrogen bonds (see Fig. 3 and Table S6, Supplementary material). Every cation is involved in a ring pattern, described by graph-set motif $R_2^2(8)$, containing N1-H1…O3[a] and N2-H2A…O2[a] hydrogen bonds with donor-acceptor distances 2.629(3) and 2.896(3) Å), respectively. Cations with neighbouring anions also form chains (graph-set motif $C_2^2(9)$) involving N2-H2B…O3[c] and C2-H2…O1[d] hydrogen bonds with donor-acceptor distances 2.827(3) and 3.326(3) Å, respectively. The crystal structure of **2-AmpH$_2$PO$_3$** does not contain any π-π stacking interaction within the generally used definition (see, e.g., refs. [48, 49]).

Comparing the results of the crystal structure determination of **2-AmpH$_2$PO$_3$** at 273 K [23] and the presented measurement at 150 K, the volume of the unit cell follows expected trends and decreases with temperature – i.e. 373.97(6) Å$^3$ and 367.25(3) Å$^3$ for 273 K and 150 K, respectively. Surprisingly, the lattice parameter ***c*** is slightly longer at 150 K (4.7606(2) Å) compared to the results at 273 K (4.7517(4) Å). This finding reflects that the length of this parameter is governed by a strong hydrogen bond O1-H10…O2[b], which is not affected by the decrease in temperature movement of the atoms. The remaining two lattice parameters decrease with a temperature of about 1% of their length.

The more precise low-temperature measurement also allows us to evaluate the quality of **2-AmpH$_2$PO$_3$** crystal. When the chirality of the crystal results exclusively from the symmetry operations of the space group, the possibility of twinning by inversion needs to be investigated. The absolute structure parameter (Flack parameter) 0.06(10) obtained in work [23] at 273 K does not exclude the occurrence of the inversion twins due to its large estimated standard deviation (esd); our value -0.05(3) at 150 K is witnessing the pure chirality character of the crystal and agrees with the observed morphology of its large specimens (see Chapter 2.4.).

The results of our crystallographic study demonstrate that both polymorphs **2-AmpHSO$_4$ (I)** and **2-AmpHSO$_4$ (II)** crystallise in the monoclinic space group $P2_1/c$. However, their crystal packing is entirely different.

The crystal structure of **2-AmpHSO$_4$ (I)** (presented for the first time in ref. [50]) consists of alternating 2-aminopyrimidium(1+) cations and hydrogen sulfate anions interconnected by an extensive network of hydrogen bonds of N-H…O (donor-acceptor distance ranging from 2.702(2) to 3.009(2) Å) and O-H…N type (donor-acceptor distance equal to 2.652(2) Å). Moreover, weak hydrogen bonds C3-H3…O3[c] and C4-H4…O2[c] were also found in the structure (see Fig. 4 and Table S7, Supplementary material). Interionic hydrogen bonding of cations and anions leads to ring pattern formation described by graph-set descriptors $R_2^2(8)$ and $R_1^2(4)$. Chains characterized by graph-set descriptor $C_2^2(8)$ are also involved in the 3D packing of this polymorph.

In contrast, the crystal structure of the **2-AmpHSO$_4$ (II)** polymorph (presented for the first time in ref. [51]) is based on chains formed along the ***b***-axis (graph-set descriptor $C_1^1(4)$**)** by hydrogen sulfate anions (via O-H…O hydrogen bonds with O…O distance 2.591(1) Å, see Table S8, Supplementary material). These chains are interconnected by N-H…O interactions (donor-acceptor distance ranging from 2.747(1) to 3.167(2) Å) with centrosymmetric dimers of 2-aminopyrimidium(1+) cations (graph-set descriptor $R_2^2(8)$) – see Fig. 5. Resulting 3D arrangement involves also three types of C-H…O hydrogen bonds with donor-acceptor distance ranging from 3.208(2) to 3.370(2) Å.

The asymmetric unit of monoclinic (space group $P2_1/n$) crystals of **(2-Amp)$_2$SO$_4$H$_2$O** contains two 2-aminopyrimidinium(1+) cations, sulfate anion and a water molecule, see Fig. S5, Supplementary material. The crystal structure (presented for the first time in ref. [50]) contains chains (graph-set motif $C_2^2(6)$) of alternating sulfate anions and water molecules interconnected by O-H…O hydrogen bonds (donor-acceptor distance 2.733(1) and 2.796(1) Å) along the ***b*** axis (see Fig. S6, Supplementary material). The sulfate anions in the chains interact with pairs of symmetry-independent cations (see Fig. S7, Supplementary material) by pairs of N-H…O hydrogen bonds (donor-acceptor distances ranging from 2.637(1) Å to 2.983(1) Å) to form the two ring patterns, which can be described by the $R_2^2(8)$ graph-set descriptor. The resulting 3D crystal structure (see Fig. 6) also incorporates several cation…water interactions (i.e., N-H…O, C-H…O and C-H…N hydrogen bonds) – see Table S9, Supplementary material. The crystals of **(2-Amp)$_2$SO$_4$H$_2$O** are isostructural with the previously reported bis(2-aminopyrimidinium(1+)) selenate monohydrate [52].

Discussed salts of **2-Amp** with sulfuric acid – i.e., **2-AmpHSO$_4$(I)** (this work and [50]), **2-AmpHSO$_4$(II)** (this work and [51]), **(2-Amp)$_2$SO$_4$H$_2$O** (this work and [50]) together with bis(2-aminopyrimidinium(1+)) sulfate **(2-Amp)$_2$SO$_4$** [53] represent an exciting group of molecular crystals with different structural roles of anions in the crystal packing. In the case of **(2-Amp)$_2$SO$_4$** and **(2-Amp)$_2$SO$_4$H$_2$O**, the anions act only as acceptors of hydrogen bonds of the type N-H...O and C-H...O (cation...anion) and O-H...O (water...anion). In the structures **2-AmpHSO$_4$(I)** and **2-AmpHSO$_4$(II)**, protonated anions have a more complex function - they are involved not only as the acceptors mentioned above but also act as donors of hydrogen bonds of the type O-H...N (anion...cation) in the case of **2-AmpHSO$_4$(I)** and of the type O-H...O (anion...anion) in the case of **2-AmpHSO$_4$(II)**.

*3.2 Thermal behaviour*

Crystals of **2-AmpH$_2$PO$_3$** are stable in air up to a melting point of 163 °C. Subsequent DSC recordings showed no thermal effect in the region from -160 °C to 150 °C. Melting of **2-AmpHSO$_4$ (I)** was observed at 154 °C and DSC recordings in the region -160 °C to 135 °C did not show any thermal effects and confirmed that there is no phase transition to the **2-AmpHSO$_4$ (II)** phase. This conclusion agrees with the crystal structure determination results – cooling/heating of the crystal leads only to small thermal contraction/expansion of the unit cell parameters. Unfortunately, due to the minimal number of crystals of **2-AmpHSO$_4$ (II)** polymorph obtained, a study of the thermal behaviour could not be performed for this phase. The melting point of **(2-Amp)$_2$SO$_4$H$_2$O** was determined (visual detection) to be 186 °C. On the other hand, DSC recordings (see Fig. S8, Supplementary material) in the range -160 °C to 180 °C exhibit several weak exothermic effects at 82 °C ($\Delta H$ = 22.4 J/g), 102 °C ($\Delta H$ = 13.0 J/g) and 146 °C ($\Delta H$ = 4.0 J/g). These effects occur only during heating runs and can be attributed to the gradual dehydration of the compound, which starts before melting is observed.

*3.3. Vibrational spectra*

The vibrational spectra of **2-AmpH$_2$PO$_3$**, both polymorphs of **2-AmpHSO$_4$** and (**2-Amp)$_2$SO$_4$H$_2$O**, are depicted in Figs. 7-11. The assignment of the observed bands (see Tables 2-5) is based on the quantum-chemical calculation concerning **2-Amp**(1+) cation (see Table S12, Supplementary material), the detailed analysis of **2-AmpCl½H$_2$O** spectrum (see Table S13 and Fig. S9, Supplementary material) and the literature concerning vibrational manifestation of the involved inorganic anions [54-58]. Following the previously published assignment of vibrational spectra of the **2-Amp** molecule [20], the formation of the **2-Amp**(1+) cation in solution was studied using Raman spectroscopy – see Fig. S10, Supplementary material. The obtained experimental results are consistent with quantum-chemical calculation and last but not least, the confirmation of the formation of the **2-Amp**(1+) cation is also provided by the results of the X-ray structural analysis of **2-AmpCl½H$_2$O**, which crystallized as the only product from the studied equimolar aqueous solution of **2-Amp** and hydrochloric acid.

The assignment of the bands of stretching vibrations of N-H and O-H groups involved in hydrogen bonds is based on correlation curves [59, 60] concerning the position of the vibrational bands and appropriate hydrogen bond lengths. In the case of **2-AmpH$_2$PO$_3$**, the results of solid-state DFT computations for the vibrational spectra assignment were also used - see Chapter 3.5.

The number of expected normal modes of the title crystals was determined by the nuclear site group analysis [61]. The crystals of **2-AmpH$_2$PO$_3$** belong to the space group $P2_1$ ($C_2^2$, No. 4) with 19 atoms ($Z$=2) per asymmetric unit, which form one 2-aminopyrimidinium(1+) cation and one hydrogen phosphite anion. All atoms occupy two-fold Wyckoff positions $a(C_1)$. The polymorphs of **2-AmpHSO$_4$ (I)** and **2-AmpHSO$_4$ (II)** belong to the space group $P2_1/c$ ($C_{2h}^5$, No. 14). The asymmetric units with 19 atoms ($Z$=4) are formed by one 2-aminopyrimidinium(1+) cation and one hydrogen sulfate anion. All atoms occupy four-fold Wyckoff positions $e(C_1)$. The crystals of (**2-Amp)$_2$SO$_4$H$_2$O** belong to the space group $P2_1/n$ ($C_{2h}^5$, No. 14, cell choice 2) with 34 atoms ($Z$=4) per asymmetric unit, which form two 2-aminopyrimidinium(1+) cations, one sulfate anion and one water molecule. All atoms occupy four-fold Wyckoff positions $e(C_1)$.

The symmetry analysis of the optical vibrational modes (see Table 6) gave for **2-AmpH$_2$PO$_3$** 11*A*(IR,Ra) + 10*B*(IR,Ra) representations for the external modes and 45*A*(IR,Ra) + 45*B*(IR,Ra) representations for the internal modes; 12$A_g$(IR,Ra) + 11$A_u$(IR,Ra) + 12$B_g$(IR,Ra) + 10$B_u$(IR,Ra) representations for the external modes and 45$A_g$(IR,Ra) + 45$A_u$(IR,Ra) + 45$B_g$(IR,Ra) + 45$B_u$(IR,Ra) representations for the internal modes for both polymorphs of **2-AmpHSO$_4$**; 24$A_g$(IR,Ra) + 23$A_u$(IR,Ra) + 24$B_g$(IR,Ra) + 22$B_u$(IR,Ra) for the external modes and 78$A_g$(IR,Ra) + 78$A_u$(IR,Ra) + 78$B_g$(IR,Ra) + 78$B_u$(IR,Ra) for the internal modes for (**2-Amp)$_2$SO$_4$H$_2$O**.

*3.3.1. Vibrational bands associated with hydrogen bonds*

The broad strong to medium-intensity bands (ranging from 3400 to 2400 cm$^{-1}$) in the IR spectrum of **2-AmpH$_2$PO$_3$** were assigned to the stretching modes of NH groups participating in the hydrogen bonds of the N-H...O type (donor-acceptor distances of 2.629(3)-2.896(3) Å). Structured medium-intensity IR bands in the 2900-1800 cm$^{-1}$ region are related to the stretching vibrations of OH groups involved in O-H…O hydrogen bonds interconnecting anions with a donor-acceptor distance of 2.572(2) Å. These manifestations form characteristic “ABC bands” with the main maxima at 2550 cm$^{-1}$ (“A band”), 2280 cm$^{-1}$ (“B band”) and 1930 cm$^{-1}$ (“C band”). The “ABC bands” were previously observed in the IR spectra of several hydrogen-bonded systems (including hydrogen phosphites – see refs. [56, 62]), and their explanation continues to attract the attention of spectroscopists (see e.g., refs. [63, 64]).

The manifestations of the NH stretching modes related to N-H...O hydrogen bonds in the crystal structure of **2-AmpHSO$_4$ (I)** (donor-acceptor distances of 2.702(2)-3.009(2) Å) were recorded as strong to medium-intensity IR bands in the 3400-2100 cm$^{-1}$ region. The bands of stretching vibrations of OH groups involved in O-H…N hydrogen bonds (donor-acceptor distance of 2.652(2) Å) can be expected in the 3100-2900 cm$^{-1}$ region.

Structured medium-intensity bands recorded in the IR spectrum of **2-AmpHSO$_4$ (II)** in the 3450-2300 cm$^{-1}$ region correspond to the stretching modes of NH groups participating in the hydrogen bonds of the N-H...O type (donor-acceptor distances of 2.748(1)-3.167(2) Å). The presence of N-H…N hydrogen bonds with a donor-acceptor distance of 2.999(2) Å is reflected by the bands of stretching NH modes located in the 3450-3100 cm$^{-1}$ region. The vibrational bands of OH groups involved in O-H…O hydrogen bonds interconnecting anions with donor-acceptor distance of 2.591(1) Å were recorded in the 2900-2300 cm$^{-1}$ region.

The broad strong to medium-intensity bands ranging from 3400 to 2200 cm$^{-1}$ in the IR

spectrum of **(2-Amp)$_2$SO$_4$H$_2$O** can be assigned to the stretching vibrations of NH groups involved in the hydrogen bonds of the N-H...O type (donor-acceptor distances of 2.637(1)-2.983(1) Å). The bands related to the stretching modes of OH groups participating in O-H…O hydrogen bonds (donor-acceptor distances of 2.733(1)-2.796(1) Å) appear in the 3400-3100 cm$^{-1}$ region.

### *3.3.2. Vibrational bands associated with **2-Amp(1+)** cation*

The IR spectra of the studied salts exhibit, in addition to strong to medium-intensity bands of stretching vibrations of NH$_x$ groups participating in hydrogen bonds (see Chapter 3.3.1.), several characteristic manifestations of **2-Amp**(1+) cations. The maxima of strong bands related to mixed modes of νC-NH$_2$, νrg, δNH$_x$ and δrg vibrations were recorded in the 1690-1645 cm$^{-1}$ region. The manifestations of mixed νrg, δNH$_x$ and δCH vibrations appear as strong bands with maxima in the 1642-1624 cm$^{-1}$ region. The strong to medium-intensity bands associated with mixed modes of δCH, νrg and δNH vibrations were located in the 1371-1344 cm$^{-1}$ region. The maxima of bands related to mixed bending δCH and stretching νrg vibrations were observed in the 1228-1199 cm$^{-1}$ region, and the maxima related to bending δrg vibrations in the 578-575 cm$^{-1}$ region.

The manifestations of stretching C-H modes were recorded in the Raman spectra in the region ranging from 3130 to 3020 cm$^{-1}$. The maxima of the other characteristic bands (medium to strong intensity) are located in the 1645-1625 cm$^{-1}$ region (mixed modes of νrg, δNH$_x$ and δCH vibrations) and 1556-1541 cm$^{-1}$ region (mixed modes of νrg, δCH, δNH$_x$ and δNCN vibrations). The maxima of the bands related to mixed δCH and νrg vibrations were observed in the 1230-1198 cm$^{-1}$ region. Very strong bands recorded in Raman spectra in the 890-870 cm$^{-1}$ region represent manifestations of mixed νC-NH$_2$, νrg, δNH$_x$ and δrg vibrations. The bands with the maxima located in the 584-579 cm$^{-1}$ region belong to characteristic δrg vibrations.

For the assignment of the remaining **2-Amp**(1+) cation vibrational bands in the studied salts, see Tables 2-5.

### *3.3.3. Vibrational bands associated with anions*

The correlation diagrams concerning the internal modes of anions and their labelling for **2-AmpH$_2$PO$_3$**, both polymorphs of **2-AmpHSO$_4$** and **(2-Amp)$_2$SO$_4$H$_2$O** are presented in Tables S14-S16 (Supplementary material).

The medium-intensity band in the Raman spectrum of **2-AmpH$_2$PO$_3$** located at 2412 cm$^{-1}$ and the analogous strong band in the IR spectrum at 2408 cm$^{-1}$ correspond to the characteristic P-H stretching vibration of hydrogen phosphite anion. The bands of δPOH vibrations (overlapping with cation modes) were recorded at 1228 cm$^{-1}$ in both spectra. The manifestations of $\nu_{as}PO_2$ modes, which overlap with δCH and νrg vibrations, were located at 1131 cm$^{-1}$ and 1124 cm$^{-1}$ in the Raman and IR spectra, respectively. The bands of symmetric $\nu_sPO_2$ vibrations were recorded at 1039 cm$^{-1}$ (Raman) and 1043 cm$^{-1}$, with the shoulder at 1054 cm$^{-1}$ (IR). The manifestations of γPH and δPH modes were observed in both spectra at ~1025 cm$^{-1}$ and ~1015 cm$^{-1}$, respectively. The medium-intensity Raman band at 926 cm$^{-1}$ and strong IR band at 927 cm$^{-1}$ correspond to stretching νPO(H) vibrations. The bands recorded in both spectra at ~550 cm$^{-1}$ and ~520 cm$^{-1}$ represent the manifestations of $\rho PO_2$ and $\delta PO_2$ (overlapping with cation vibrations) modes, respectively. The medium-intensity bands at 433 cm$^{-1}$ (Raman) and 426 cm$^{-1}$ (IR) were assigned to δPO(H) vibrations overlapping with δCNC modes.

As expected, the nature of the vibrational manifestations of $HSO_4^-$ anions in both polymorphs of **2-AmpHSO$_4$** is quite similar. However, there are significant differences in the spectra regarding band intensities and shapes (see Fig. 10), which are also affected by the somewhat different overlap with the cation modes. Medium-intensity bands of δSOH vibrations (in the case of phase **I** overlapping with cation δCH, νrg, δNH modes) were recorded in the 1350-1330 cm$^{-1}$ region in the IR spectra. The characteristic intense bands of $\nu_{as}SO_3$ vibrations are present in the IR spectra as asymmetric slightly split (due to overlap with cation modes) doublets located in the 1260-1130 cm$^{-1}$ region. The corresponding Raman bands exhibit much lower intensities, with some exception for the medium-intensity band recorded at 1234 cm$^{-1}$ in the spectrum of phase **I**. The manifestations of the $\nu_sSO_3$ vibrations, which overlap with mixed δrg, νrg, δNH and γCH, γrg modes of cation, were localized in the 1035-990 cm$^{-1}$ region as strong bands in both spectra of both polymorphs. A pair of strong bands in the 860-800 cm$^{-1}$ region in the IR spectra (weak bands in the Raman spectra) represent manifestations of νSO(H) stretching vibrations. The bands recorded in all available spectra in the 610-570 cm$^{-1}$ region were assigned to anion bending modes $\delta SO_3$ overlapping with cation δrg vibrations. The manifestations of $\delta(H)OSO_3$ bending vibrations (overlapping with δCNC cation modes) are represented by the bands present in the 440-415 cm$^{-1}$ region of both polymorphs.

Two strong bands recorded in the IR spectrum of **(2-Amp)$_2$SO$_4$H$_2$O** at 1132 and 1109 cm$^{-1}$ represent originally triply degenerate $\nu_3SO_4$ vibrations. The corresponding weak Raman band (overlapping with cation deformation modes) was located at 1120 cm$^{-1}$. The strong Raman band at 974 cm$^{-1}$ (medium-intensity IR band at 979 cm$^{-1}$) corresponds to symmetric stretching $\nu_1SO_4$ vibration. The manifestations of originally triply degenerate $\nu_4SO_4$ vibrations (partially overlapping with δrg mode) were recorded in the 645-590 cm$^{-1}$ region. Two bands present in both spectra at 446 and 434 cm$^{-1}$ (partially overlapping with δCNC mode) belong to originally doubly degenerate $\nu_2SO_4$ vibrations.

### *3.4. UV-Vis-NIR spectrum of* ***2-AmpH$_2$PO$_3$***

The spectrum in Fig. 12 represents the UV-Vis-NIR absorption of the **2-AmpH$_2$PO$_3$** single-crystal plate. The spectrum shows the excellent optical transparency of this polar crystal from the NIR region to the edge of the UV region (down to at least 370 nm).

### *3.5. Solid-state computation results for* ***2-AmpH$_2$PO$_3$***

To find a suitable computational method for modelling the spectral and optical properties of crystalline **2-AmpH$_2$PO$_3$**, we first carried out a detailed comparison of the calculated vibrational bands with the recorded Raman and IR spectra. The presented graphical comparisons (Figs. S11 and S12, Supplementary material) confirm the considerable influence of the computational method on the agreement of the calculated band positions and intensities with the experimental results.

The comparison of the calculated modes and recorded Raman spectra (see Fig. S11, Supplementary material) shows mainly differences in the intensities of the lattice modes and the positions and/or intensities of the bands of stretching vibrations of the C-H, N-H and O-H groups. The results obtained by the “B3LYP” method exhibit an overestimation of the lattice modes intensities and an overestimation of the positions of stretching modes of the O-H and N-H groups involved in the O-H…O and N-H…O hydrogen bonds (2800-2600 cm$^{-1}$ region). The results obtained by the “B3LYP-advanced” method exhibit an overestimation of the intensities of stretching modes of the C-H, N-H, O-H and P-H groups (3300-2400 cm$^{-1}$ region) and also an overestimation of the positions of stretching modes of the O-H and N-H groups involved in the O-H…O and N-H…O hydrogen bonds (2800-2600 cm$^{-1}$ region). On the other hand, this method generally provides the best match for the Raman spectrum in the fingerprint region (see Fig. S11, Supplementary material). The results of the “PBESOL0” method are characterized by the excellent agreement obtained for the positions of the stretching modes of the C-H, N-H, O-H and P-H groups (3300-2400 cm$^{-1}$ region) and an overestimation of lattice modes intensities.

A comparison of the calculated modes and recorded IR spectra (see Fig. S12, Supplementary material) shows somewhat lower agreement than in the case of Raman spectra (see Fig. S11, Supplementary material). In general, an underestimation of the intensity of the P-H stretching modes (2400 $cm^{-1}$) and low-frequency modes can be observed for all computational methods. Some problems with the position of the calculated modes are also evident for the most intense bands in the recorded IR spectrum in the 1100-900 $cm^{-1}$ region (mainly concerning vibrational manifestations of the anion). It is again evident that the best agreement for the calculated stretching vibrations positions of the C-H, N-H, O-H and P-H groups was obtained with the "PBESOL0" method. Within the trio of computational methods compared, this method also provides the best match for the IR spectrum in the fingerprint region (see Fig. S12, Supplementary material).

The solid-state DFT computational studies based on crystal structure determination concluded that the **2-AmpH$_2$PO$_3$** crystal is optically anisotropic biaxial negative material. The values of the calculated refractive indices and static ($\lambda=\infty$) nonlinear susceptibility tensor $\chi^{(2)}$ components are presented in Table S17, Supplementary material. The results indicate a general underestimation of the refractive indices values compared to the experimental data (roughly extrapolated for $\lambda=\infty$) - see Chapter 3.6. The resulting set of refractive indices obtained by the "B3LYP" and "PBESOL0" methods seems to be closest to the expected values derived from experimental measurements. When comparing the results of the $\chi^{(2)}$ components calculations, the largest differences are seen for $\chi^{(2)}_{xxy}$ and $\chi^{(2)}_{xyz}$ components obtained by the "B3LYP" and "PBESOL0" methods.

### *3.6. Linear optical properties of* ***2-AmpH$_2$PO$_3$***

From the determination of the refractive indices and their wavelength dependence of crystals of **2-AmpH$_2$PO$_3$**, anomalous dispersion of the refractive indices becomes evident. Sellmeier coefficients of the three principal refractive indices are given in Table 7, and the measured (and corrected) refractive indices, together with the fitted Sellmeier functions, are plotted in Fig. 13. While the birefringence $\Delta n_1 = n_2^0 - n_1^0$ shows only little wavelength dependence, the refractive indices $n_2^0$ and $n_3^0$ approach for increasing wavelengths and become equal at 1065 nm, thus resulting in optically uniaxial behaviour at this wavelength (point of uniaxiality). At wavelengths > 1065 nm $n_3^0$, which is the largest principal refractive index $n_\gamma$ of **2-AmpH$_2$PO$_3$** below 1065 nm, changes to the medium principal refractive index $n_\beta$, while $n_2^0$ becomes $n_\gamma$.

This anomalous dispersion behaviour gives rise to anomalous interference colours of crystal samples, which are visible in a polarization microscope between crossed polarizers.

Due to the anomalous dispersion of the refractive indices, a strong wavelength-dependent variation is also found for the angle between the optic axes $2V_\gamma$, which increases from ~125° at 400 nm to 180° at 1065 nm at the point of uniaxiality (i.e., $2V_\alpha = 180° - 2V_\gamma = 0$) of the crystal, see Fig. 14a. This dispersion of the optic axes is clearly visible in an interference figure (conoscopic illumination in a polarizing microscope) between crossed polarizers along the acute bisectrix, which is the direction of $\boldsymbol{e}_1^0$ in the case of **2-AmpH$_2$PO$_3$**, see inset of Fig. 14a. On the other hand, the orientation of the principal axes of the optical indicatrix of **2-AmpH$_2$PO$_3$** is rather fixed with respect to the frame of the axes $\{\boldsymbol{e}_\mathrm{i}\}$ and of the crystallographic axes, with a small variation between $\psi = 42.6°$ at 400 nm and $\psi = 41.5°$ at 1080 nm, see Fig. 14b.

From the refractive indices and their dispersion possibilities for collinear phase matching for SHG were analysed within the transparency range of **2-AmpH$_2$PO$_3$**. The crystals allow both, type I and type II phase matching (where type I refers to *ss-f* interaction, with *s* = slow wave and *f* = fast wave, and type II to *sf-f* interaction). A stereographic projection (i.e. a Hobden plot [65]) of collinear SHG phase matching loci for selected wavelengths in crystals of **2-AmpH$_2$PO$_3$** is given in Fig. 15.

### *3.7. SHG measurements and determination of SHG tensor coefficients of **2-AmpH$_2$PO$_3$***

The performed SHG powder measurements for the non-centrosymmetric **2-AmpH$_2$PO$_3$** samples yielded (powdered KDP used as reference material) a relative SHG efficiency $d_{(\mathbf{2\text{-}AmpH_2PO_3})} = 1.18\ d_{(\mathbf{KDP})}$ for 800 nm laser irradiation. The phase purity of the polycrystalline samples prepared was controlled by powder X-ray diffraction (diffraction data are included in Table S2, Supplementary material). A qualitative check for the phase-matchability of the compound was performed using particle-size-dependent measurements. The particle-size dependence of the SHG signal (see Fig. 16) indicates the phase-matchability of **2-AmpH$_2$PO$_3$**. This result confirms the expected SHG phase matching calculated based on the determined refractive indices (see Chapter 3.6.). Moreover, the optical damage threshold was determined for the polycrystalline sample (150-100 m fraction) using 800 nm and 1000 nm laser lines. The studied material was stable up to the highest achievable peak laser power for our experimental setup, which was 220 TWm$^{-2}$ and 70 TWm$^{-2}$ for 800 nm and 1000 nm irradiation, respectively.

The promising results from the phase matching analysis and the SHG powder measurements of **2-AmpH$_2$PO$_3$**, together with the availability of single crystals of optical quality and sufficient size, gave the motivation to determine preliminarily coefficients $d_{ijk}^{SHG}$ of the SHG tensor using the Maker fringe technique. In point group 2, the SHG tensor possesses eight independent coefficients, e.g. [66] (here and in the following contracted Voigt notation is used):

$$\begin{pmatrix} 0 & 0 & 0 & d_{14} & 0 & d_{16} \\ d_{21} & d_{22} & d_{23} & 0 & d_{25} & 0 \\ 0 & 0 & 0 & d_{34} & 0 & d_{36} \end{pmatrix}$$

If Kleinman symmetry assumption [67] is valid in point group 2, four independent tensor coefficients would remain, with $d_{21} = d_{16}$, $d_{22}$, $d_{23}$ = $d_{34}$ and $d_{14} = d_{25} = d_{36}$.

For **2-AmpH$_2$PO$_3$** $d_{21}$ = 0.25 pm/V, $d_{22}$ = 0.85 pm/V, $d_{23}$ = 0.15 pm/V and $d_{14}$ = 0.25 pm/V. While for $d_{14}$ Kleinman symmetry holds (and thus $d_{14} \approx d_{25} \approx d_{36}$), this is not the case for $d_{21}$ and $d_{23}$. Here, the corresponding coefficients $d_{16}$ and $d_{34}$ are approximately more than two times smaller. The coefficient $d_{22}$ is remarkably large and amounts more than two times $d_{36}$ of KDP (with $d_{36}$(KDP) = 0.39 pm/V [66, 68]); unfortunately, $d_{22}$ is not suitable for phase-matching geometries in the principal planes $(\boldsymbol{e}_i^0, \boldsymbol{e}_j^0)$.

## 4. Conclusion

Low-temperature crystal structures have been refined for four inorganic salts of 2-aminopyrimidine - i.e. **2-AmpH$_2$PO$_3$** (space group $P2_1$), **(2-Amp)$_2$SO$_4$H$_2$O** (space group $P2_1/n$) and two polymorphs of **2-AmpHSO$_4$** (both phases **I** and **II** belong to space group $P2_1/c$). The results obtained, which improved the conclusions of previous structural studies [23, 50, 51], exhibit the crucial role of hydrogen bonding in the crystal packing of molecular crystals. The studied polymorphs of **2-AmpHSO$_4$** with the same crystal symmetry and similar unit cell parameters differ entirely in the overall crystal packing. The crystal structure of **2-AmpHSO$_4$ (I)** consists of alternating **2-Amp**(1+) cations and hydrogen sulfate anions interconnected by an extensive network of N-H…O hydrogen bonds in contrast to the crystal structure of the **2-AmpHSO$_4$ (II**) polymorph, which is based on chains of hydrogen sulfate anions (formed via O-H…O hydrogen bonds) interconnected by N-H…O interactions with centrosymmetric dimers of **2-Amp**(1+) cations.

The vibrational spectra of the title compounds were recorded and successfully assigned by combining quantum-chemical computations, complemented by the Potential Energy

Distribution analysis, concerning isolated **2-Amp**(1+) cation and the results of the correlation analysis involving hydrogen phosphite, hydrogen sulfate and sulfate anions. Recorded spectra of polycrystalline samples do not exhibit the expected level of factor group splitting, on the other hand, the comparison of the spectra of two **2-AmpHSO$_4$** polymorphs confirmed the unique sensitivity of vibrational spectroscopy in distinguishing closely related crystalline phases. The assignment of the spectra of polar **2-AmpH$_2$PO$_3$** crystals has been extended by the utilization of solid-state DFT computations, leading to the conclusion that of the trio of computational approaches considered, the “PBESOL0” method provides the best agreement with the experimental data.

For **2-AmpH$_2$PO$_3$**, the only non-centrosymmetric material among the salts studied, large single crystals were successfully grown, and the quality and size of the crystals achieved allowed precise measurements of linear optical properties. In addition, a study of SHG was performed using the Maker fringes method. The crystals show an interesting case of anomalous dispersion of the refractive indices, resulting in a point of uniaxiality. Furthermore, the refractive indices and their dispersion allow phase-matching for collinear SHG processes of both, type I and type II in a broad wavelength range.

Powder measurements confirmed significant SHG efficiency of **2-AmpH$_2$PO$_3$** ($d_{(\mathbf{2\text{-}AmpH_2PO_3})}$ = 1.18 $d_{(\mathbf{KDP})}$ for 800 nm laser irradiation) and excellent resistance to optical damage (resisting at least 220 TWm$^{-2}$ and 70 TWm$^{-2}$ for 800 nm and 1000 nm irradiation, respectively).

Last but not least, the potential of the **2-AmpH$_2$PO$_3$** crystals as a nonlinear optical material for combined $\chi(2) + \chi(3)$ processes should be mentioned. The sharp, intense bands recorded in the Raman spectrum at 876 and 2412 cm$^{-1}$ (representing manifestations of symmetric vibrational modes of the cation and stretching P-H modes of the anion, respectively) could also be active in stimulated Raman scattering (SRS) processes, including cascaded $\chi(2)\leftrightarrow\chi(3)$ processes, similar to that observed in the case of guanylurea(1+) hydrogen phosphite (GUHP) [27, 28].

## CRediT authorship contribution statement

**Irena Matulková**: Writing – review & editing, Writing – original draft, Conceptualization, Investigation, Visualization, Validation. **Ladislav Bohatý**: Writing – review & editing, Writing – original draft, Investigation, Methodology, Conceptualization, Visualization. **Petra Becker**: Writing – review & editing, Writing – original draft, Investigation, Methodology,

Conceptualization, Visualization. **Ivana Císařová**: Investigation, Writing – original draft. **Róbert Gyepes**: Investigation, Writing – original draft. **Michaela Fridrichová**: Investigation, Writing – original draft. **Jan Kroupa**: Investigation, Writing – original draft. **Petr Němec**: Investigation, Writing – original draft, Funding acquisition. **Ivan Němec**: Conceptualization, Methodology, Writing – original draft, Supervision, Investigation, Writing – review & editing, Validation, Visualization, Funding acquisition.

## Declaration of competing interest

The authors declare that they have no known competing financial interests or personal relationships that could have appeared to influence the work reported in this paper.

## Data availability

Data will be made available on request.

## Acknowledgements

We gratefully acknowledge the assistance provided by the Advanced Multiscale Materials for Key Enabling Technologies project, supported by the Ministry of Education, Youth, and Sports of the Czech Republic. Project No. CZ.02.01.01/00/22_008/0004558 (https://raid.org/10.82841/58215ce5), Co-funded by the European Union. This work was also supported by the CUCAM Centre of Excellence (OP VVV “Excellent Research Teams” project No. CZ.02.1.01/0.0/0.0/15_003/0000417). CzechNanoLab project LM2023051 funded by MEYS CR is gratefully acknowledged for the financial support of the Measurements at LNSM Research Infrastructure.
The authors gratefully thank Dr. Carlos V. Melo for editing the manuscript.

## Appendix A. Supplementary data

Supplementary data to this article can be found online at http:

**Table 1a.** Basic crystallographic data and structure refinement details for **2-AmpH$_2$PO$_3$** and (**2-Amp)$_2$SO$_4$H$_2$O** crystals.

| Identification code | **2-HAMPH$_2$PO$_3$** | **(2-HAMP)$_2$SO$_4$H$_2$O** |
|---|---|---|
| Empiric formula | $C_4 H_8 N_3 O_3 P$ | $C_8 H_{14} N_6 O_5 S$ |
| Formula weight | 177.10 | 306.31 |
| Temperature (K) | 150(2) | 150(2) |
| $a$ (Å) | 4.4925(2) | 13.8890(2) |
| $b$ (Å) | 17.2391(8) | 6.54100(10) |
| $c$ (Å) | 4.7606(2) | 14.0780(3) |
| $\alpha$ (°) | 90 | 90 |
| $\beta$ (°) | 95.070(3) | 91.0510(11) |
| $\gamma$ (°) | 90 | 90 |
| Volume (Å$^3$) | 367.25(3) | 1278.74(4) |
| $Z$ | 2 | 4 |
| Calculated density (Mg/m$^3$) | 1.602 | 1.591 |
| Crystal system | Monoclinic | Monoclinic |
| Space group | $P2_1$ | $P2_1/n$ |
| Absoption coeficient (mm$^{-1}$) | 0.336 | 0.286 |
| $F$(000) | 184 | 640 |
| Crystal size (mm) | 0.3 x 0.3 x 0.2 | 0.40 x 0.37 x 0.37 |
| Diffractometer and radiation | Nonius Kappa CCD, Mo $\lambda$ = 0.71073 Å | |
| Scan technique | $\omega$ and $\psi$ scans to fill the Ewald sphere | |
| Completeness to $\theta$ | 27.47 99.6 % | 27.48 99.9 % |
| Range of h, k and l | -5 → 5, -22 → 22, -6 → 6 | -17 → 18, -8 → 8, -18 → 18 |
| $\theta$ Range for data collection (°) | 2.36 to 27.47 | 2.04 to 27.48 |
| Reflection collected/unique ($R_{int}$) | 8445 / 1669 (0.0195) | 24932 / 2928 (0.0161) |
| No. of observed reflection | 1636 | 2677 |
| Criterion for observed reflection | $I > 2\sigma(I)$ | |
| Absorption correction | multi-scan | multi-scan |
| Function minimized | $\Sigma$ w$(F_o^2 - F_c^2)^2$ | |
| Parameters refined | 100 | 181 |
| $R$; w$R$ ($I$>2$\sigma$($I$)) | 0.0240; 0.0657 | 0.0291; 0.0820 |
| $R$; w$R$ (all data) | 0.0249; 0.0648 | 0.0321; 0.0844 |
| Value of $S$ | 1.103 | 1.053 |
| Max. and min. heights in final $\Delta\rho$ map (eÅ$^{-3}$) | 0.157 and -0.260 | 0.223 and -0.519 |
| Weighting scheme | w = $[\sigma^2(F_o^2) + aP^2 + bP]^{-1}$ $P = (F_o^2 + 2F_c^2)/3$ | |
| | $a$ = 0.0364 | $a$ = 0.0467 |
| | $b$ = 0.0788 | $b$ = 0.5630 |

**Table 1b.** Basic crystallographic data and structure refinement details for **2-AmpHSO4 (I)** and **2-AmpHSO4(II)** polymorphs.

| Identification code | **2-HAMPHSO$_4$ (I)** | **2-HAMPHSO$_4$ (II)** |
|---|---|---|
| Empiric formula | $C_4 H_7 N_3 O_4 S$ | $C_8 H_{14} N_6 O_5 S$ |
| Formula weight | 193.19 | 193.19 |
| Temperature (K) | 150(2) | 120(2) |
| *a* (Å) | 5.86700(10) | 8.2198(4) |
| *b* (Å) | 8.2530(3) | 5.1376(2) |
| *c* (Å) | 15.5520(5) | 18.5781(9) |
| $\alpha$ (°) | 90 | 90 |
| $\beta$ (°) | 106.173(2) | 112.688(2) |
| $\gamma$ (°) | 90 | 90 |
| Volume (Å$^3$) | 723.23(4) | 723.84(6) |
| *Z* | 4 | 4 |
| Calculated density (Mg/m$^3$) | 1.774 | 1.773 |
| Crystal system | Monoclinic | Monoclinic |
| Space group | $P2_1/c$ | $P2_1/n$ |
| Absoption coeficient (mm$^{-1}$) | 0.427 | 0.426 |
| *F*(000) | 400 | 400 |
| Crystal size (mm) | 0.60 x 0.30 x 0.13 | 0.65 x 0.10 x 0.07 |
| Diffractometer and radiation | Nonius Kappa CCD, Mo $\lambda$ = 0.71073 Å | Bruker D8 Venture, Mo λ = 0.71073 Å |
| Scan technique | $\omega$ and $\psi$ scans to fill the Ewald sphere | |
| Completeness to $\theta$ | 27.49 100 % | 27.48 99.8 % |
| Range of h, k and l | -7 → 7, -10 → 10, -20 → 20 | -10 → 10, -6 → 6, -24 → 24 |
| $\theta$ Range for data collection (°) | 2.73 to 27.49 | 2.67 to 27.48 |
| Reflection collected/unique ($R_{int}$) | 14053 / 1663 (0.033) | 43457 / 1664 (0.0082) |
| No. of observed reflection | 1428 | 1648 |
| Criterion for observed reflection | $I > 2\sigma(I)$ | |
| Absorption correction | multi-scan | multi-scan |
| Function minimized | $\Sigma$ w$(F_o^2 - F_c^2)^2$ | |
| Parameters refined | 110 | 109 |
| *R;* w*R* ($I > 2\sigma(I)$) | 0.0291; 0.0795 | 0.0241; 0.0661 |
| *R;* w*R* (all data) | 0.0367; 0.0833 | 0.0242; 0.0661 |
| Value of *S* | 1.087 | 1.083 |
| Max. and min. heights in final Δρ map (eÅ$^{-3}$) | 0.197 and -0.491 | 0.353 and -0.434 |
| Weighting scheme | w = $[\sigma^2(F_o^2) + aP^2 + bP]^{-1}$ $P = (F_o^2 + 2F_c^2)/3$ | |
| | $a$ = 0.0464 | $a$ = 0.0293 |
| | $b$ = 0.2335 | $b$ = 0.6213 |

**Table 2.** Recorded FTIR and Raman maxima ($cm^{-1}$) of **2-AmpH$_2$PO$_3$** and their assignment.

| FTIR | Raman | Assignment | FTIR | Raman | Assignment |
|---|---|---|---|---|---|
| 76m | 81m | External modes | 1228s | 1228m | δPOH, δCH, νrg |
| 116m | 96s | | 1302w | 1304m | δCH, νrg, $\delta NH_x$ |
| 144m | 135s | | 1359s | 1361m | δCH, νrg, δNH |
| 167mb | | | 1434s | | $\nu C\text{-}NH_2$, νrg, $\delta NH_x$, δCH |
| 202m | | γrg | 1482mb | 1494w | δCH, νrg, $\delta NH_x$ |
| 214mb | 215m | | | 1509w | ? |
| 402m | 404m | γrg, γCH, $\gamma NH_x$ | 1535sh | 1534sh | νrg, δCH, $\delta NH_x$, δNCN |
| 426mb | 433mb | δPO(H), δCNC | 1542m | 1542m | |
| 458m | 459m | δCNC | 1560m | 1575w | |
| 475w | 476m | | 1625s | 1626m | νrg, $\delta NH_x$, δCH |
| 523w | 523w | $\delta PO_2$, γrg, γCH, $\gamma NH_x$ | 1651s | | $\nu C\text{-}NH_2$, νrg, $\delta NH_x$, δrg |
| 555s | 549w | $\rho PO_2$ | 1690s | | |
| 578m | 579s | δrg | 1930mb | | νO-H(…O) |
| 637m | 638s | | 1981m | | |
| 688mb | | γNH, $\tau NH_2$ | 2061m | | |
| 760wb | | ? | 2280mb | | |
| 785w | | γrg, $\gamma CN_3$ | 2408s | 2412m | νPH |
| 801m | 802w | γCH | 2550mb | | νO-H(…O), νN-H(…O) |
| 855wb | | ? | 2703m | | |
| 874w | 876vs | $\nu_s$rg, $\delta_s$rg, $\nu C\text{-}NH_2$ | 2767m | | |
| 927s | 926m | νPO(H) | 2808m | | |
| 982s | 985m | δrg, νrg, δNH | 2980m | 2980vw | |
| 1015s | 1017m | δPH | 3027sh | 3026w | νCH, νN-H(…O) |
| 1022s | 1026m | γPH | | 3038w | |
| 1043s | 1039s | $\nu_s PO_2$ | 3064s | 3071w | |
| 1054sh | | | 3121m | 3123m | |
| 1077sh | 1082s | $\rho NH_2$, νrg, δrg | 3145m | | νN-H(…O) |
| 1124s | 1131m | $\nu_{as} PO_2$, δCH, νrg | 3246w | | |
| 1199s | 1200m | δCH, νrg | 3362w | | |

*Note:* Abbreviations and symbols: vs, very strong; s, strong; m, medium; w, weak; b, broad; sh, shoulder; rg, ring; ν, stretching; δ, deformation or in-plane bending; γ, out-of-plane bending; ρ, rocking; τ, torsion; $_s$, symmetric; $_{as}$, antisymmetric.

**Table 3.** Recorded FTIR and Raman maxima ($cm^{-1}$) of **2-AmpHSO$_4$ (I)** and their assignment.

| FTIR | μ-FTIR | Raman | Assignment | FTIR | μ-FTIR | Raman | Assignment |
|---|---|---|---|---|---|---|---|
| | | 68m | External modes | 1157s | 1157s | 1158w | $\nu_{as}SO_3$ |
| | | 118s | | 1212s | 1215s | | δCH, νrg |
| | | 145s | | | | 1222m | $\nu_{as}SO_3$, δCH, νrg |
| | | 160s | | 1246s | 1233s | 1234m | $\nu_{as}SO_3$ |
| | | 180sh | | 1286sh | 1291w | 1291w | δCH, νrg, $\delta NH_x$ |
| | | 222w | γrg | 1344m | 1346m | 1348vw | δCH, νrg, δNH, δSOH |
| 406w | | 404m | γrg, γCH, $\gamma NH_x$ | 1432sh | 1436sh | 1430vw | $\nu C\text{-}NH_2$, νrg, $\delta NH_x$, δCH |
| 424w | | 424m | $\delta(H)OSO_3$, δCNC | 1485m | 1489w | 1493w | δCH, νrg, $\delta NH_x$ |
| 436s | | 438m | | 1552m | 1554w | 1556m | νrg, δCH, $\delta NH_x$, δNCN |
| 456s | | 457m | δCNC | 1620s | 1621w | 1625m | $\delta NH_x$, νrg |
| 513s | | 515w | γrg, γCH, $\gamma NH_x$ | 1642m | | 1645w | νrg, $\delta NH_x$, δCH |
| 575s | | 578s | $\delta SO_3$, δrg | 1664m | | 1666w | $\nu C\text{-}NH_2$, νrg, $\delta NH_x$, δrg |
| 591s | | 592m | | 1675sh | 1674w | 1690w | |
| 627m | | | δrg | 1846w | 1847w | | Combination modes |
| 656m | | 656m | | 1921w | 1922w | | |
| 688m | 683w | | γNH, $\tau NH_2$ | | 2470mb | | νN-H(…O) |
| 779m | 776w | | γrg, $\gamma CN_3$ | 2500mb | | | |
| 808sh | | | γCH | | 2683mb | | |
| 821s | 820m | 819w | νSO(H) | | 2798mb | | |
| 852s | 851s | 852w | | 2933m | 2938w | | νO-H(…N), νN-H(…O) |
| 882m | 882w | 887vs | $\nu_s rg$, $\delta_s rg$, $\nu C\text{-}NH_2$ | | | 3055w | νCH, νO-H(…N), νN-H(…O) |
| 932m | 928w | 934w | ? | 3103m | 3114s | 3116w | |
| 998s | 1001s | 1009s | $\nu_s SO_3$, δrg, νrg, δNH | 3138m | 3150s | | νN-H(…O) |
| 1035s | 1034s | 1034s | $\nu_s SO_3$, γCH, γrg | 3295m | | | |
| 1082m | 1083w | 1083s | $\rho NH_2$, νrg, δrg | | 3327mb | | |
| 1121m | 1124w | 1120m | δCH, $\delta NH_x$ | | | | |

**Table 4.** Recorded FTIR and Raman maxima ($cm^{-1}$) of **2-AmpHSO4 (II)** and their assignment.

| μ-FTIR | Raman | Assignment | μ-FTIR | Raman | Assignment |
|---|---|---|---|---|---|
| | 90vs | External modes | | 1207vw | δCH, νrg |
| | 97sh | | 1225s | 1224w | $\nu_{as}SO_3$, δCH, νrg |
| | 136vs | | 1235sh | 1249w | |
| | 158sh | | 1302m | 1300w | δCH, νrg, $\delta NH_x$ |
| | 215w | γrg | 1328mb | | δSOH |
| | 399vw | γrg, γCH, $\gamma NH_x$ | 1352m | 1359w | δCH, νrg, δNH |
| | 421m | $\delta(H)OSO_3$, δCNC | 1431w | 1428w | $\nu C\text{-}NH_2$, νrg, $\delta NH_x$, δCH |
| | 437m | | 1469m | | δCH, νrg, $\delta NH_x$ |
| | 451sh | δCNC | | 1481w | |
| | 574w | $\delta SO_3$, δrg | 1548m | 1542w | νrg, δCH, $\delta NH_x$, δNCN |
| | 584m | | 1617sh | 1622w | $\delta NH_x$, νrg |
| | 609w | $\delta SO_3$ | 1633s | 1635 sh | νrg, $\delta NH_x$, δCH |
| | 647m | δrg | | 1660vw | $\nu C\text{-}NH_2$, νrg, $\delta NH_x$, δrg |
| 706mb | | γNH, $\tau NH_2$ | 1671s | | |
| 777s | 783w | γrg, $\gamma CN_3$ | 2537mb | | νO-H(…O), νN-H(…O) |
| 813s | 810vw | νSO(H) | 2710m | | |
| 846s | 847w | | 2757mb | | |
| 872sh | 879s | $\nu_s$rg, $\delta_s$rg, $\nu C\text{-}NH_2$ | 2810mb | | |
| 905wb | | ? | 2937m | | νN-H(…O) |
| 1000s | 995sh | $\nu_s SO_3$, δrg, νrg, δNH | 3036m | 3043w | νCH, νN-H(…O) |
| 1010s | 1013s | | 3109m | 3110m | |
| 1038m | 1046w | γCH, γrg | 3140sh | | νN-H(…O), νN-H(…N) |
| 1074m | 1081m | $\rho NH_2$, νrg, δrg | 3279m | | |
| 1127sh | 1135w | δCH, $\delta NH_x$ | 3342m | | |
| 1149s | 1166vw | $\nu_{as}SO_3$ | 3397m | | |

**Table 5.** Recorded FTIR and Raman maxima ($cm^{-1}$) of (**2-Amp)$_2$SO$_4$H$_2$O** and their assignment.

| FTIR | Raman | Assignment | FTIR | Raman | Assignment |
|---|---|---|---|---|---|
| | 65vs | External modes | | 1009vw | γCH, γrg |
| | 84vs | | 1035m | | |
| | 101s | | 1050m | 1053w | $\rho NH_2$, νrg, δrg |
| | 172w | | 1081sh | 1081m | |
| | 203w | γrg | 1109s | | $\nu_3SO_4$ |
| | 216w | | | 1120w | $\nu_3SO_4$, δCH, $\delta NH_x$ |
| | 392w | γrg, γCH, $\gamma NH_x$ | 1132s | | $\nu_3SO_4$ |
| | 402w | | 1146sh | 1148w | |
| 434w | 434w | $\nu_2SO_4$, δCNC | | 1198vw | δCH, νrg |
| 446vw | 446w | $\nu_2SO_4$ | 1225m | 1230m | |
| 467vw | 463w | δCNC | 1295w | 1293w | δCH, νrg, $\delta NH_x$ |
| 507vw | 507w | γrg, γCH, $\gamma NH_x$ | 1347m | 1349w | δCH, νrg, δNH |
| 517vw | 517w | | 1371m | 1373w | |
| 577m | 578s | δrg | 1436m | | $\nu C\text{-}NH_2$, νrg, $\delta NH_x$, δCH |
| 597w | 597vw | $\nu_4SO_4$ | 1470m | 1476m | δCH, νrg, $\delta NH_x$ |
| 617m | | | 1540m | 1541m | νrg, δCH, $\delta NH_x$, δNCN |
| 624m | 626sh | | 1624s | 1628m | νrg, $\delta NH_x$, δCH |
| 640w | 637w | $\nu_4SO_4$, δrg | 1645s | | $\nu C\text{-}NH_2$, νrg, $\delta NH_x$, δrg, $\delta H_2O$ |
| 655wb | | δrg | 1685sh | 1675w | $\nu C\text{-}NH_2$, νrg, $\delta NH_x$, δrg |
| 713wb | | γNH, $\tau NH_2$ | 1934wb | | Combination modes |
| 785m | | γrg, $\gamma CN_3$ | 2008wb | | |
| 805m | 801w | γCH | 2500mb | | νN-H(…O) |
| 872w | 877vs | $\nu_s$rg, $\delta_s$rg, $\nu C\text{-}NH_2$ | 3031sb | 3036w | νCH, νN-H(…O) |
| 941w | 942sh | ? | 3109m | 3090w | |
| 955m | | γCH | 3126m | 3129w | |
| 969m | | | 3150m | | νO-H(…O), νN-H(…O) |
| 979m | 974s | $\nu_1SO_4$ | 3253mb | | |

**Table 6**. The results of the nuclear site group analysis for **2-AmpH$_2$PO$_3$**, both polymorphs of **2-AmpHSO$_4$** and (**2-Amp)$_2$SO$_4$H$_2$O**.

| Compound | | **2-AmpH$_2$PO$_3$** $P2_1$ ($C_2^2$) | | **2-AmpHSO$_4$** $P2_1/c$ ($C_{2h}^5$) | | | | **(2-Amp)$_2$SO$_4$H$_2$O** $P2_1/n$ ($C_{2h}^5$) | | | |
|---|---|---|---|---|---|---|---|---|---|---|---|
| Representations | | $A$ | $B$ | $A_g$ | $A_u$ | $B_g$ | $B_u$ | $A_g$ | $A_u$ | $B_g$ | $B_u$ |
| External modes | Acoustic | 1 | 2 | 0 | 1 | 0 | 2 | 0 | 1 | 0 | 2 |
| | Translational | 5 | 4 | 6 | 5 | 6 | 4 | 12 | 11 | 12 | 10 |
| | Librational | 6 | 6 | 6 | 6 | 6 | 6 | 12 | 12 | 12 | 12 |
| Internal modes | | 45 | 45 | 45 | 45 | 45 | 45 | 78 | 78 | 78 | 78 |
| Total | | 57 | 57 | 57 | 57 | 57 | 57 | 102 | 102 | 102 | 102 |
| Activity | IR | z | x, y | | z | | x, y | | z | | x, y |
| | Raman | $x^2, y^2, z^2$, xy | xz, yz | $x^2, y^2, z^2$, xy | | xz, yz | | $x^2, y^2, z^2$, xy | | xz, yz | |

**Table 7**. Sellmeier coefficients of the principal refractive indices of **2-AmpH2PO3**, calculated from measured and corrected refractive index data using equation (1). λ is in μm. $\xi^2$ is the sum of the squares of the residuals.

| | P1 | P2 | P3 | P4 | $\xi^2$ |
|---|---|---|---|---|---|
| $n_1^0$ | 2.1561(7) | 0.0131(3) | 0.035(2) | 0.0106(5) | $1.2 \cdot 10^{-9}$ |
| $n_2^0$ | 2.7464(5) | 0.0320(2) | 0.0404(4) | 0.0190(3) | $7.9 \cdot 10^{-10}$ |
| $n_3^0$ | 2.753(2) | 0.0363(6) | 0.0970(1) | 0.030(2) | $2.9 \cdot 10^{-8}$ |

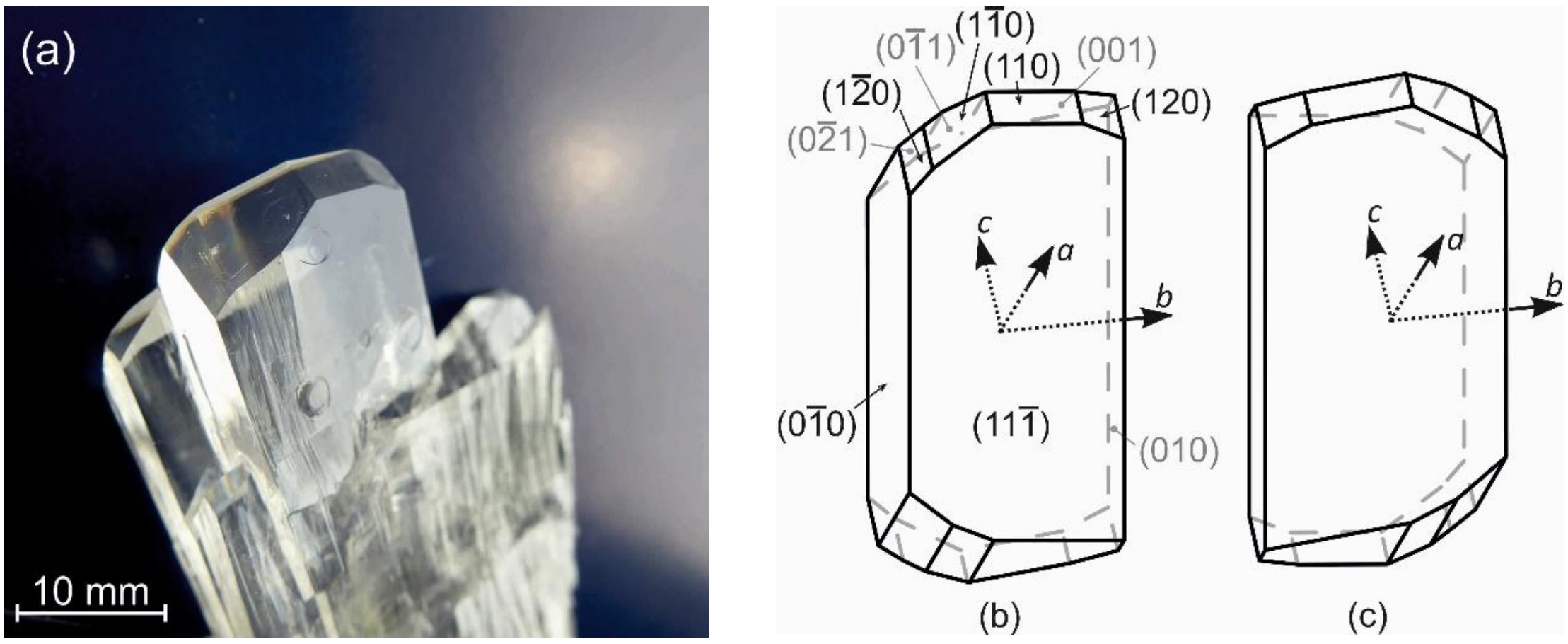

**Fig. 1.** (a) Example of grown crystals of **2-AmpH2PO3**. (b) Typical morphology of the grown crystals of **2-AmpH2PO3**, shown in (a) and used in the present work, with indicated crystallographic axes and face indices. Note that in the point group 2 of **2-AmpH2PO3** crystals of opposite handedness also occur, in (c) their typical morphology is given.

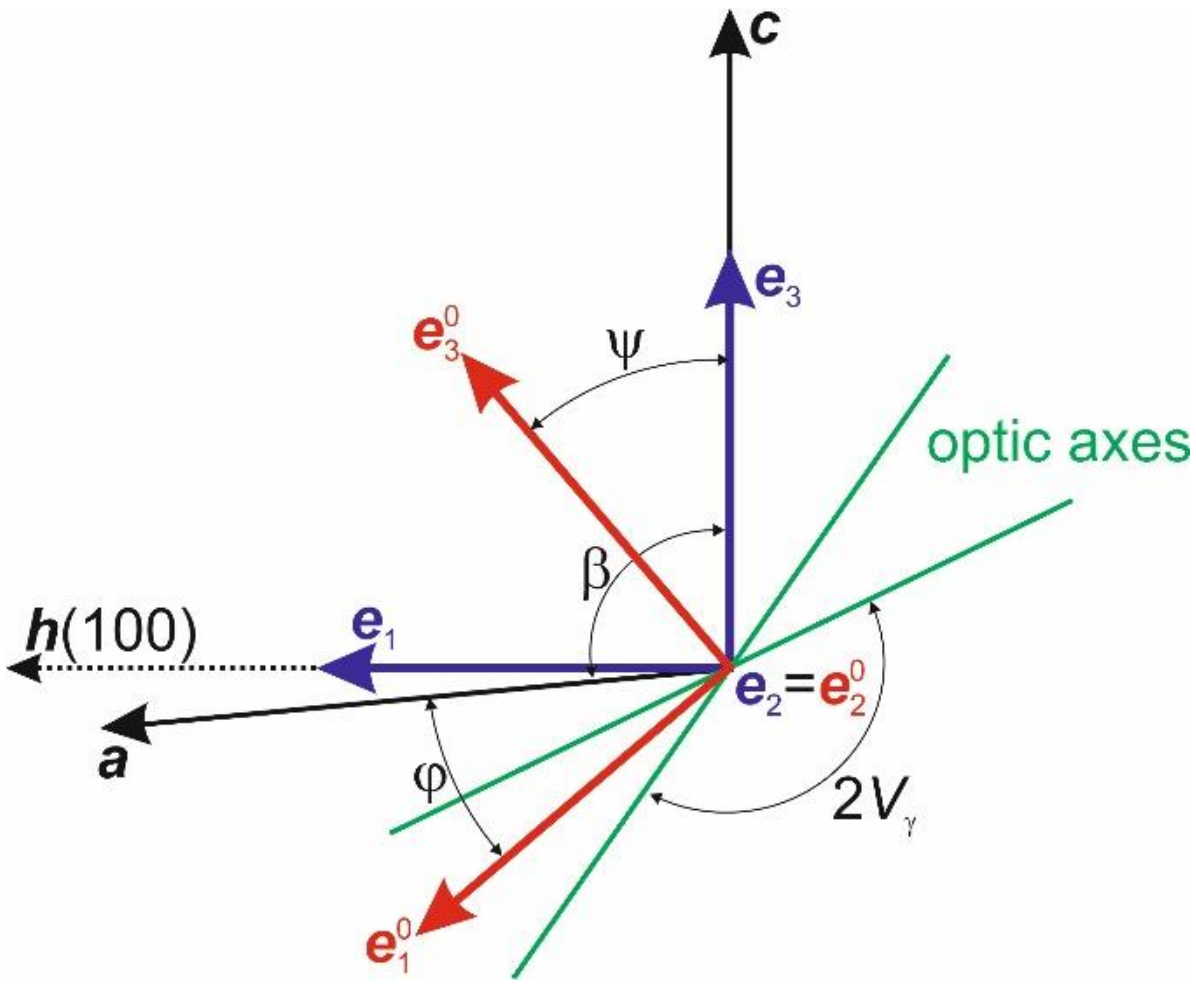

**Fig. 2.** Mutual relation of the crystallographic axes ***a***, ***b***, ***c*** (with ***b*** || ***e***2), the Cartesian reference system ("crystal physical axes") { $\boldsymbol{e}_\mathrm{i}$ }, the principal axes of the optical indicatrix { $\boldsymbol{e}_\mathrm{i}^0$ } and the optic axes for crystals of **2-AmpH2PO3**. ***h***(100) denotes the face normal of the crystal face

(100), $\beta$ is the monoclinic angle $\angle(\boldsymbol{a}, \boldsymbol{c})$ and $2V_{\gamma}$ is the angle between the optic axes (in the case of **2-AmpH$_2$PO$_3$** $\boldsymbol{e}_3^0$ is the obtuse bisectrix).

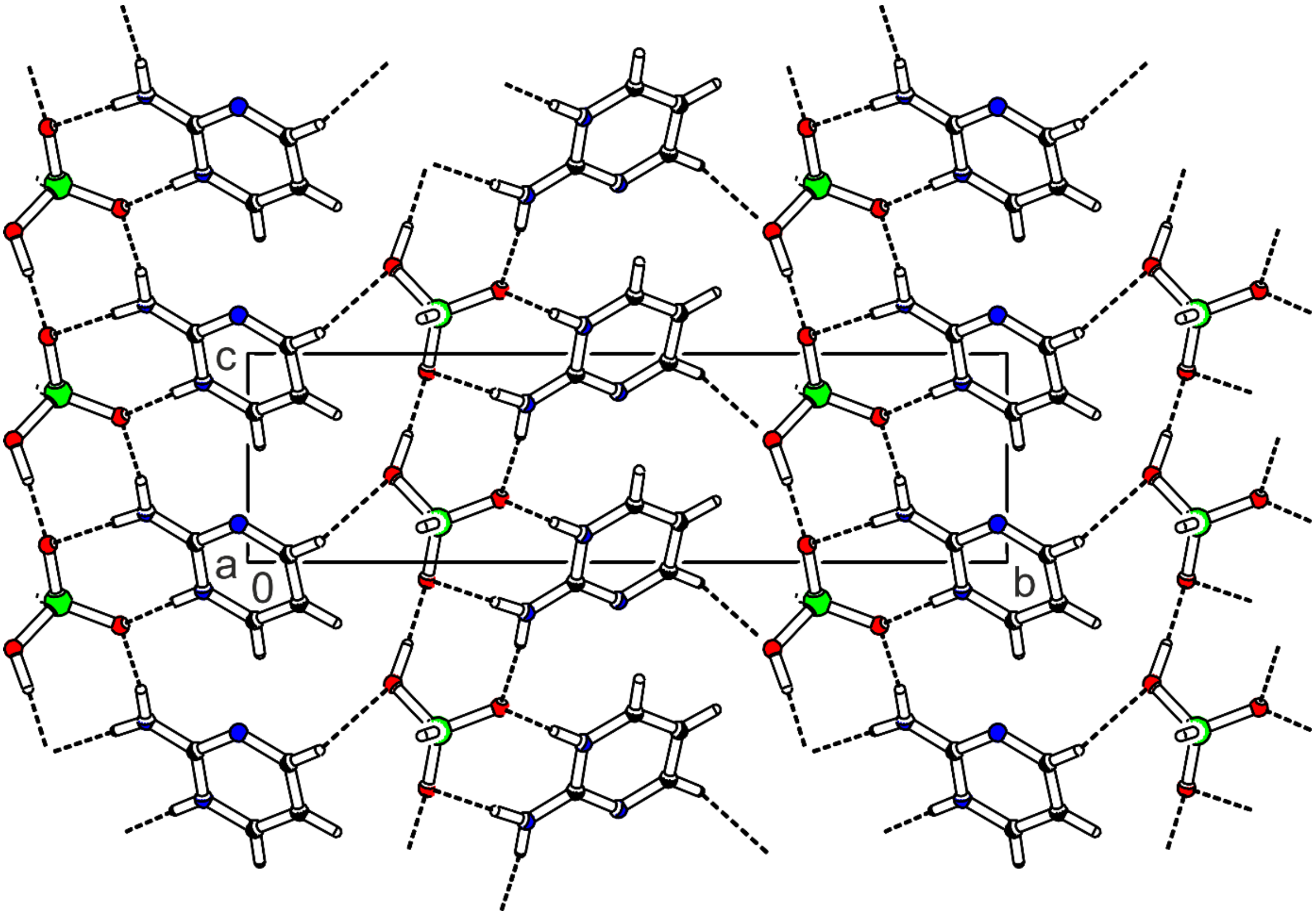

**Fig. 3**. Packing scheme of the **2-AmpH$_2$PO$_3$**. Dashed lines indicate hydrogen bonds.

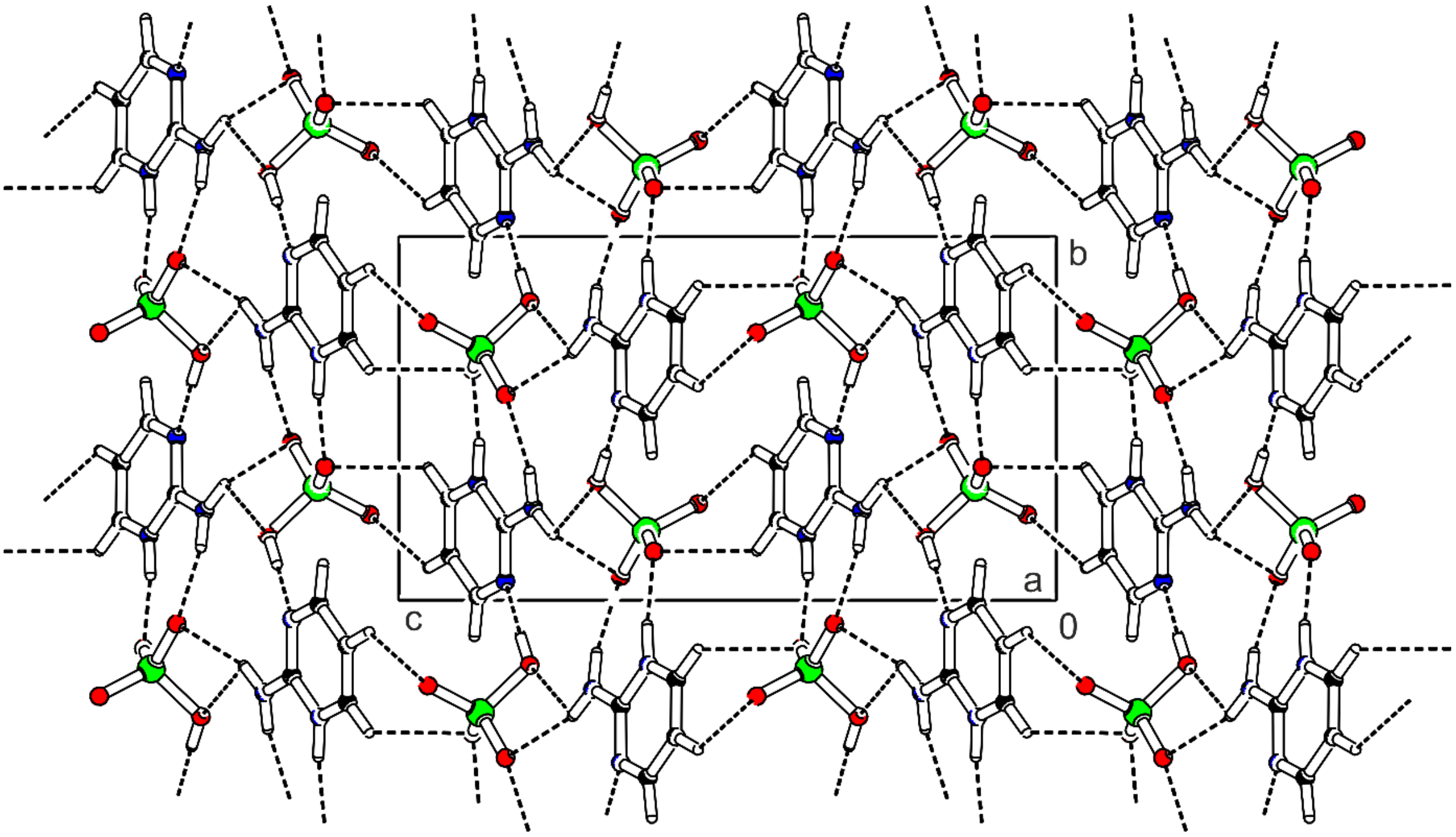

**Fig. 4.** Packing scheme of the **2-AmpHSO$_4$ (I)** polymorph. Dashed lines indicate hydrogen bonds.

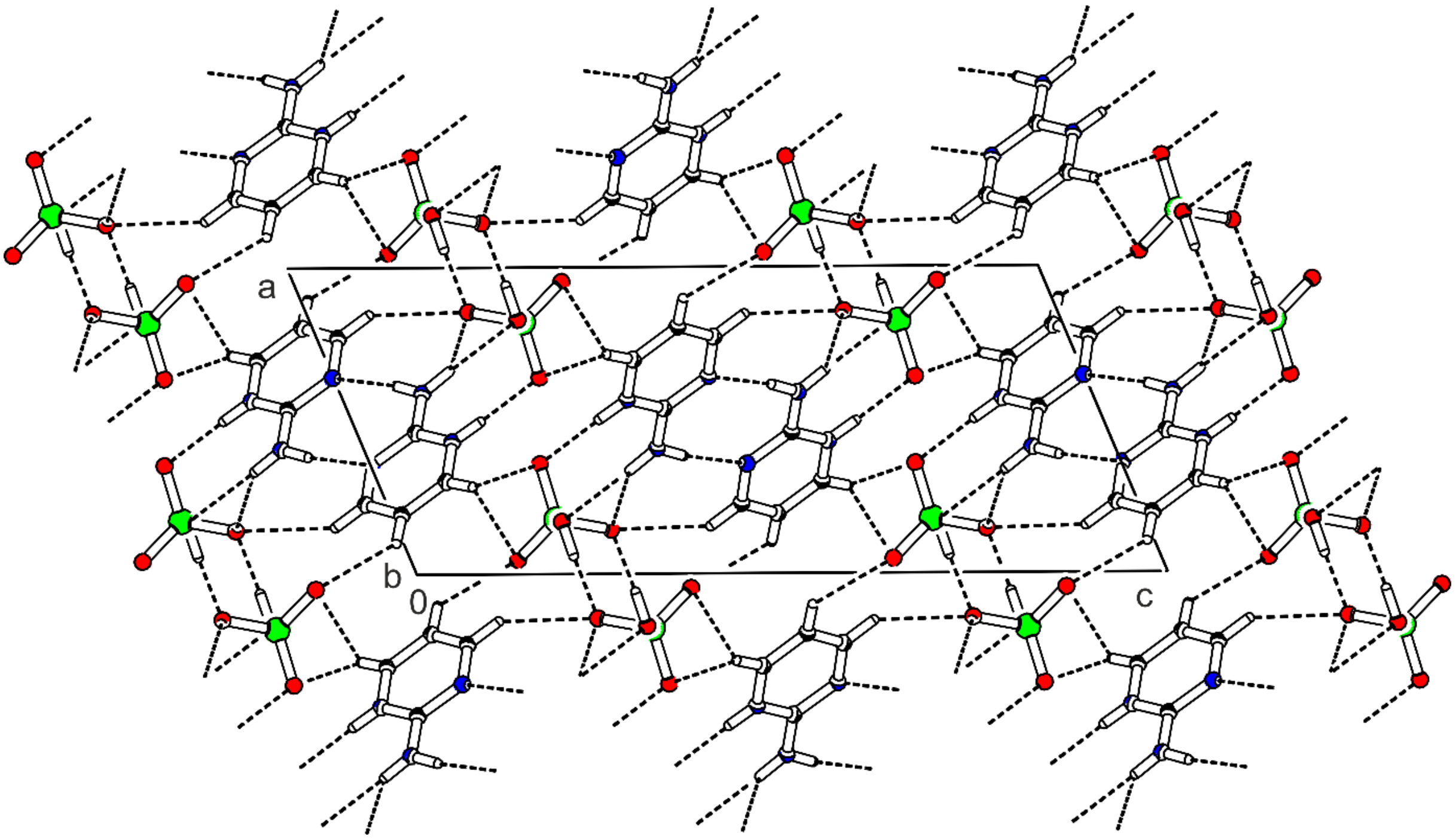

**Fig. 5.** Packing scheme of the **2-AmpHSO$_4$ (II)** polymorph. Dashed lines indicate hydrogen bonds.

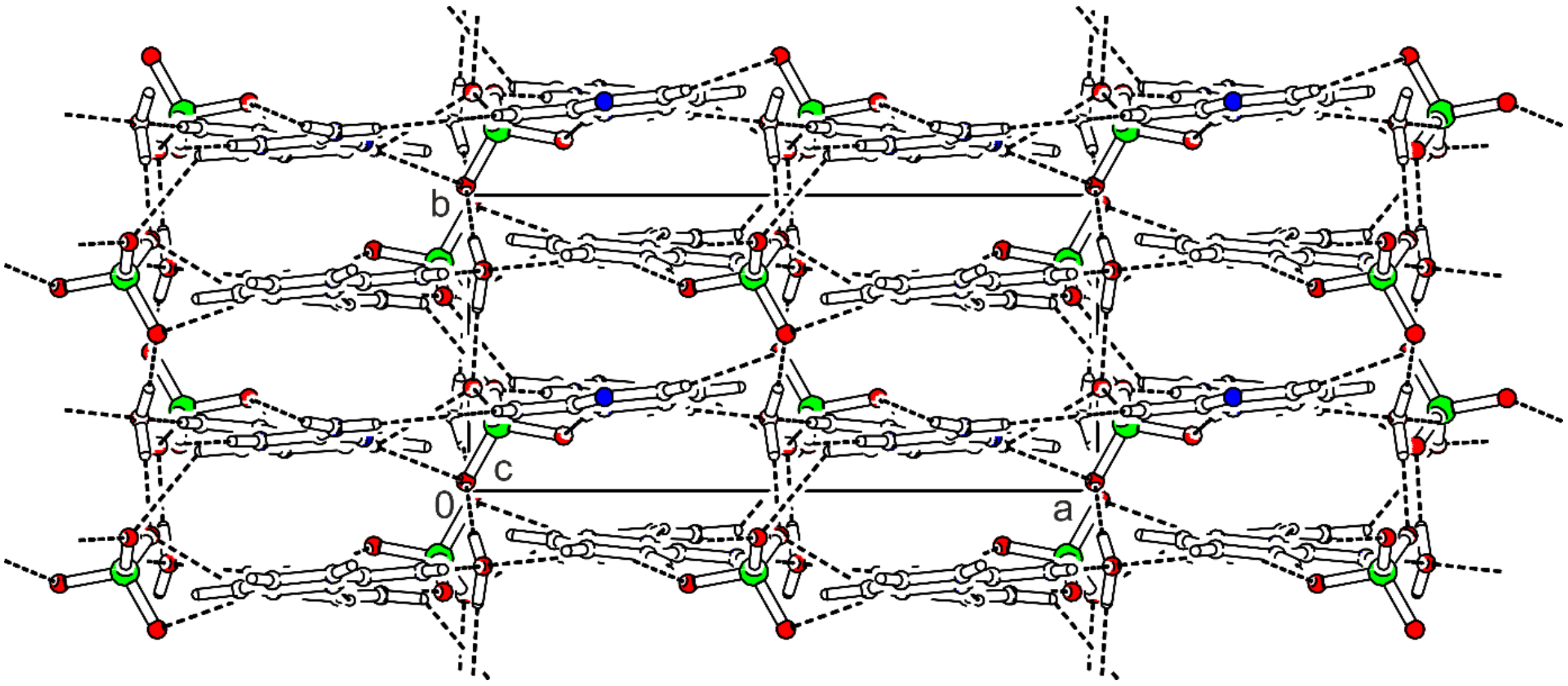

**Fig. 6.** Packing scheme of the **(2-Amp)$_2$SO$_4$H$_2$O**. Dashed lines indicate hydrogen bonds.

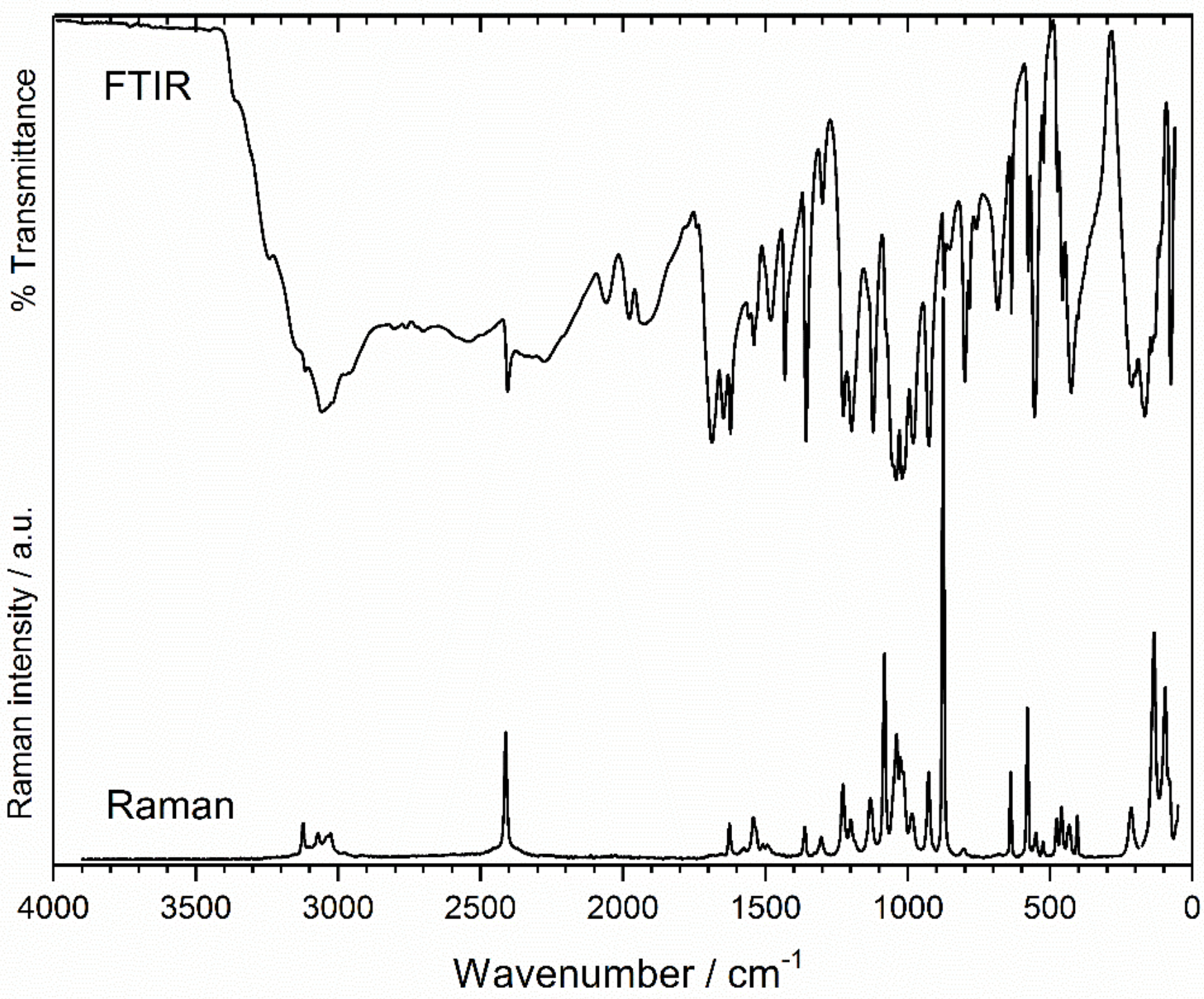

**Fig. 7.** FTIR (compiled from nujol and fluorolube mulls and PE pellet) and Raman spectra of **2-AmpH$_2$PO$_3$** crystals. The Raman spectra were recorded using laser excitation of 1064 nm (3900-200 cm$^{-1}$ region) and 780 nm (200-50 cm$^{-1}$ region).

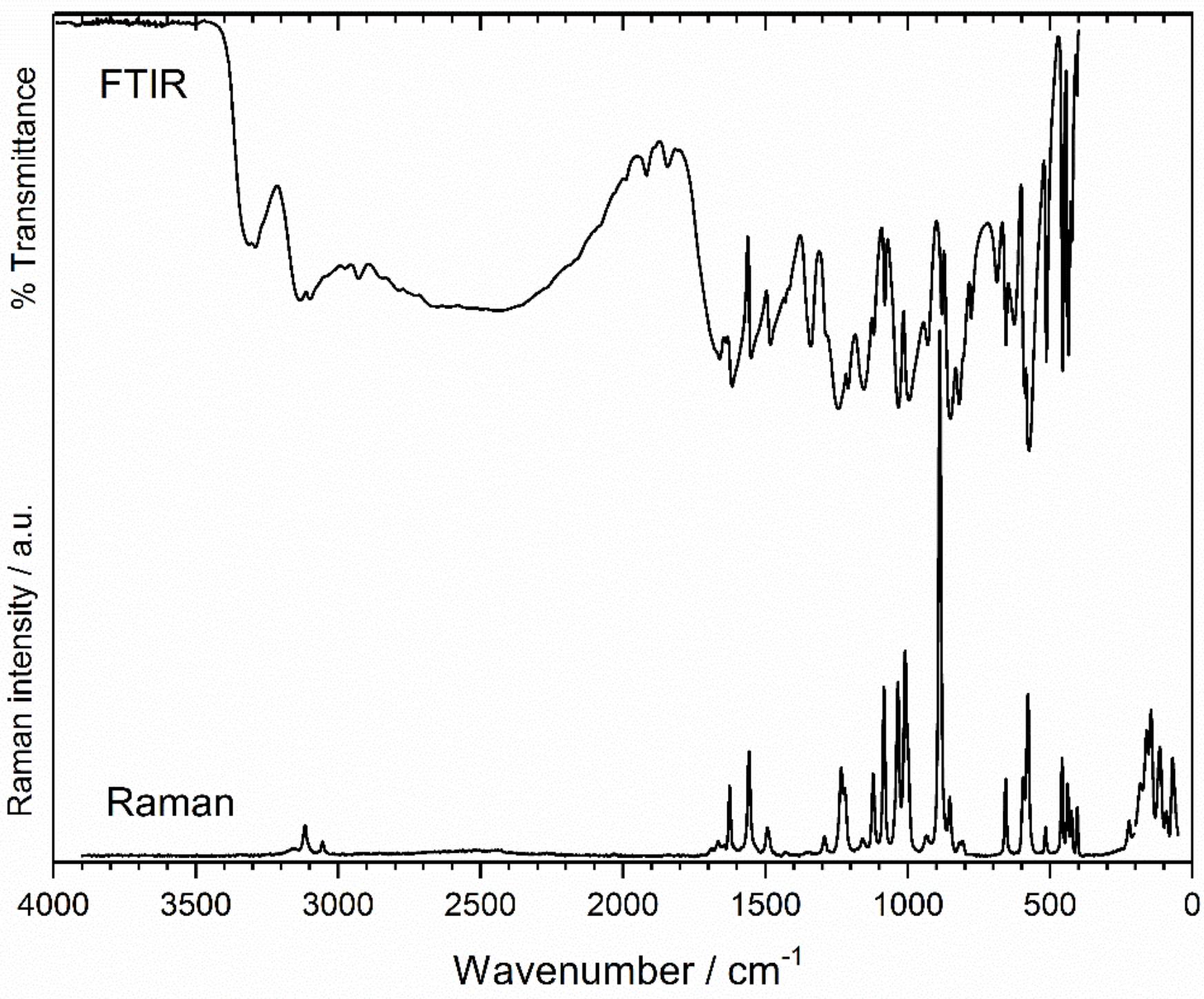

**Fig. 8.** FTIR (compiled from nujol and fluorolube mulls) and Raman spectra of **2-AmpHSO4** (phase **I**) crystals. The Raman spectra were recorded using laser excitation of 1064 nm (3900-200 $cm^{-1}$ region) and 780 nm (200-50 $cm^{-1}$ region).

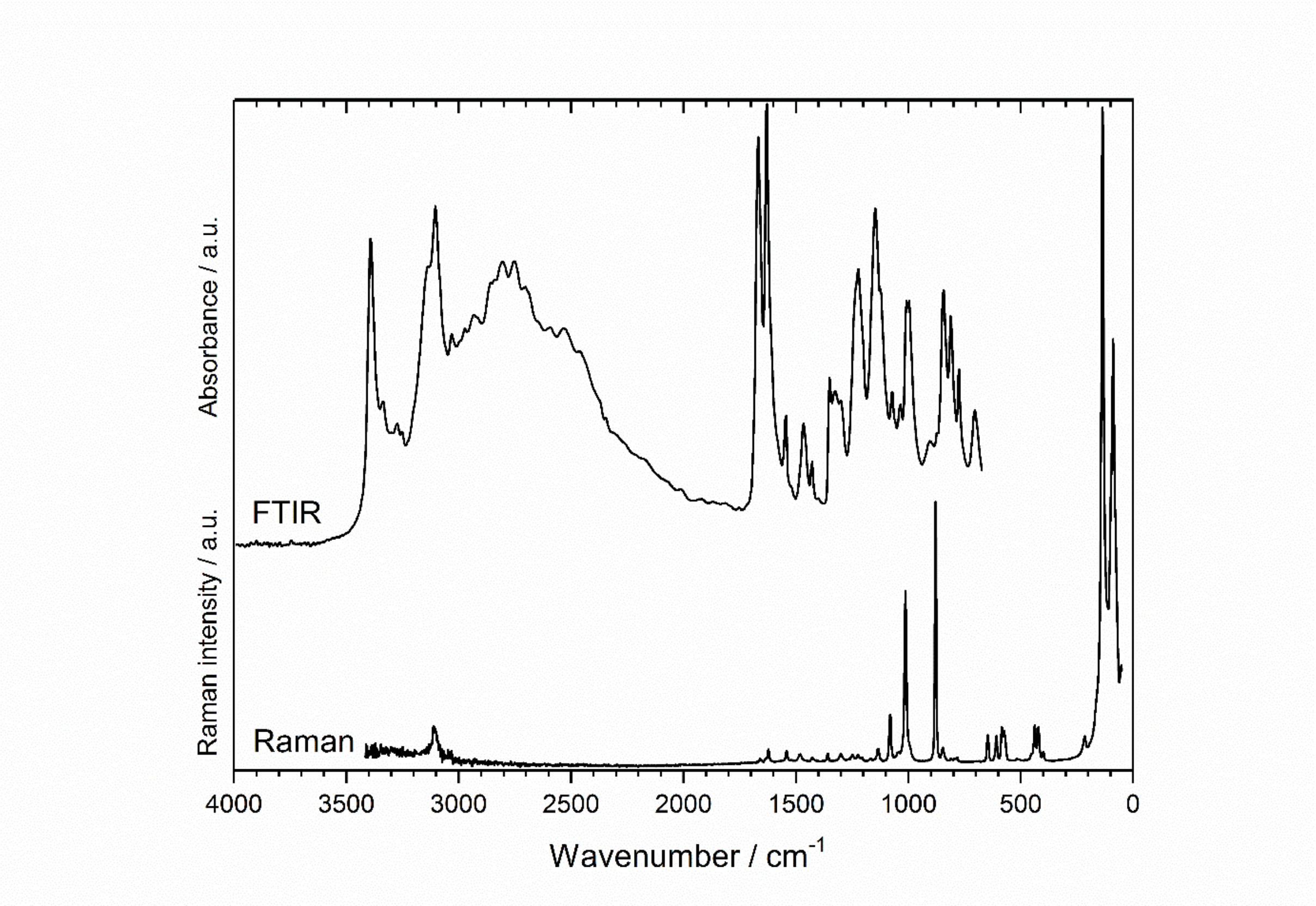

**Fig. 9**. Micro-FTIR (ATR) and micro-Raman spectra of **2-AmpHSO4** (phase **II**) crystals. The Raman spectrum was recorded using 780 nm laser excitation.

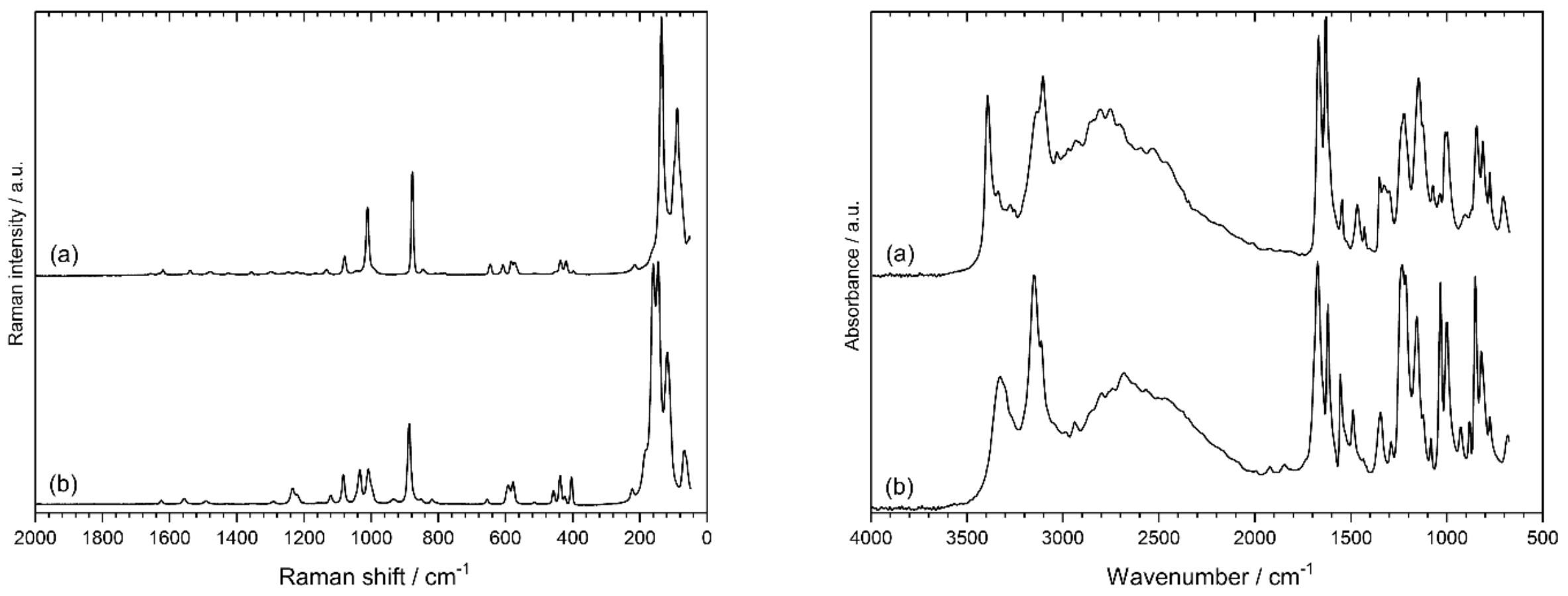

**Fig. 10.** Comparison of micro-Raman and micro-FTIR (ATR) spectra of **2-AmpHSO4** polymorphs. Spectra **(a)** and **(b)** depict **2-AmpHSO4** phase **II** and **I**, respectively. The Raman spectra were recorded using 780 nm laser excitation.

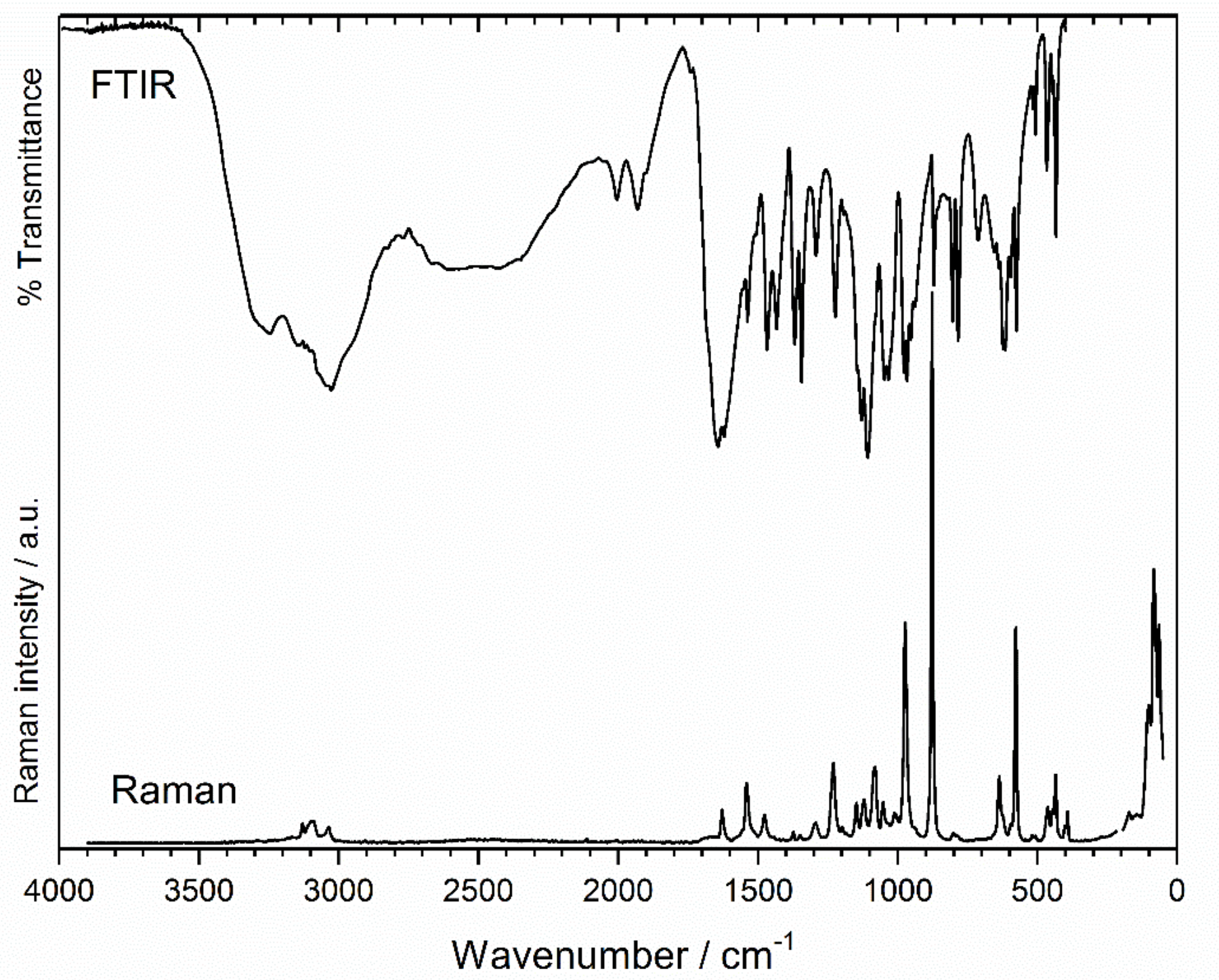

**Fig. 11.** FTIR (compiled from nujol and fluorolube mulls) and Raman spectra of (**2-Amp)$_2$SO$_4$H$_2$O** crystals. The Raman spectra were recorded using laser excitation of 1064 nm (3900-200 cm$^{-1}$ region) and 780 nm (200-50 cm$^{-1}$ region).

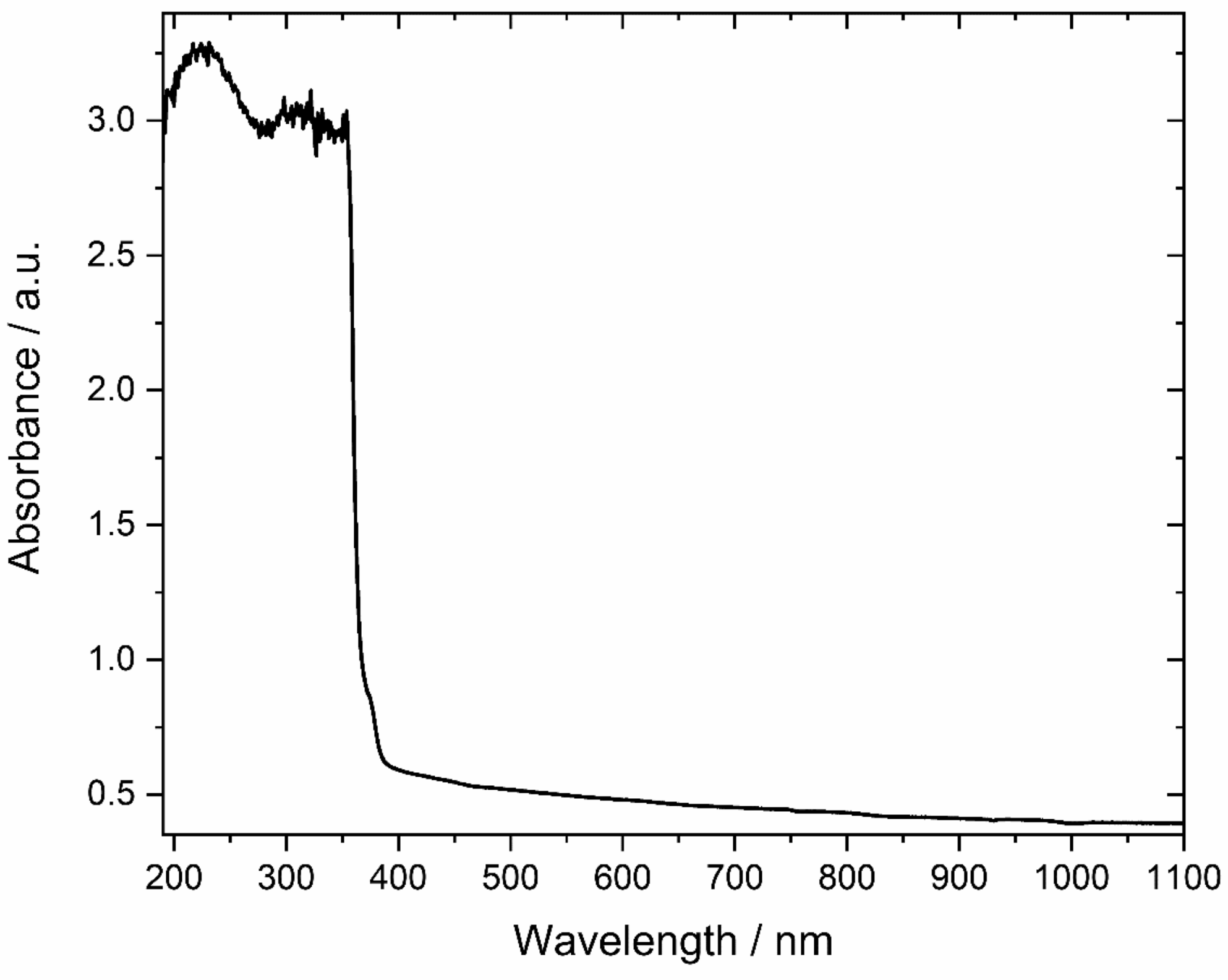

**Fig. 12.** UV-Vis-NIR absorption spectrum of **2-AmpH$_2$PO$_3$** single crystal plate.

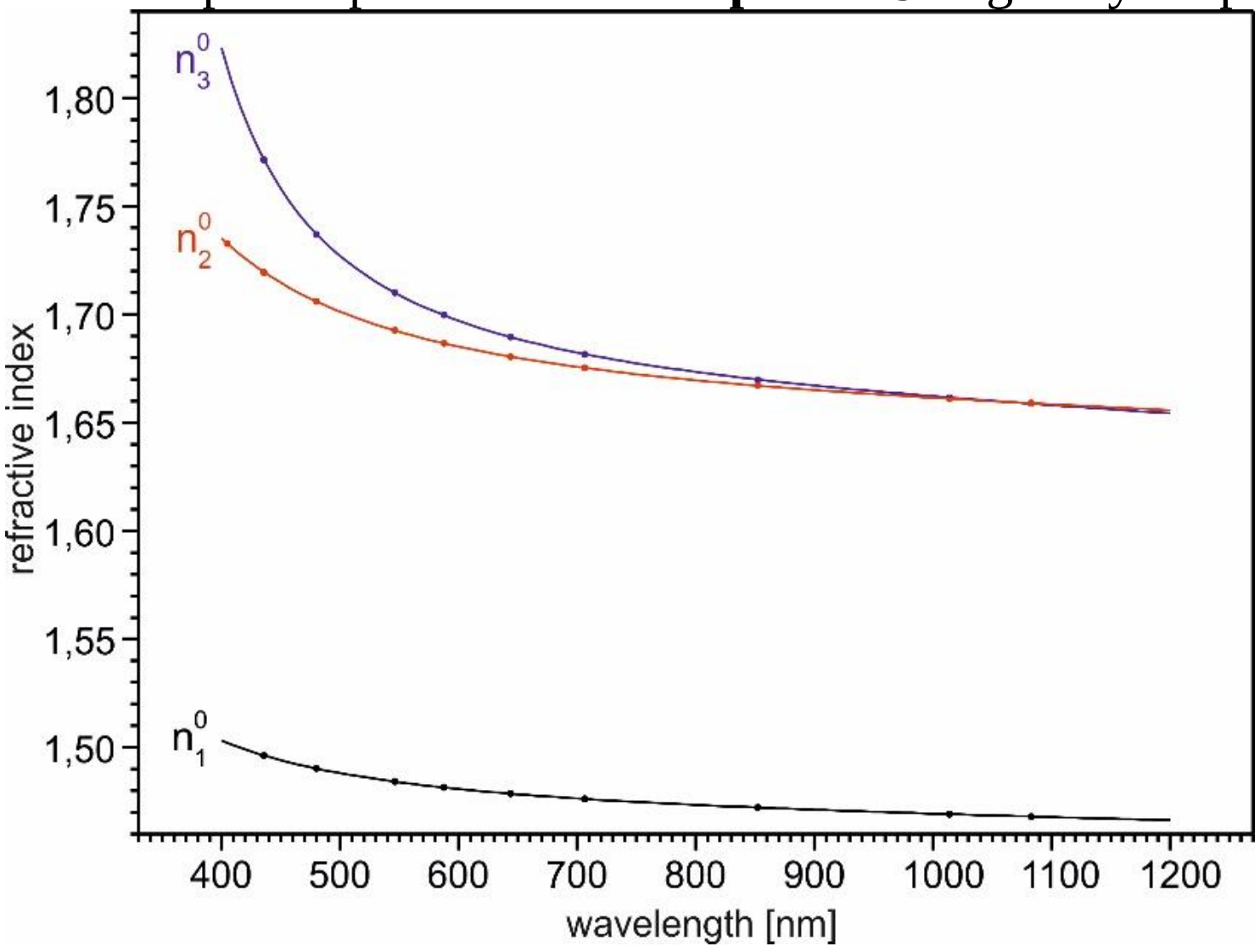

**Fig. 13.** Principal refractive indices and their dispersion of **2-AmpH$_2$PO$_3$**, together with fitted Sellmeier curves. Markers represent measured (and corrected) refractive index data, lines give the Sellmeier fits.

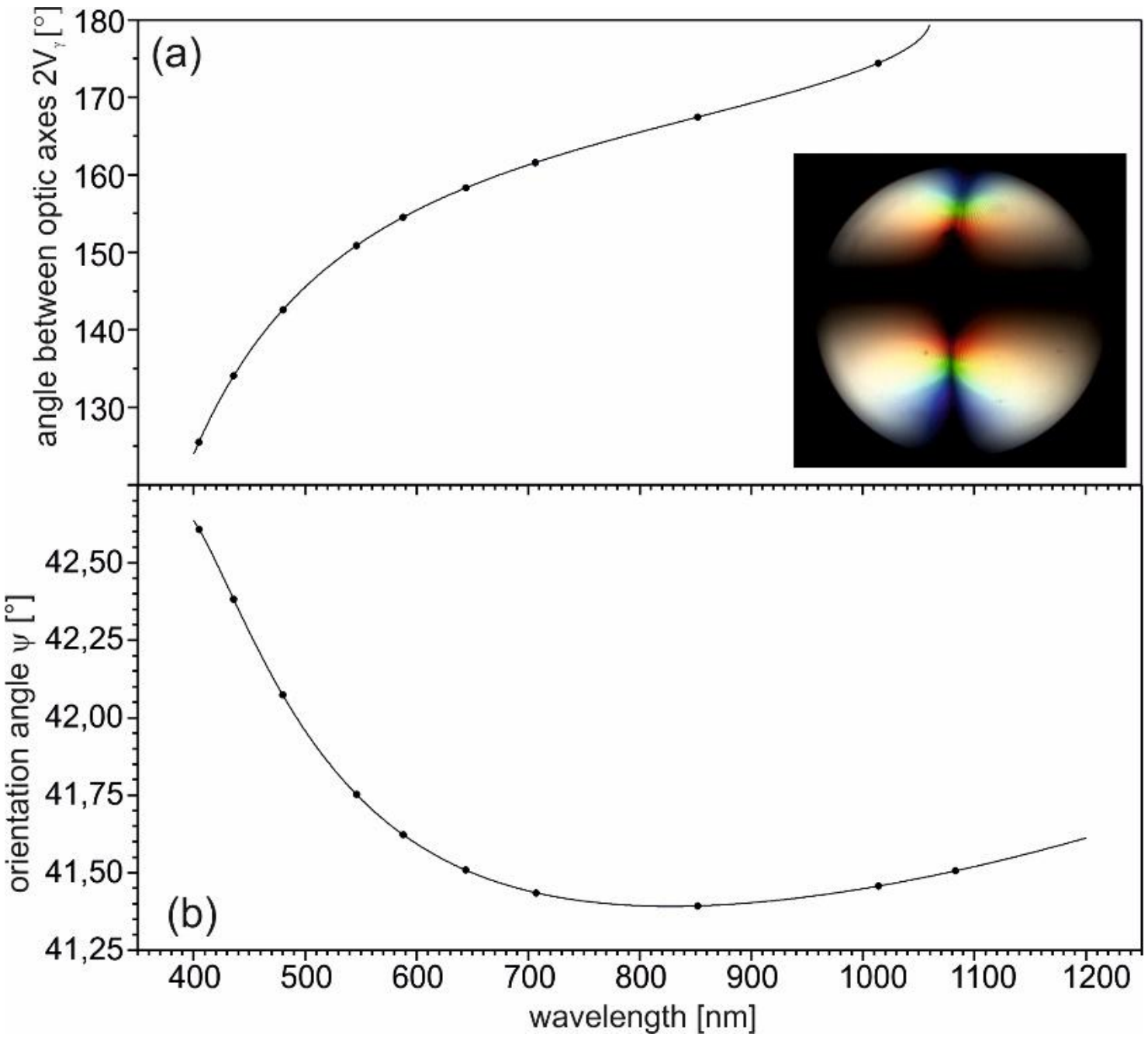

**Fig. 14.** (a) Wavelength dependence of the angle between the optic axes $2V_\gamma$ calculated from the refractive indices. Markers represent angles calculated from the measured (and corrected) refractive index data, lines represent calculated values using the Sellmeier fits. Inset: Interference figure of a crystal plate with plate normal $\vec{e}_1^{\,0}$ (i.e., the acute bisectrix) of **2-AmpH$_2$PO$_3$** (showing thus angle $2V_\alpha$ = 180° - $2V_\gamma$), taken with a polarizing microscope with crossed polarizers in conoscopic illumination and linearly polarized white light. The pronounced dispersion of the angle $2V_\alpha$ (and $2V_\gamma$) becomes evident from the marked colour fringes. (b) Wavelength dependence of the orientation angle $\psi$ between the principal axis of the indicatrix $\boldsymbol{e}_3^0$ and the axis $\boldsymbol{e}_3$ (|| $\boldsymbol{c}$). Markers represent angles calculated from the measured (and corrected) refractive index data, lines represent calculated values using the Sellmeier fits.

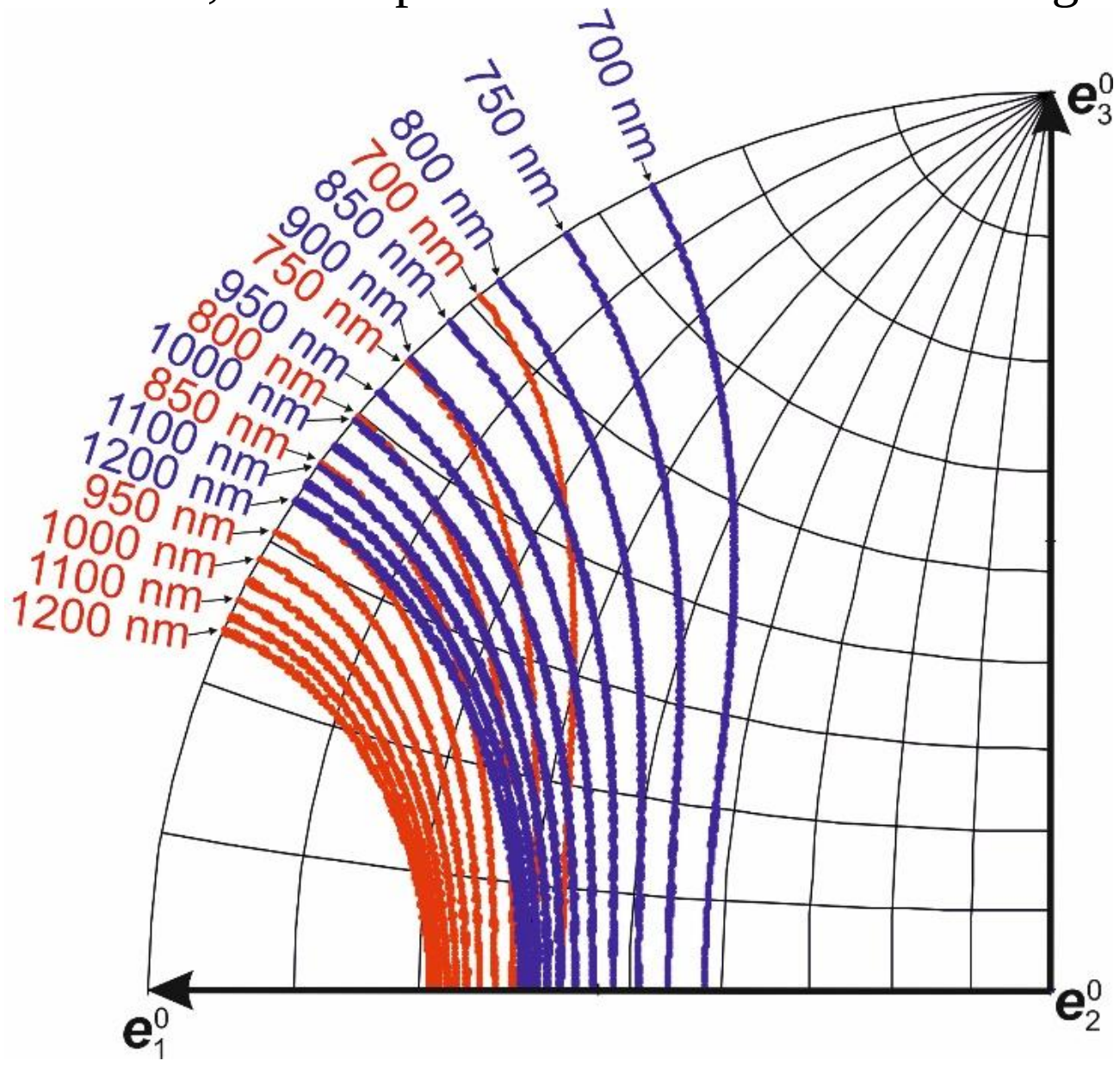

**Fig. 15.** Stereographic projection of collinear SHG phase matching loci in crystals of **2-AmpH$_2$PO$_3$** for selected wavelengths of the fundamental wave. Type I phase matching is indicated by red colour, type II phase matching by blue colour. The directions $\boldsymbol{e}_i^0$ are the directions of the principal axes of the optical indicatrix of **2-AmpH$_2$PO$_3$**.

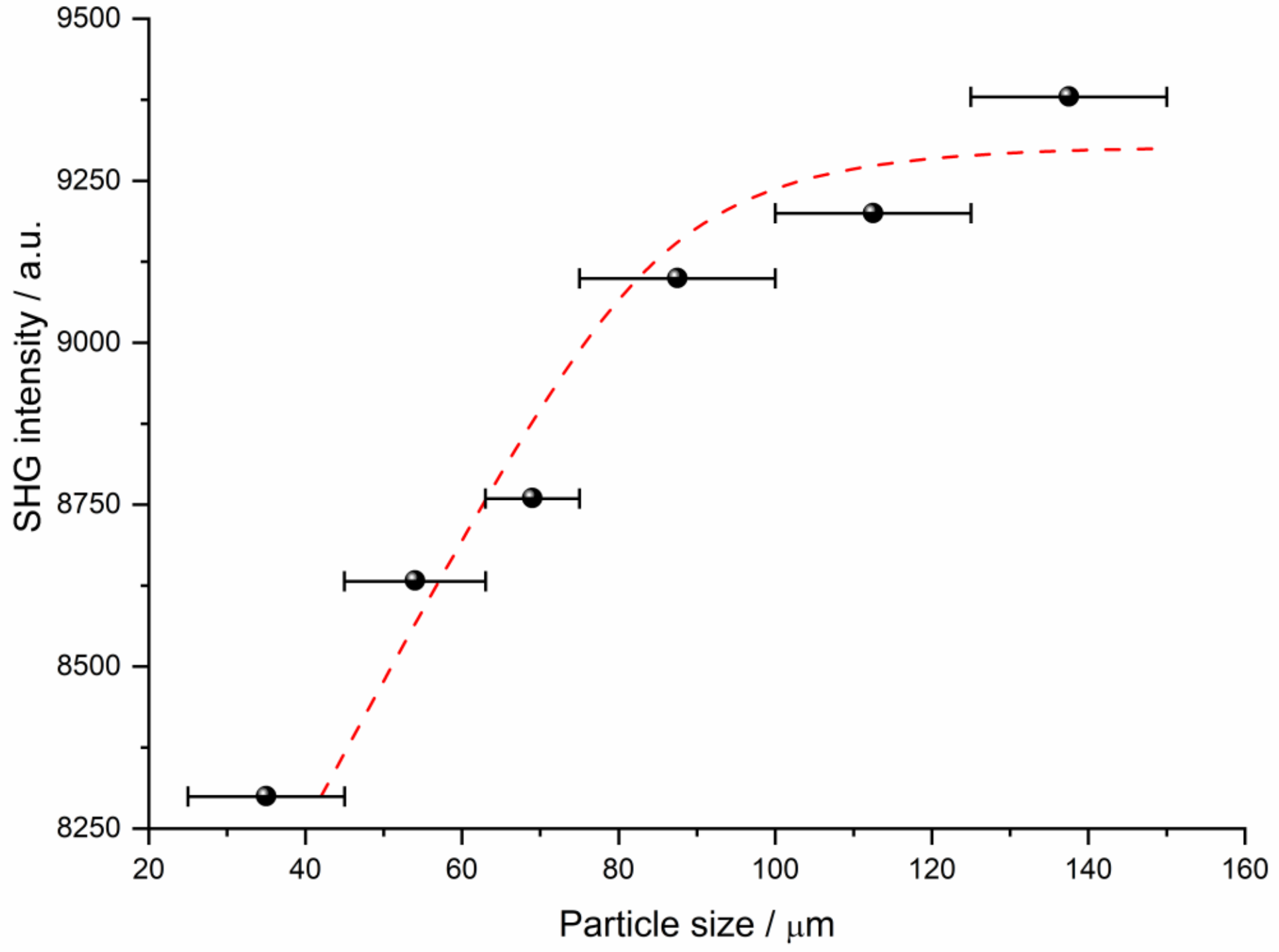

**Fig. 16.** Phase-matching curve (*i.e.*, particle size *vs.* SHG intensity) for **2-Amp$H_2PO_3$** (800 nm fundamental wavelength). Black horizontal segments represent particle size intervals. The red dashed curve drawn is to guide the eye and is not fit for the data.

**Supplementary Material**

for

**Extended study of crystal structures, optical properties and vibrational spectra of polar 2-aminopyrimidinium hydrogen phosphite and three centrosymmetric salts - bis(2-aminopyrimidinium) sulfate monohydrate and two 2-aminopyrimidinium hydrogen sulfate polymorphs**

Irena Matulková[a], Ladislav Bohatý[b], Petra Becker[b], Ivana Císařová[a], Róbert Gyepes[c], Michaela Fridrichová[a], Jan Kroupa[d], Petr Němec[e] and Ivan Němec[a*]

[a] *Charles University, Faculty of Science, Department of Inorganic Chemistry, Hlavova 8, 128 40 Prague 2, Czech Republic.*

[b] *University of Cologne, Institute of Geology and Mineralogy, Section Crystallography, Zülpicher Str. 49b, 50674 Köln, Germany.*

[c] *Czech Academy of Sciences, J. Heyrovsky Institute of Physical Chemistry, Department of Molecular Electrochemistry and Catalysis, Dolejškova 2155, 182 00 Prague 8, Czech Republic.*

[d] *Czech Academy of Sciences, Institute of Physics, Na Slovance 2, 182 21 Prague 8, Czech Republic.*
[e] *Charles University, Faculty of Mathematics and Physics, Department of Chemical Physics and Optics, Ke Karlovu 3, 121 16 Prague 2, Czech Republic.*

* Corresponding author. E-mail address: ivan.nemec@natur.cuni.cz

## Experimental details

*Vibrational spectroscopy*

FTIR spectra were recorded on a Thermo Fisher Scientific Nicolet Magna 6700 FTIR spectrometer (2 $cm^{-1}$ resolution, Happ-Genzel apodization) in the 400-4000 $cm^{-1}$ region using transmission (nujol and fluorolube mulls, KBr windows) and DRIFTS (samples mixed with KBr) techniques. The FAR IR spectra of **2-Amp$H_2PO_3$** were recorded down to 60 $cm^{-1}$ (4 $cm^{-1}$ resolution) in the PE pellets.

Micro-FTIR spectra of **2-Amp$HSO_4$** polymorphs were recorded by ATR technique on a Thermo Fisher Scientific Nicolet iN10 FTIR microscope using Ge crystal in the 675−4000 $cm^{-1}$ region (4 $cm^{-1}$ resolution, Norton−Beer strong apodization). Standard ATR correction (Thermo Nicolet Omnic 9.2 software [S1]) was applied to the recorded spectra.

FT Raman spectra of the powdered samples were recorded on a Thermo Fisher Scientific Nicolet 6700 FTIR spectrometer equipped with the Nicolet Nexus FT Raman module (2 $cm^{-1}$ resolution, Happ-Genzel apodization, 1064 nm Nd:$YVO_4$ laser excitation, 450 mW power at the sample) in the 100–3900 $cm^{-1}$ region.

Raman spectra of microcrystalline samples and aqueous **2-Amp** and **2-Amp**(1+) solutions were collected on a Thermo Scientific DXR Raman Microscope interfaced to an Olympus microscope (objectives 4x, 10x and 50x) in the 30–1800 $cm^{-1}$ spectral region (400 lines/mm and 830 lines/mm gratings) using frequency-stabilized 780 nm single mode diode laser excitation. The spectrometer was calibrated using a software-controlled calibration procedure employing multiple neon emission lines (wavelength calibration), multiple polystyrene Raman bands (laser frequency calibration) and standardized white light sources (intensity calibration).

Raman spectra of microcrystalline samples were also collected on a dispersive confocal Raman microscope MonoVista CRS+ (Spectroscopy & Imaging GmbH, Germany) interfaced to an Olympus microscope (objectives 20x and 50x) using a 785 nm diode excitation laser (10 mW laser power, 40–3800 $cm^{-1}$ spectral range, 300 lines/mm grating). The spectrometer was wavelength- and intensity-calibrated using a software-controlled auto-alignment and calibration procedure with mercury and Ne–Ar lamps.

*Quantum Chemical Computations*

The quantum chemical computations concerning the **2-Amp**(1+) cation were performed (Gaussian 09W software [36]) using the closed-shell restricted density functional theory (DFT) method using Becke's three-parameter hybrid functional [S2] combined with the Lee-Yang-Parr correlation functional (B3LYP) [S3] with the 6-311+G(d,p) basis set, applying tight convergence criteria and an ultrafine integration grid. Geometry optimization of the isolated **2-Amp**(1+) cation was followed by vibrational frequency calculations using the same method and a basis set. Theoretical Raman intensities of computed normal modes were calculated (RAINT programme [37]) for a 1064 nm excitation wavelength, taking Raman scattering activities from Gaussian output. The assignment of the computed normal vibrational modes is based on the visualization of the atom motions in the GaussView programme [38] and performed PED analysis using the VEDA4 programme [39] (described in detail in paper [S4]).

Solid-state DFT computational studies of **2-Amp$H_2PO_3$** focused on vibrational spectra and optical properties were carried out using the CRYSTAL17 program [42]. Three approaches differing in functional and basis sets were selected. The computation named "B3LYP" employed the B3LYP functional, the 6-31+G(d,p) basis for all oxygen atoms and the 6-31G(d) basis set for all other atoms. For sampling the Brillouin zone, the Pack–Monkhorst net used 8, and the Gilat net used 16 points. The numerical integration used the extra-extra-large grid (keyword XXLGRID) of the program.

The computation named "B3LYP-advanced" employed the B3LYP functional, the pob_TZVP basis set [S5] for the phosphorus atoms, the 8-411d11G basis set [S6] for the oxygen atoms and the 6-31G(d) basis set for all other atoms. For sampling the Brillouin zone, the Pack–Monkhorst and Gilat nets were each of 8 points.

The computation named "PBESOL0" employed the PBESOL0 functional and used the POB-TZVP basis set for all atoms. The choice of this computational approach was inspired by its previous successful use in the theoretical study of guanylurea(1+) hydrogen phosphite (GUHP) properties [S7]. For sampling the Brillouin zone, the Pack–Monkhorst and Gilat nets were each of 12 points. Dispersion effects were accounted for by including the Grimme DFT-D3 correction [S8], damped with the Becke-Johnson damping (BJ) [S9] ($s_6$=1.000, $a_1$=0.4466, $s_8$=2.9491, $a_2$=6.1742) and the Axildor-Teller-Muto three-body dispersion correction. The basis set superposition error (BSSE) was corrected by geometrical counterpoise involving automatic parameter setup [S10].

In all three cases, derivatives needed for computing IR and Raman spectra were obtained by the coupled-perturbed Kohn–Sham analytical approach [S11, S12]. Unit cell parameters were kept at experimental values in all cases. The starting atomic coordinates were those obtained from diffraction experiments; they were optimized using tightened convergence criteria. Spectral properties were computed on the obtained stationary-point geometries. Theoretical Raman spectra were corrected for experimental wavelength 1064 nm and temperature 293 K. Vibrational frequencies obtained from computations were scaled uniformly by an empirical factor of 0.97 (both B3LYP computation) or 0.96 (PBESOL0 computation).

*SHG measurements*

Initial SHG measurements on **2-AmpH$_2$PO$_3$** were performed using the modified Kurtz-Perry powder method [45]. The samples were irradiated with 160 fs laser pulses generated at an 82 MHz repetition rate by a Ti: sapphire laser (MaiTai, Spectra-Physics) at 800 nm. For quantitative determination of the SHG efficiency, the intensity of back-scattered laser light at 400 nm generated in the sample was measured by a grating spectrograph with a diode array (USB2000+, Ocean Optics) and the signal was compared with that produced by a potassium dihydrogen phosphate (KDP) standard. The first experiments were performed on a powdered sample (100-150 μm particle size) loaded into a 5 mm glass cell by using a mechanical vibrator. The measurements were repeated in different areas of the same sample, and the results were averaged. This experimental procedure minimizes the signal fluctuations induced by sample packing. Subsequently, the measurements were performed also with size-fractioned samples (particle size: 25-45, 45-63, 63-75, 75-100, 100-125, and 125-150 μm). Lastly, the optical damage threshold experiments with 800 and 1000 nm laser pulses were performed.

The standard Maker fringe method [46] was used for the determination of the individual components of the second order nonlinear optical tensor $[d_{ijk}^{SHG}]$ of single crystal samples of **2-AmpH$_2$PO$_3$**. SHG measurements with plane parallel samples placed on a computer-driven rotational stage were performed using a Q-switched Nd-YAG laser (6 ns pulses at 20 Hz, λ = 1064 nm). For the quantitative determination of the SHG efficiency, the intensity of the filtered SHG signal at 532 nm generated in the sample was measured by a photomultiplier with a boxcar averager, and the signal was compared with that generated in KDP [66]. The samples were single crystalline polished plates of 3x2 mm$^2$ to 5x4 mm$^2$ area and thickness ranging from 0.5 mm to 1 mm, oriented either perpendicularly to the particular principal axis of the optical indicatrix or at 45° relative to these axes for non-diagonal components determination.

**Additional references:**

**Table S1.** The list of crystal structures of inorganic salts (cocrystals) containing **2-Amp** molecule or **2-Amp**(1+) cation.

| Compound | Space group | Temp (K) | R-factor | CCDC code | Reference |
|---|---|---|---|---|---|
| **2-Amp** | *Pbca* | 107 | 0.046 | AmpYRM01 | Acta Chem. Scand., 33 (1979) 715. |
| **2-Amp** | *Pcab* | 295 | 0.048 | AmpYRM10 | Acta Crystallog., B32 (1976) 607. |
| **2-Amp** | *Pbca* | 90 | 0.030 | AmpYRM11 | R. Sparrow (Private Communication) |
| **2-Amp Br** | $P2_1/c$ | 81 | 0.030 | IPICAL | J. Coord. Chem., 56 (2003) 1425. |
| **2-Amp** $\mathbf{BF_4}$ | $P2_1/n$ | 293 | 0.043 | CEDDAR | Solid State Science, 8 (2006) 86. |
| $\mathbf{(2\text{-}Amp)_3\,(H_3BO_3)_2}$ | $P322_1$ | 150 | 0.027 | COLHIX | Crystals, 9 (2019) 403. |
| **2-Amp** $\mathbf{(H_3BO_3)_2}$ | $C2/c$ | 150 | 0.029 | COLHOD | Crystals, 9 (2019) 403. |
| **2-Amp** $\mathbf{H_2PO_3}$ | $P2_1$ | 273 | 0.031 | SILLOS | Chem. Mater., 34 (2022) 1976. |
| **2-Amp** $\mathbf{H_2PO_4\ H_2O}$ | *P*-1 | 293 | 0.042 | UPALON | Acta Crystallogr., E67 (2011) o970. |
| **2-Amp** $\mathbf{H_2PO_4\ H_2O}$ | *P*-1 | 298 | 0.040 | UPALON01 | J. Chem. Cryst., 42 (2012) 276. |
| **2-Amp** $\mathbf{H_2PO_4\ H_2O}$ | *P*-1 | 293 | 0.028 | UPALON02 | J. Mol. Struct., 1074 (2014) 107. |
| **2-Amp** $\mathbf{NO_3}$ | $C2/c$ | 293 | 0.039 | HUFSUX | Acta Crystallogr., E66 (2010) o127. |
| $\mathbf{(2\text{-}Amp)_2\ SO_4}$ | $P2_1/n$ | 293 | 0.032 | CEGFEB | Acta Crystallogr., E68 (2012) o2925. |
| $\mathbf{(2\text{-}Amp)_2\ SO_4\ H_2O}$ | $P2_1/n$ | 296 | 0.035 | HANRUN | J. Mol. Struct., 1257 (2022) 132530. |
| $\mathbf{(2\text{-}Amp)_2\ SeO_4\ H_2O}$ | $P2_1/n$ | 295 | 0.026 | NENHOF | Struct. Chem., 23 (2012) 307. |
| **2-Amp** $\mathbf{HSO_4}$ **(I)** | $P2_1/n$ | 273 | 0.032 | UPASUA01 | J. Mol. Struct., 1257 (2022) 132530. |
| **2-Amp** $\mathbf{HSO_4}$ **(II)** | $P2_1/c$ | 293 | 0.056 | UPASUA | Acta Crystallogr., E67 (2011) o1013. |
| **2-Amp** $\mathbf{ClO_4}$ | $P2_1/n$ | 298 | 0.048 | VAGSEC | Z. Kristallogr.- N. Cryst. Struct., 217 (2002) 501. |
| $\mathbf{(2\text{-}Amp)_2}$ **Amp** $\mathbf{ClO_4}$ | $P2/c$ | 100 | 0.031 | CEDDEV | Solid State Science, 8 (2006) 86. |
| $\mathbf{(2\text{-}Amp)_2\ Cr_2O_7}$ | *P*-1 | 293 | 0.023 | KIJZAF | Acta Crystallogr., E63 (2007) m2336. |

**Table S2**. Experimental powder diffraction data for **2-Amp$H_2PO_3$**.

| 2 Theta (°) | d (Å) | Intensity (%) | 2 Theta (°) | d (Å) | Intensity (%) |
|---|---|---|---|---|---|
| 10.13 | 8.74 | 12 | 30.69 | 2.91 | 2 |
| 18.77 | 4.73 | 4 | 33.17 | 2.70 | 1 |
| 19.45 | 4.56 | 5 | 36.18 | 2.48 | 3 |
| 20.32 | 4.37 | 15 | 40.56 | 2.22 | 1 |
| 21.35 | 4.16 | 2 | 41.24 | 2.19 | 1 |
| 22.12 | 4.02 | 25 | 42.69 | 2.12 | 1 |
| 24.24 | 3.67 | 6 | 45.50 | 1.99 | 1 |
| 24.88 | 3.58 | 1 | 46.05 | 1.97 | 1 |
| 25.97 | 3.43 | 37 | 47.58 | 1.91 | 1 |
| 26.46 | 3.37 | 100 | 49.23 | 1.85 | 1 |
| 27.92 | 3.20 | 75 | 53.33 | 1.72 | 1 |
| 28.39 | 3.14 | 4 | 54.43 | 1.69 | 1 |
| 28.98 | 3.08 | 2 | 55.79 | 1.65 | 3 |
| 30.18 | 2.96 | 3 | 57.64 | 1.60 | 2 |

**Table S3**. Experimental powder diffraction data for **2-Amp$HSO_4$ (I)**.

| 2 Theta (°) | d (Å) | Intensity (%) | 2 Theta (°) | d (Å) | Intensity (%) |
|---|---|---|---|---|---|
| 11.80 | 7.50 | 10 | 27.34 | 3.26 | 2 |
| 15.94 | 5.56 | 2 | 27.97 | 3.19 | 3 |
| 16.71 | 5.31 | 3 | 28.24 | 3.16 | 1 |
| 18.46 | 4.81 | 4 | 28.86 | 3.09 | 15 |
| 18.90 | 4.70 | 3 | 29.58 | 3.02 | 1 |
| 19.88 | 4.47 | 12 | 30.93 | 2.89 | 2 |
| 21.06 | 4.22 | 1 | 31.93 | 2.80 | 9 |
| 21.48 | 4.14 | 8 | 32.20 | 2.78 | 2 |
| 22.02 | 4.04 | 7 | 32.79 | 2.73 | 1 |
| 22.29 | 3.99 | 4 | 33.01 | 2.71 | 2 |
| 23.70 | 3.75 | 100 | 37.42 | 2.40 | 2 |
| 24.54 | 3.63 | 32 | 38.44 | 2.34 | 1 |
| 26.61 | 3.35 | 2 | 44.71 | 2.03 | 2 |
| 26.82 | 3.32 | 2 | 50.59 | 1.80 | 1 |

**Table S4**. Experimental powder diffraction data for **$(2\text{-Amp})_2SO_4H_2O$**.

| 2 Theta (°) | d (Å) | Intensity (%) | 2 Theta (°) | d (Å) | Intensity (%) |
|---|---|---|---|---|---|
| 8.85 | 9.99 | 10 | 25.58 | 3.48 | 6 |
| 12.57 | 7.04 | 6 | 26.99 | 3.30 | 81 |
| 16.17 | 5.48 | 29 | 27.73 | 3.22 | 28 |
| 17.73 | 5.00 | 41 | 28.52 | 3.13 | 17 |
| 18.08 | 4.91 | 100 | 28.79 | 3.10 | 5 |
| 18.51 | 4.79 | 11 | 28.96 | 3.08 | 3 |
| 19.55 | 4.54 | 4 | 30.00 | 2.98 | 4 |
| 19.80 | 4.48 | 8 | 30.47 | 2.93 | 9 |
| 20.03 | 4.43 | 11 | 30.62 | 2.92 | 9 |
| 20.25 | 4.38 | 6 | 31.27 | 2.86 | 5 |
| 22.29 | 3.99 | 16 | 32.65 | 2.74 | 3 |
| 22.56 | 3.94 | 4 | 35.24 | 2.55 | 5 |
| 23.73 | 3.75 | 2 | 38.76 | 2.32 | 3 |
| 24.19 | 3.68 | 14 | 49.59 | 1.84 | 4 |

**Table S5**. Experimental powder diffraction data for **$2\text{-AmpCl½}H_2O$**.

| 2 Theta (°) | d (Å) | Intensity (%) | 2 Theta (°) | d (Å) | Intensity (%) |
|---|---|---|---|---|---|
| 11.40 | 7.76 | 28 | 28.76 | 3.10 | 100 |
| 12.48 | 7.09 | 6 | 29.45 | 3.03 | 4 |
| 17.57 | 5.05 | 4 | 30.74 | 2.91 | 2 |
| 20.60 | 4.31 | 8 | 32.25 | 2.78 | 2 |
| 21.52 | 4.13 | 10 | 33.00 | 2.71 | 5 |
| 21.92 | 4.05 | 4 | 33.24 | 2.70 | 5 |
| 22.83 | 3.90 | 6 | 33.90 | 2.64 | 3 |
| 23.91 | 3.72 | 3 | 34.45 | 2.60 | 2 |
| 24.31 | 3.66 | 11 | 36.28 | 2.48 | 2 |
| 25.45 | 3.50 | 13 | 37.30 | 2.41 | 1 |
| 26.92 | 3.31 | 3 | 39.31 | 2.29 | 1 |
| 27.13 | 3.29 | 13 | 40.44 | 2.23 | 3 |
| 27.33 | 3.26 | 5 | 43.21 | 2.09 | 1 |
| 27.68 | 3.22 | 2 | 48.34 | 1.88 | 1 |

**Table S6**. Selected bond lengths (Å) and angles (°) for **2-AmpH$_2$PO$_3$**.

| Bond/Angle | Value (Å/°) | Angle | Value (°) |
|---|---|---|---|
| C1-N2 | 1.311(3) | C4-C3-C2 | 116.7(2) |
| C1-N3 | 1.355(3) | N1-C4-C3 | 119.9(2) |
| C1-N1 | 1.363(3) | C4-N1-C1 | 121.1(2) |
| C2-N3 | 1.330(3) | C4-N1-H1 | 119.8 |
| C2-C3 | 1.390(4) | C1-N1-H1 | 119.2 |
| C3-C4 | 1.363(4) | C1-N2-H2A | 124.5 |
| C4-N1 | 1.348(3) | C1-N2-H2B | 117.9 |
| O1-P1 | 1.5703(18) | H2A-N2-H2B | 117.1 |
| O1-H1O | 0.9707 | C2-N3-C1 | 117.3(2) |
| O2-P1 | 1.5012(16) | P1-O1-H1O | 118.4 |
| O3-P1 | 1.5074(17) | O2-P1-O3 | 115.89(10) |
| P1-H1P | 1.3165 | O2-P1-O1 | 108.63(10) |
| N2-C1-N3 | 119.5(2) | O3-P1-O1 | 110.62(10) |
| N2-C1-N1 | 119.8(2) | O2-P1-H1P | 109.9 |
| N3-C1-N1 | 120.7(2) | O3-P1-H1P | 108.4 |
| N3-C2-C3 | 124.1(2) | O1-P1-H1P | 102.6 |

Hydrogen-bonds

| D-H…A | d (D-H) | d (A…H) | d (D…A) | <(DHA) |
|---|---|---|---|---|
| N1-H1…O3[a] | 0.87 | 1.76 | 2.629(3) | 177 |
| O1-H10…O2[b] | 0.97 | 1.60 | 2.572(2) | 175 |
| N2-H2A…O2[a] | 0.89 | 2.00 | 2.896(3) | 179 |
| N2-H2B…O3[c] | 0.94 | 1.89 | 2.827(3) | 172 |
| C2-H2…O1[d] | 0.95 | 2.43 | 3.326(3) | 157 |

*Note*. Equivalent positions: [a] 1+x, y, z; [b] x, y, 1+z; [c] x, y, -1+z; [d] -x, ½+y, -z. Abbreviations: A, acceptor; D donor.

**Table S7**. Selected bond lengths (Å) and angles (°) for **2-AmpHSO$_4$ (I)**.

| Bond/Angle | Value (Å/°) | Angle | Value (°) |
|---|---|---|---|
| S1-O3 | 1.4368(12) | O4-S1-O2 | 112.60(7) |
| S1-O4 | 1.4560(11) | O3-S1-O1 | 102.46(6) |
| S1-O2 | 1.4622(11) | O4-S1-O1 | 106.15(7) |
| S1-O1 | 1.5630(11) | O2-S1-O1 | 107.80(7) |
| O1-H1O | 0.9039 | S1-O1-H1O | 111.4 |
| N1-C4 | 1.347(2) | C4-N1-C1 | 121.35(13) |
| N1-C1 | 1.3554(19) | C2-N3-C1 | 117.96(14) |
| N2-C1 | 1.3215(19) | N1-C4-C3 | 119.78(15) |
| N3-C2 | 1.3335(19) | N2-C1-N3 | 119.83(14) |
| N3-C1 | 1.3486(19) | N2-C1-N1 | 119.56(13) |
| C4-C3 | 1.364(2) | N3-C1-N1 | 120.61(13) |
| C2-C3 | 1.390(2) | N3-C2-C3 | 123.35(15) |
| O3-S1-O4 | 112.42(7) | C4-C3-C2 | 116.92(14) |
| O3-S1-O2 | 114.43(7) | | |

Hydrogen bonds

| D-H…A | d (D-H) | d (A…H) | d (D…A) | <(DHA) |
|---|---|---|---|---|
| N1-H1…O2[a] | 0.95 | 1.76 | 2.7017(17) | 173 |
| O1-H10…N3[a] | 0.90 | 1.75 | 2.6517(17) | 176 |
| N2-H2A…O4[a] | 0.89 | 2.02 | 2.9009(18) | 168 |
| N2-H2B…O1[b] | 0.89 | 2.57 | 2.9671(18) | 108 |
| N2-H2B…O4[b] | 0.89 | 2.15 | 3.0085(18) | 162 |
| C3-H3…O3[c] | 1.04 | 2.54 | 3.349(2) | 135 |
| C4-H4…O2[c] | 0.95 | 2.45 | 3.146(2) | 129 |

*Note*. Equivalent positions: [a] 1-x, -1/2+y, 1/2-z; [b] 1+x, y, z; [c] x, 1/2-y, 1/2+z. Abbreviations: A, acceptor; D donor.

**Table S8**. Selected bond lengths (Å) and angles (°) for **2-AmpHSO$_4$ (II)**.

| Bond/Angle | Value (Å/°) | Angle | Value (°) |
|---|---|---|---|
| S1-O4 | 1.4393 | O2-S1-O1 | 110.77 (5) |
| S1-O2 | 1.4553 (9) | O4-S1-O3 | 107.96 (6) |
| S1-O1 | 1.4669 (9) | N4-C2-N1 | 120.32 |
| S1-O3 | 1.5748 | N4-C2-N3 | 118.95 |
| O3-H3 | 0.9174 | N1-C2-N3 | 120.73 |
| N1-C6 | 1.3482 | O2-S1-O3 | 103.70 (6) |
| N1-C2 | 1.3508 | O1-S1-O3 | 106.20 (5) |
| N3-C4 | 1.3264 | S1-O3-H3 | 107.4 |
| N3-C2 | 1.3557 | C6-N1-C2 | 121.75 |
| N4-C2 | 1.3198 | C4-N3-C2 | 117.38 (11) |
| C4-C5 | 1.3960 | N3-C4-C5 | 124.04 (12) |
| C5-C6 | 1.3663 | C6-C5-C4 | 116.60 (12) |
| O4-S1-O2 | 114.13 (6) | N1-C6-C5 | 119.48 (12) |
| O4-S1-O1 | 113.27 (6) | | |

Hydrogen bonds

| D-H…A | d (D-H) | d (A…H) | d (D…A) | <(DHA) |
|---|---|---|---|---|
| O3-H3···O1[a] | 0.92 | 1.68 | 2.5908 (13) | 174 |
| N1-H1···O2 | 0.86 | 1.91 | 2.7478 (14) | 168 |
| N4-H4A···N3[b] | 0.89 | 2.11 | 2.9988 (16) | 174 |
| N4-H4B···O1 | 0.87 | 2.60 | 3.1673 (15) | 124 |
| N4-H4B···O3[c] | 0.87 | 2.31 | 3.0969 (15) | 151 |

*Note*. Equivalent positions: [a] -x+2, y-1/2, -z+3/2, 1/2-z; [b] -x+1, -y+2, -z+1; [c] x, y+1, z. Abbreviations: A, acceptor; D donor.

**Table S9**. Selected bond lengths (Å) and angles (°) for **(2-Amp)$_2$SO$_4$H$_2$O**.

| Bond/Angle | Value (Å/°) | Angle | Value (°) |
|---|---|---|---|
| N11-C14 | 1.3515(17) | N12-C11-N13 | 119.28(12) |
| N11-C11 | 1.3559(17) | N12-C11-N11 | 119.76(11) |
| N12-C11 | 1.3206(17) | N13-C11-N11 | 120.96(12) |
| N13-C12 | 1.3270(17) | N13-C12-C13 | 124.69(12) |
| N13-C11 | 1.3567(16) | C14-C13-C12 | 116.47(12) |
| C12-C13 | 1.393(2) | N11-C14-C13 | 119.57(12) |
| C13-C14 | 1.3646(19) | C24-N21-C21 | 121.19(11) |
| N21-C24 | 1.3519(17) | C22-N23-C21 | 116.70(11) |
| N21-C21 | 1.3607(16) | N22-C21-N21 | 118.92(11) |
| N22-C21 | 1.3201(16) | N22-C21-N23 | 119.62(11) |
| N23-C22 | 1.3222(17) | N21-C21-N23 | 121.46(11) |
| N23-C21 | 1.3520(17) | N23-C22-C23 | 124.55(12) |
| C22-C23 | 1.4055(19) | C24-C23-C22 | 116.55(12) |
| C23-C24 | 1.3612(19) | N21-C24-C23 | 119.51(12) |
| S1-O3 | 1.4681(9) | O4-S1-O3 | 109.94(5) |
| S1-O4 | 1.4740(9) | O4-S1-O2 | 110.09(6) |
| S1-O2 | 1.4765(9) | O3-S1-O2 | 110.41(6) |
| S1-O1 | 1.4953(9) | O4-S1-O1 | 108.45(5) |
| C14-N11-C11 | 121.46(11) | O3-S1-O1 | 109.34(5) |
| C12-N13-C11 | 116.83(11) | O2-S1-O1 | 108.56(5) |

Hydrogen bonds

| D-H…A | d (D-H) | d (A…H) | d (D…A) | <(DHA) |
|---|---|---|---|---|
| O1W-H1W…O4[a] | 0.90 | 1.84 | 2.7332(13) | 169 |
| O1W-H2W…O3 | 0.93 | 1.88 | 2.7957(13) | 171 |
| N11-H11…O1 | 0.91 | 1.74 | 2.6374(14) | 169 |
| N12-H12A…O1W[b] | 0.91 | 2.03 | 2.9366(15) | 177 |
| N12-H12B…O4 | 0.89 | 2.15 | 2.9827(14) | 156 |
| N21-H21…O2[c] | 0.91 | 1.77 | 2.6845(14) | 178 |
| N22-H22A…O1[c] | 0.89 | 1.94 | 2.8240(14) | 171 |
| N22-H22B…O1W | 0.92 | 1.94 | 2.8446(15) | 168 |
| C12-H12…O3[d] | 0.95 | 2.44 | 3.2668(16) | 146 |
| C22-H22…O3[b] | 0.95 | 2.59 | 3.5037(16) | 162 |
| C22-H22…O4[b] | 0.95 | 2.52 | 3.2101(16) | 130 |
| C23-H23…N13[e] | 0.95 | 2.62 | 3.4933(17) | 154 |
| C24-H24…O2[d] | 0.95 | 2.51 | 3.2034(16) | 130 |

*Note.* Equivalent positions: [a] x, 1+y, z; [b] 1-x, 1-y, 2-z; [c] 3/2-x, 1/2+y, 3/2-z; [d] 1/2+x, 1/2-y, 1/2+z; [e] 3/2-x, 1/2+y, 5/2-z. Abbreviations: A, acceptor; D donor.

**Table S10.** Basic crystallographic data and structure refinement details for **2-AmpCl½H2O** crystals.

| Identification code | **2-AmpCl½$H_2O$** |
|---|---|
| Empiric formula | $C_8 H_{14} N_6 O Cl_2$ |
| Formula weight | 281.15 |
| Temperature (K) | 100 |
| $a$ (Å) | 8.6805(2) |
| $b$ (Å) | 9.2749(6) |
| $c$ (Å) | 9.7867(6) |
| $\alpha$ (°) | 63.347(2) |
| $\beta$ (°) | 69.318(2) |
| $\gamma$ (°) | 66.664(2) |
| Volume (Å$^3$) | 631.65(7) |
| $Z$ | 2 |
| Calculated density (Mg/m$^3$) | 1.478 |
| Crystal system | Triclinic |
| Space group | $P$-1 |
| Absoption coeficient (mm$^{-1}$) | 0.509 |
| $F$(000) | 292 |
| Crystal size (mm) | 0.17 x 0.37 x 0.14 |
| Diffractometer and radiation | Bruker D8 VENTURE Kappa Duo PHOTONIII CMOS, Mo $\lambda$ = 0.71073 Å |
| Scan technique | $\omega$ and $\psi$ scans to fill the Ewald sphere |
| Completeness to $\theta$ | 25.242 99.8 % |
| Range of h, k and l | -11 → 11,<br>-12 → 12,<br>-13 → 12 |
| $\theta$ Range for data collection (°) | 2.384 to 28.271 |
| Reflection collected/unique ($R_{int}$) | 23483 / 9960 (0.0129) |
| No. of observed reflection | 3124 |
| Criterion for observed reflection | $I > 2\sigma(I)$ |
| Absorption correction | multi-scan |
| Function minimized | $\Sigma$ w$(F_o^2 - F_c^2)^2$ |
| Parameters refined | 154 |
| $R$; w$R$ ($I$>2$\sigma(I)$) | 0.0221; 0.0584 |
| $R$; w$R$ (all data) | 0.0227; 0.0588 |
| Value of $S$ | 1.05 |
| Max. and min. heights in final Δρ map (eÅ$^{-3}$) | 0.38 and -0.24 |
| Weighting scheme | w = $[\sigma^2(F_o^2) + aP^2 + bP]^{-1}$<br>$P = (F_o^2 + 2F_c^2)/3$<br>$a$ = 0.0225<br>$b$ = 0.3637 |

**Table S11**. Selected bond lengths (Å) and angles (°) for **2-AmpCl½$H_2O$**.

| Bond/Angle | Value (Å/°) | Angle | Value (°) |
|---|---|---|---|
| N11-C14 | 1.3520(13) | C12-N13-C11 | 116.90(9) |
| N11-C11 | 1.3524(13) | N12-C11-N13 | 119.35(9) |
| N12-C11 | 1.3279(13) | N12-C11-N11 | 118.87(9) |
| N13-C12 | 1.3237(13) | N13-C11-N11 | 121.78(9) |
| N13-C11 | 1.3452(13) | N13-C12-C13 | 123.94(10) |
| C12-C13 | 1.4006(15) | C14-C13-C12 | 117.11(9) |
| C13-C14 | 1.3615(14) | N11-C14-C13 | 119.01(9) |
| N21-C24 | 1.3479(14) | C24-N21-C21 | 121.51(9) |
| N21-C21 | 1.3544(13) | C22-N23-C21 | 117.54(9) |
| N22-C21 | 1.3167(14) | N22-C21-N21 | 119.98(9) |
| N23-C22 | 1.3261(13) | N22-C21-N23 | 119.42(9) |
| N23-C21 | 1.3561(13) | N21-C21-N23 | 120.60(9) |
| C22-C23 | 1.3999(14) | N23-C22-C23 | 123.94(10) |
| C23-C24 | 1.3607(15) | C24-C23-C22 | 116.47(10) |
| C14-N11-C11 | 121.25(9) | N21-C24-C23 | 119.93(10) |

Hydrogen bonds

| D-H…A | d (D-H) | d (A…H) | d (D…A) | <(DHA) |
|---|---|---|---|---|
| N11-H11...Cl1 | 0.87 | 2.17 | 3.0163(9) | 164.2 |
| N12-H12A...N23 | 0.79 | 2.29 | 3.0646(13) | 167.7 |
| N21-H21...Cl2[a] | 0.81 | 2.26 | 3.0281(9) | 158.3 |
| N22-H22B...O1W | 0.86 | 1.96 | 2.8121(13) | 174.4 |
| N22-H22A...Cl2[b] | 0.88 | 2.33 | 3.1881(10) | 164.7 |
| O1W-H1W...Cl1[c] | 0.88 | 2.32 | 3.1669(9) | 163.4 |
| O1W-H2W...Cl2 | 0.9 | 2.25 | 3.1446(9) | 170.5 |

*Note*. Equivalent positions: [a] *x*, *y*+1, *z*; [b] -*x*, -*y*+1, -*z*+1; [c] -*x*, -*y*, -*z*+2. Abbreviations: A, acceptor; D donor.

**Table S12.** The internal modes definition for **2-Amp**(1+) cation. The output from VEDA4 programme.

Average max. Potential Energy <EPm> = 65.000
TED Above 100 Factor TAF=0.197
Average coordinate population 2.364
Most complex coordinate No. 8 , population = 5

| **Coord. No.** | **Coef.** | **Mode Type** | **Atom Nos** | **Atom Types** | **Struct. Par. value** | **Freq. to which the coord. participates and PED%** |
|---|---|---|---|---|---|---|
| s 1 | 1.00 | STRE | 4 10 | NH | 1.014560 | f3559 **91** |
| s 2 | 1.00 | STRE | 5 6 | NH | 1.011040 | f3695 **99** |
| | -1.00 | | 5 7 | NH | 1.008560 | |
| s 3 | 1.00 | STRE | 5 6 | NH | 1.011040 | f3575 **90** |
| | 1.00 | | 5 7 | NH | 1.008560 | |
| s 4 | 1.00 | STRE | 2 3 | CH | 1.081992 | f3239 **18** f3224 **81** |
| s 5 | 1.00 | STRE | 8 9 | CH | 1.080339 | f3239 **80** f3224 **18** |
| s 6 | 1.00 | STRE | 12 13 | CH | 1.085896 | f3181 **99** |
| s 7 | 1.00 | STRE | 2 8 | CC | 1.363464 | f1665 **58** f1143 **17** |
| s 8 | 1.00 | STRE | 11 12 | NC | 1.319836 | f1705 **41** f1637 **14** f1386 **10** |
| | -1.00 | | 11 1 | NC | 1.338103 | |
| | 1.00 | | 5 1 | NC | 1.332059 | |
| | 1.00 | | 4 2 | NC | 1.365227 | |
| | -1.00 | | 4 1 | NC | 1.367610 | |
| s 9 | -1.00 | STRE | 4 2 | NC | 1.365227 | f1318 **39** f1239 **21** |
| | -1.00 | | 11 1 | NC | 1.338103 | |
| | 1.00 | | 2 8 | CC | 1.363464 | |
| | 1.00 | | 11 12 | NC | 1.319836 | |
| s 10 | -1.00 | STRE | 4 2 | NC | 1.365227 | f1558 **40** f1074 **14** f1017 **16** |
| | 1.00 | | 11 1 | NC | 1.338103 | |
| | 1.00 | | 11 12 | NC | 1.319836 | |
| s 11 | 1.00 | STRE | 2 8 | CC | 1.363464 | f1437 **37** f1386 **18** f1074 **13** |
| | -1.00 | | 5 1 | NC | 1.332059 | |
| | -1.00 | | 11 1 | NC | 1.338103 | |
| | 1.00 | | 11 12 | NC | 1.319836 | |
| s 12 | 1.00 | STRE | 4 1 | NC | 1.367610 | f1003 **12** f881 **64** |
| | 1.00 | | 5 1 | NC | 1.332059 | |
| | 1.00 | | 11 1 | NC | 1.338103 | |
| | 1.00 | | 4 2 | NC | 1.365227 | |
| s 13 | 1.00 | BEND | 6 5 1 | HNC | 117.50 | f1017 **58** |
| | -1.00 | | 1 11 12 | CNC | 118.41 | |
| | 1.00 | | 8 2 4 | CCN | 119.18 | |
| s 14 | -1.00 | BEND | 6 5 7 | HNH | 118.45 | f1705 **28** f1665 **16** f1437 **21** |
| | 1.00 | | 10 4 2 | HNC | 118.64 | |

s 15 1.00 BEND 6 5 1 HNC 117.50 f1074 **37** f1003 **14** f582 **12**
-1.00 8 2 4 CCN 119.18
1.00 1 11 12 CNC 118.41
1.00 1 4 2 CNC 121.25
s 16 1.00 BEND 6 5 7 HNH 118.45 f1637 **65** f1558 **12**
1.00 10 4 2 HNC 118.64
s 17 1.00 BEND 3 2 8 HCC 124.62 f1665 **15** f1318 **10** f1239 **43** f1143 **10**
s 18 1.00 BEND 3 2 8 HCC 124.62 f1475 **44** f1437 **11** f1143 **25**
1.00 9 8 12 HCC 121.92
s 19 1.00 BEND 13 12 11 HCN 115.93 f1558 **21** f1386 **45** f1318 **19**
s 20 1.00 BEND 1 4 2 CNC 121.25 f1705 **10** f582 **68**
1.00 1 11 12 CNC 118.41
-1.00 4 1 11 NCN 120.76
s 21 1.00 BEND 5 1 11 NCN 118.97 f415 **73**
s 22 1.00 BEND 1 11 12 CNC 118.41 f647 **74**
1.00 8 2 4 CCN 119.18
1.00 4 1 11 NCN 120.76
s 23 1.00 BEND 1 4 2 CNC 121.25 f1475 **20** f1003 **51**
1.00 8 2 4 CCN 119.18
1.00 4 1 11 NCN 120.76
s 24 1.00 TORS 10 4 2 8 HNCC -180.00 f706 **82**
s 25 1.00 TORS 1 4 2 8 CNCC 0.00 f993 **11** f493 **55** f162 **10**
-1.00 OUT 5 4 11 1 NNNC 0.00
s 26 1.00 TORS 1 4 2 8 CNCC 0.00 f1023 **11** f389 **61**
-1.00 2 4 1 11 CNCN 0.00
-1.00 12 11 1 4 CNCN 0.00
1.00 OUT 5 4 11 1 NNNC 0.00
s 27 1.00 TORS 3 2 8 12 HCCC -180.00 f993 **70** f493 **12**
-1.00 9 8 12 11 HCCN -180.00
s 28 1.00 TORS 9 8 12 11 HCCN -180.00 f817 **74**
1.00 3 2 8 12 HCCC -180.00
s 29 1.00 TORS 13 12 11 1 HCNC -180.00 f1023 **75**
s 30 1.00 TORS 1 4 2 8 CNCC 0.00 f780 **78**
1.00 12 11 1 4 CNCN 0.00
1.00 OUT 5 4 11 1 NNNC 0.00
s 31 1.00 TORS 7 5 1 4 HNCN 0.00 f371 **82**
-1.00 1 4 2 8 CNCC 0.00
1.00 12 11 1 4 CNCN 0.00

| | | | | | | | | | | |
|---|---|---|---|---|---|---|---|---|---|---|
| | | -1.00 | OUT | 5 | 4 | 11 | 1 | NNNC | 0.00 | |
| s 32 | 1.00 | TORS | | 2 | 4 | 1 | 11 | CNCN | 0.00 | f162 **78** |
| | 1.00 | OUT | | 5 | 4 | 11 | 1 | NNNC | 0.00 | |
| s 33 | 1.00 | TORS | | 6 | 5 | 1 | 4 | HNCN | -180.00 | f535 **85** |
| | | -1.00 | | 1 | 4 | 2 | 8 | CNCC | 0.00 | |
| | 1.00 | OUT | | 5 | 4 | 11 | 1 | NNNC | 0.00 | |

****

12 STRE modes: 1 2 3 4 5 6 7 8 9 10 11 12
11 BEND modes: 13 14 15 16 17 18 19 20 21 22 23
10 TORS modes: 24 25 26 27 28 29 30 31 32 33
9 CH modes: 4 5 6 17 18 19 27 28 29

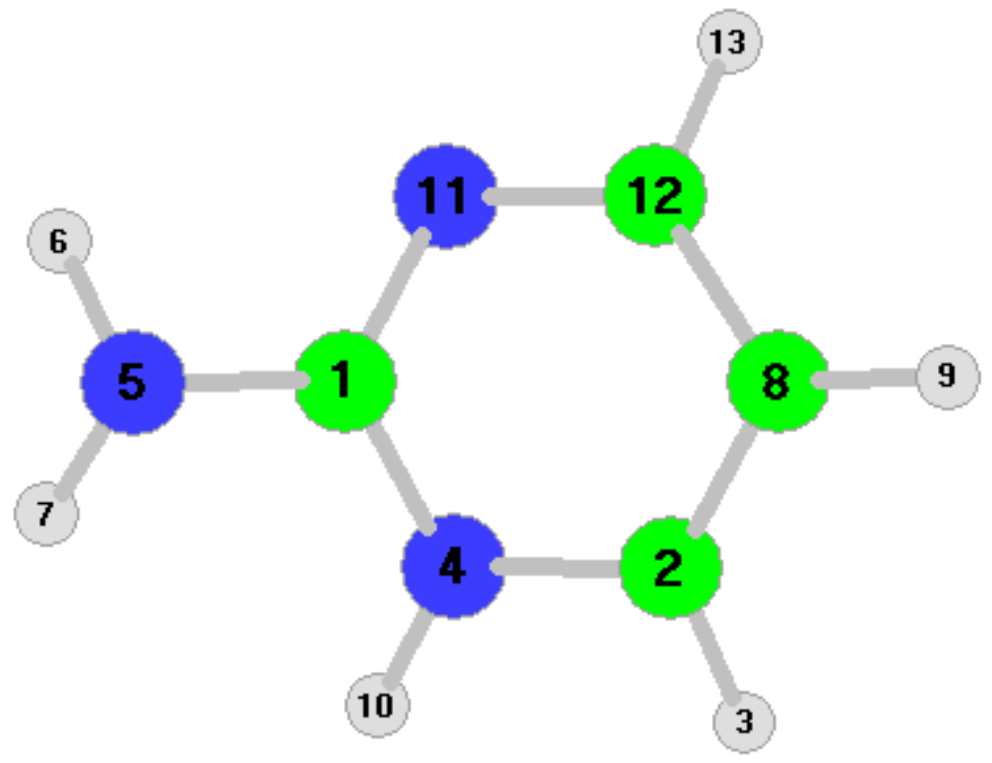

**Atom numbering of 2-Amp**(1+) **cation**

**Table S13.** Calculated and recorded vibrational frequencies (cm$^{-1}$) of **2-Amp**(1+) cation.

| Computed vibrational frequencies (cm$^{-1}$) | Dual scaling[a] (cm$^{-1}$) | WLS scaling[b] (cm$^{-1}$) | Relative intensity IR/Raman[c] | Assignment | Recorded 2-AmpCl½$H_2O$ (cm$^{-1}$) FTIR | Raman |
|---|---|---|---|---|---|---|
| | | | | External modes | | 104 s |
| | | | | | | 129 vs |
| 162 | 165 | 163 | 0/0 | γrg | | 190 m |
| | | | | | | 204 m |
| 371 | 378 | 372 | 12/0 | γ$NH_x$, γrg | | 391 m |
| 389 | 396 | 390 | 6/0 | γrg, γCH, γ$NH_x$ | | 396 m |
| 415 | 423 | 416 | 1/1 | δCNC | 431 m | 431 m |
| | | | | | 461 w | 461 m |
| 493 | 502 | 493 | 13/0 | γrg, γCH, γ$NH_x$ | 515 m | 505 w |
| 535 | 545 | 535 | 5/0 | τ$NH_2$, γrg | 534 m | |
| 582 | 593 | 582 | 0/5 | δrg | 580 m | 583 s |
| 647 | 659 | 646 | 0/2 | δrg | 639 m | 638 s |
| 706 | 719 | 704 | 12/0 | γNH, τ$NH_2$ | 679 m | 681 vwb |
| 780 | 795 | 777 | 3/0 | γrg, γ$CN_3$ | 779 m | 780 w |
| 818 | 833 | 814 | 2/0 | γCH | 791 m | 793 w |
| 881 | 898 | 876 | 0/21 | $ν_s$rg, $δ_s$rg, νC-$NH_2$ | 870 m | 874 vs |
| | | | | ? | 921 m | |
| 993 | 1012 | 986 | 0/0 | γCH | | |
| 1003 | 1022 | 996 | 1/4 | δrg, νrg, δNH | 991 m | 993 s |
| 1017 | 1036 | 1009 | 0/0 | ρ$NH_2$, δrg | | |
| 1022 | 1041 | 1014 | 0/0 | γCH, γrg | 1020 w | 1012 w |
| 1074 | 1094 | 1065 | 3/5 | ρ$NH_2$, νrg, δrg | 1072 m | 1072 s |
| 1143 | 1165 | 1132 | 0/2 | δCH, δ$NH_x$ | 1112 w | 1112 s |
| 1239 | 1200 | 1225 | 2/6 | δCH, νrg | 1206 m | 1193 s |
| | | | | | 1228 m | 1231 s |
| 1318 | 1277 | 1301 | 2/5 | δCH, νrg, δ$NH_x$ | 1273 m | 1272 m |
| | | | | | 1293 m | 1294 sh |
| 1386 | 1343 | 1367 | 11/2 | δCH, νrg, δNH | 1346 s | 1349 m |
| | | | | | 1381 w | |
| 1437 | 1392 | 1416 | 1/2 | νC-$NH_2$, νrg, δ$NH_x$, δCH | 1418 m | 1410 w |
| 1475 | 1429 | 1453 | 4/4 | δCH, νrg, δ$NH_x$ | 1463 m | 1459 m |
| | | | | ? | 1509 m | 1511 w |
| 1558 | 1510 | 1532 | 6/7 | νrg, δCH, δ$NH_x$, δNCN | 1538 m | 1536 s |
| | | | | ? | | 1603 w |
| 1637 | 1586 | 1608 | 4/1 | δ$NH_x$, νrg | 1617 s | 1618 m |
| 1665 | 1613 | 1635 | 17/5 | νrg, δ$NH_x$ δCH | 1628 s | 1627 m |
| 1705 | 1652 | 1673 | 100/3 | νC-$NH_2$, νrg, δ$NH_x$, δrg | 1655 s* | 1647 m* |
| | | | | | 1670 s* | 1664 vw* |
| | | | | νNH(…O,Cl) | 2605 mb | 2600 wb |
| | | | | | 2930 mb | |
| 3181 | 3082 | 3044 | 0/60 | νCH | | 3006 m |
| 3224 | 3124 | 3083 | 1/33 | νCH | | 3032 m |
| | | | | | | 3052 w |
| | | | | | | 3088 sh |
| | | | | | | 3098 m |
| 3239 | 3138 | 3097 | 2/85 | νCH | | 3116 m |
| 3559 | 3448 | 3384 | 15/27 | νNH | 3360 mb* | 3375 w* |
| 3575 | 3464 | 3398 | 32/100 | νNH | | |
| 3695 | 3580 | 3505 | 16/26 | νNH | 3565 wb* | |

*Note:* Abbreviation and Greek symbols used for vibrational modes: rg, ring; $NH_x$, $NH_2$ and NH groups; $_s$, symmetric; n, stretching; d, deformation or in-plane bending; g, out-of-plane bending; r, rocking; t, twisting.
[a]Scaling factors 1.0189 (below 1100 cm$^{-1}$) and 0.9689 (above 1100 cm$^{-1}$) [40].

[b]According to [41]: $n_{obs}/n_{calc}$ = 1.0087 - 0.0000163·$n_{calc}$ .
[c]Raman intensities were calculated using RAINT programme [37] for a 1064 nm excitation wavelength.
*These bands are also overlapping with the manifestations of crystal water molecules.

**Table S14.** Correlation diagram of $H_2PO_3^-$ internal modes in **2-AmpH$_2$PO$_3$** crystals.

| Free ion $HPO_3^{2-}$ modes | Free ion $HPO_3^{2-}$ symmetry $C_{3v}$ | Free ion $H_2PO_3^-$ symmetry* $C_s$ | Site symmetry $C_1$ | Factor group symmetry $C_2 (Z = 2)$ |
|---|---|---|---|---|
| $\nu_1 (\nu$ PH) | $A_1$ | $A'$ ($\nu$ PH) | $A$ | $A$ (IR, Ra) + $B$ (IR, Ra) |
| $\nu_2 (\delta$ PH) | $E$ | $A'$ ($\delta$ PH) | | $A$ (IR, Ra) + $B$ (IR, Ra) |
| | | $A''$ ($\gamma$ PH) | | $A$ (IR, Ra) + $B$ (IR, Ra) |
| $\nu_3'$ ($\nu_s$ PO$_2$) | $A_1$ | $A'$ ($\nu$ PO(H)) | | $A$ (IR, Ra) + $B$ (IR, Ra) |
| $\nu_3''$ ($\nu_{as}$ PO$_2$) | $E$ | $A'$ ($\nu_s$ PO$_2$) | | $A$ (IR, Ra) + $B$ (IR, Ra) |
| | | $A''$ ($\nu_{as}$ PO$_2$) | | $A$ (IR, Ra) + $B$ (IR, Ra) |
| $\nu_4'$ ($\delta_s$ PO$_3$) | $A_1$ | $A'$ ($\delta$ PO(H)) | | $A$ (IR, Ra) + $B$ (IR, Ra) |
| $\nu_4''$ ($\delta_{as}$ PO$_3$) | $E$ | $A'$ ($\delta$ PO$_2$) | | $A$ (IR, Ra) + $B$ (IR, Ra) |
| | | $A''$ ($\rho$ PO$_2$) | | $A$ (IR, Ra) + $B$ (IR, Ra) |

*Note*. The OH group was assumed to be a single atom

**Table S15.** Correlation diagram of $HSO_4^-$ internal modes in **2-AmpHSO$_4$ (I)** and **(II)** crystals.

| Free ion $SO_4^{2-}$ modes | Free ion $SO_4^{2-}$ symmetry $T_d$ | Free ion $HSO_4^-$ symmetry* $C_{3v}$ | Site symmetry $C_1$ | Factor group symmetry $C_{2h}$ (Z = 4) |
|---|---|---|---|---|
| $\nu_1$ ($\nu_s$ SO) | $A_1$ | $A_1$ ($\nu_s$ SO$_3$) | $A$ | $A_g$ (IR, Ra) + $A_u$ (Ra) + $B_g$ (IR, Ra) + $B_u$ (IR, Ra) |
| $\nu_2$ ($\delta_d$ SO$_2$) | $E$ | $E$ ($\delta$ (H)OSO$_3$) | | 2$A_g$ (IR, Ra) + 2$A_u$ (Ra) + 2$B_g$ (IR, Ra) + 2$B_u$ (IR, Ra) |
| $\nu_3$ ($\nu_d$ SO) | $F_2$ | $A_1$ ($\nu$ SO(H)) | | $A_g$ (IR, Ra) + $A_u$ (Ra) + $B_g$ (IR, Ra) + $B_u$ (IR, Ra) |
| | | $E$ ($\nu_{as}$ SO$_3$) | | 2$A_g$ (IR, Ra) + 2$A_u$ (Ra) + 2$B_g$ (IR, Ra) + 2$B_u$ (IR, Ra) |
| $\nu_4$ ($\delta_d$ SO$_2$) | $F_2$ | $A_1$ ($\delta_s$ SO$_3$) | | $A_g$ (IR, Ra) + $A_u$ (Ra) + $B_g$ (IR, Ra) + $B_u$ (IR, Ra) |
| | | $E$ ($\delta$ SO$_3$) | | 2$A_g$ (IR, Ra) + 2$A_u$ (Ra) + 2$B_g$ (IR, Ra) + 2$B_u$ (IR, Ra) |

*Note*. The OH group was assumed to be a single atom

**Table S16.** Correlation diagram of $SO_4^{2-}$ internal modes in **(2-Amp)$_2$SO$_4$H$_2$O** crystals

| Free ion $SO_4^{2-}$ modes | Free ion $SO_4^{2-}$ symmetry $T_d$ | Site symmetry $C_1$ | Factor group symmetry $C_{2h}$ (Z = 4) |
|---|---|---|---|
| $\nu_1$ ($\nu_s$ SO) | $A_1$ | $A$ | $A_g$ (IR, Ra) + $A_u$ (Ra) + $B_g$ (IR, Ra) + $B_u$ (IR, Ra) |
| $\nu_2$ ($\delta_d$ SO$_2$) | $E$ | | 2$A_g$ (IR, Ra) + 2$A_u$ (Ra) + 2$B_g$ (IR, Ra) + 2$B_u$ (IR, Ra) |
| $\nu_3$ ($\nu_d$ SO) | $F_2$ | | 3$A_g$ (IR, Ra) + 3$A_u$ (Ra) + 3$B_g$ (IR, Ra) + 3$B_u$ (IR, Ra) |
| $\nu_4$ ($\delta_d$ SO$_2$) | $F_2$ | | 3$A_g$ (IR, Ra) + 3$A_u$ (Ra) + 3$B_g$ (IR, Ra) + 3$B_u$ (IR, Ra) |

**Table S17**. Calculated refractive indices and independent $\chi^{(2)}$ tensor components (A.U.) of **2-AmpH$_2$PO$_3$** crystal (l=¥).

| | **2-AmpH$_2$PO$_3$** | | |
|---|---|---|---|
| | **B3LYP** | **B3LYP** | **PBESOL0** |

| | advanced | | |
|---|---|---|---|
| $n_a$ | 1.411 | 1.381 | 1.423 |
| $n_b$ | 1.623 | 1.605 | 1.612 |
| $n_c$ | 1.627 | 1.611 | 1.623 |
| $\chi^{(2)}_{xxx}$ | 0 | $4{\cdot}10^{-24}$ | $2{\cdot}10^{-24}$ |
| $\chi^{(2)}_{xxy}$ | **0.006** | **0.144** | **0.185** |
| $\chi^{(2)}_{xxz}$ | 0 | $1{\cdot}10^{-24}$ | $3{\cdot}10^{-26}$ |
| $\chi^{(2)}_{xyy}$ | 0 | $-2{\cdot}10^{-18}$ | $1{\cdot}10^{-17}$ |
| $\chi^{(2)}_{xyz}$ | **-0.048** | **-0.149** | **-0.835** |
| $\chi^{(2)}_{xzz}$ | 0 | $-9{\cdot}10^{-25}$ | $5{\cdot}10^{-25}$ |
| $\chi^{(2)}_{yyy}$ | **-0.804** | **-0.887** | **-0.775** |
| $\chi^{(2)}_{yyz}$ | 0 | $-7{\cdot}10^{-17}$ | $-6{\cdot}10^{-17}$ |
| $\chi^{(2)}_{yzz}$ | **-0.429** | **-0.383** | **-0.368** |
| $\chi^{(2)}_{zzz}$ | 0 | 0 | 0 |

*Note*. Main $\chi^{(2)}$ components are marked as bold numbers.

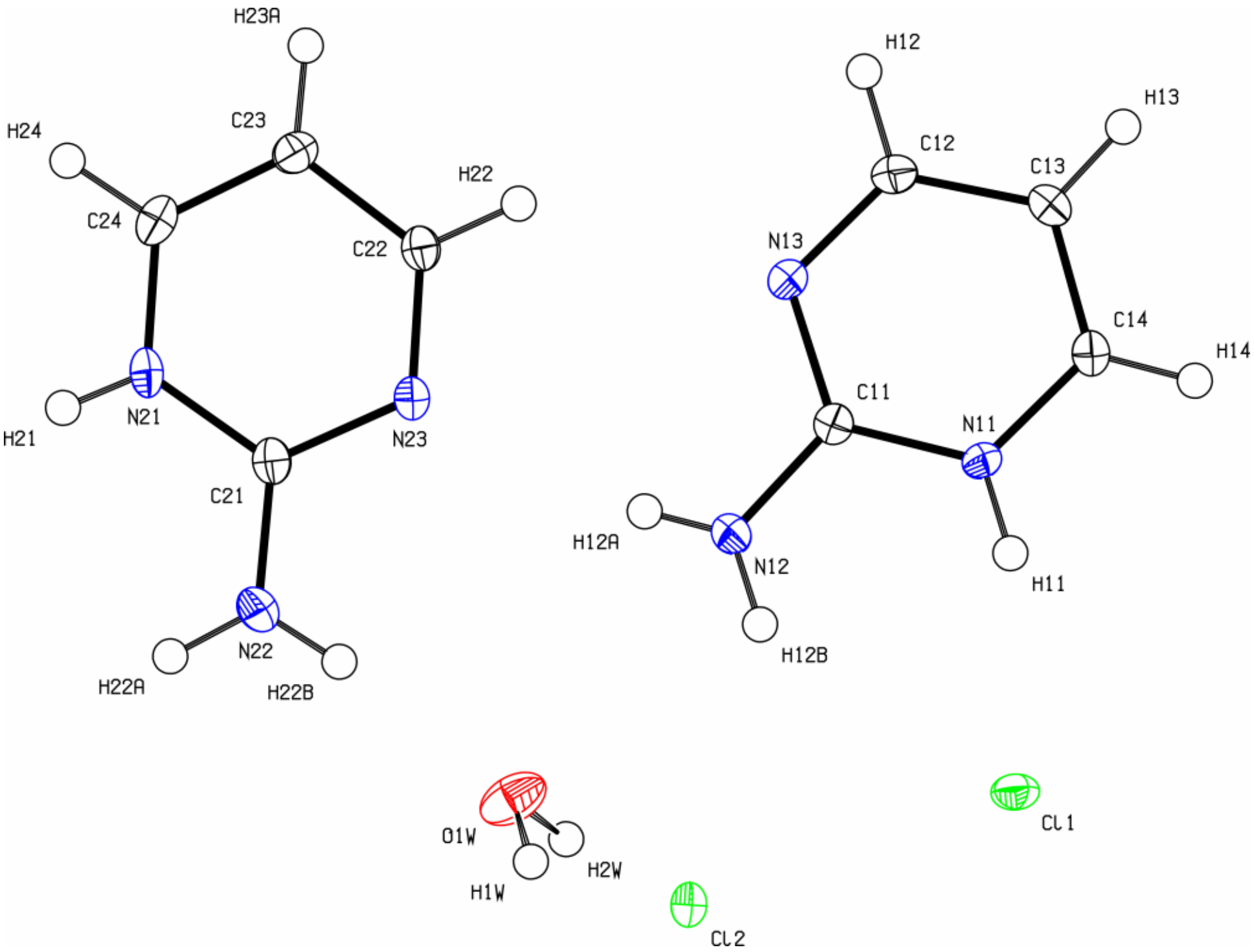

**Fig. S1.** ORTEP plot of the asymmetric unit of **2-AmpCl½H₂O** with atom numbering. The displacement parameters are shown at the 50% probability level.

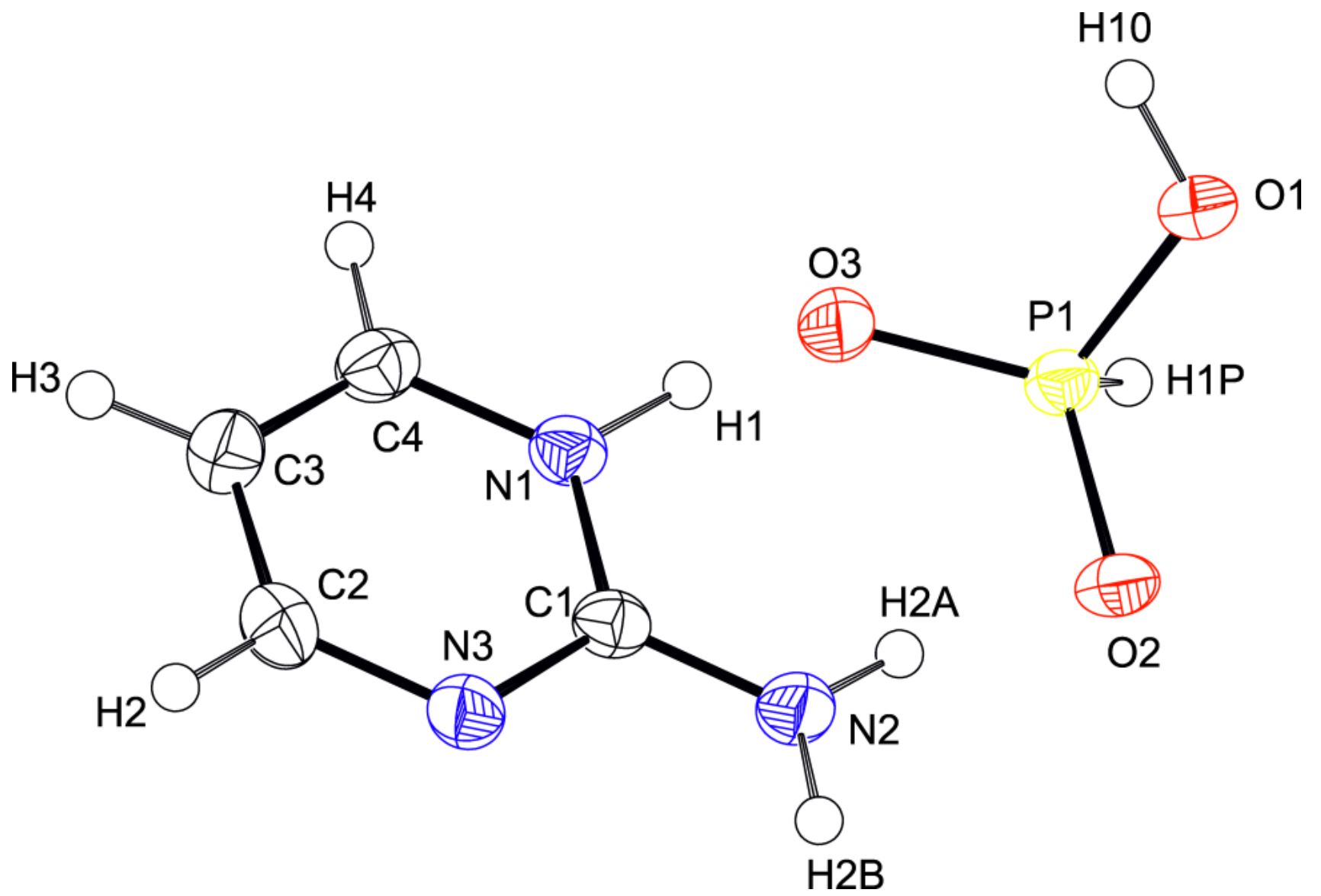

**Fig. S2.** ORTEP plot of the asymmetric unit of **2-Amp$H_2PO_3$** with atom numbering. The displacement parameters are shown at the 50% probability level.

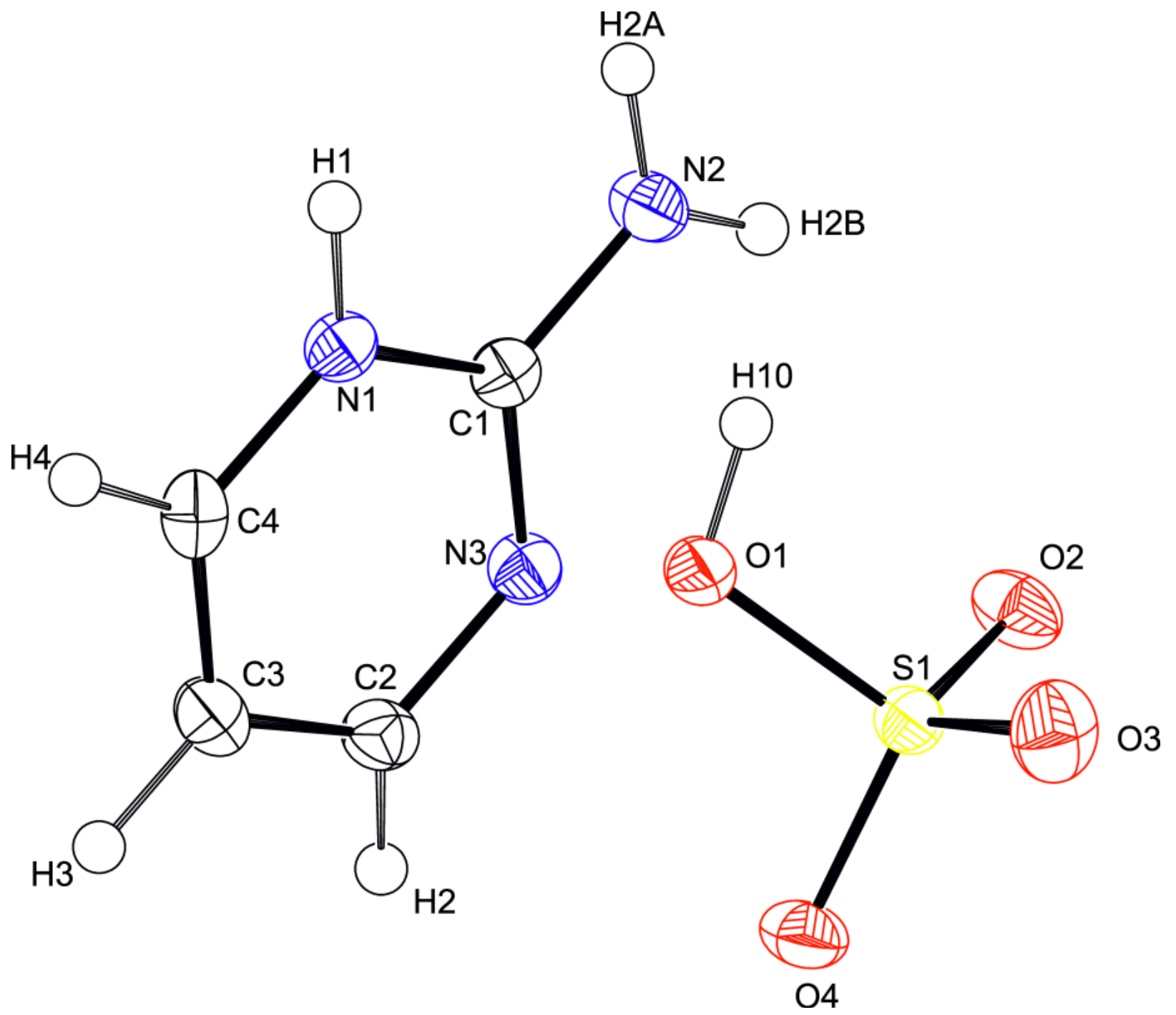

**Fig. S3.** ORTEP plot of the asymmetric unit of **2-Amp$HSO_4$ (I)** with atom numbering. The displacement parameters are shown at the 50% probability level.

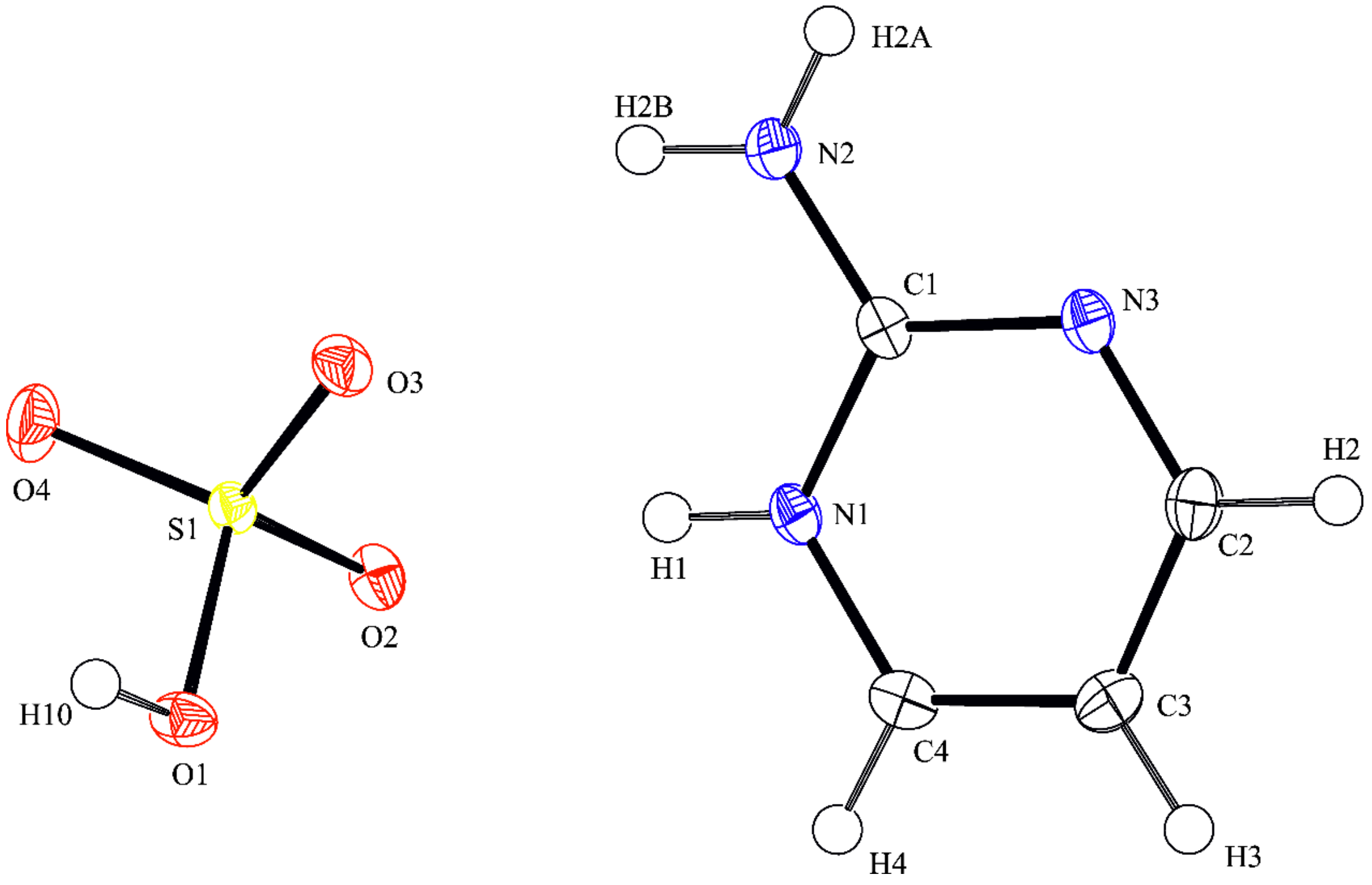

**Fig. S4.** ORTEP plot of the asymmetric unit of **2-AmpHSO$_4$ (II)** with atom numbering. The displacement parameters are shown at the 50% probability level.

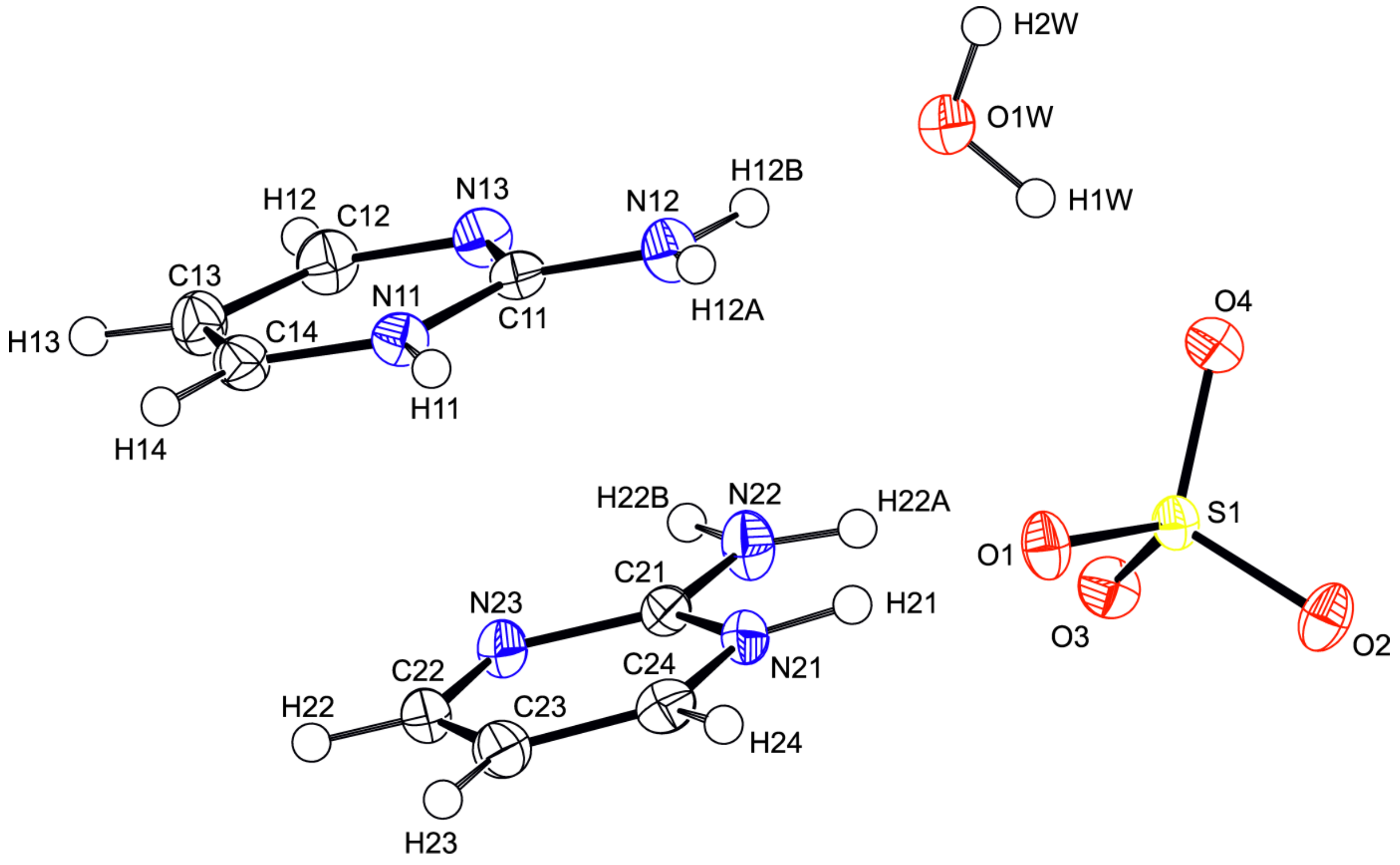

**Fig. S5.** ORTEP plot of the asymmetric unit of **(2-Amp)$_2$SO$_4$H$_2$O** with atom numbering. The displacement parameters are shown at the 50% probability level.

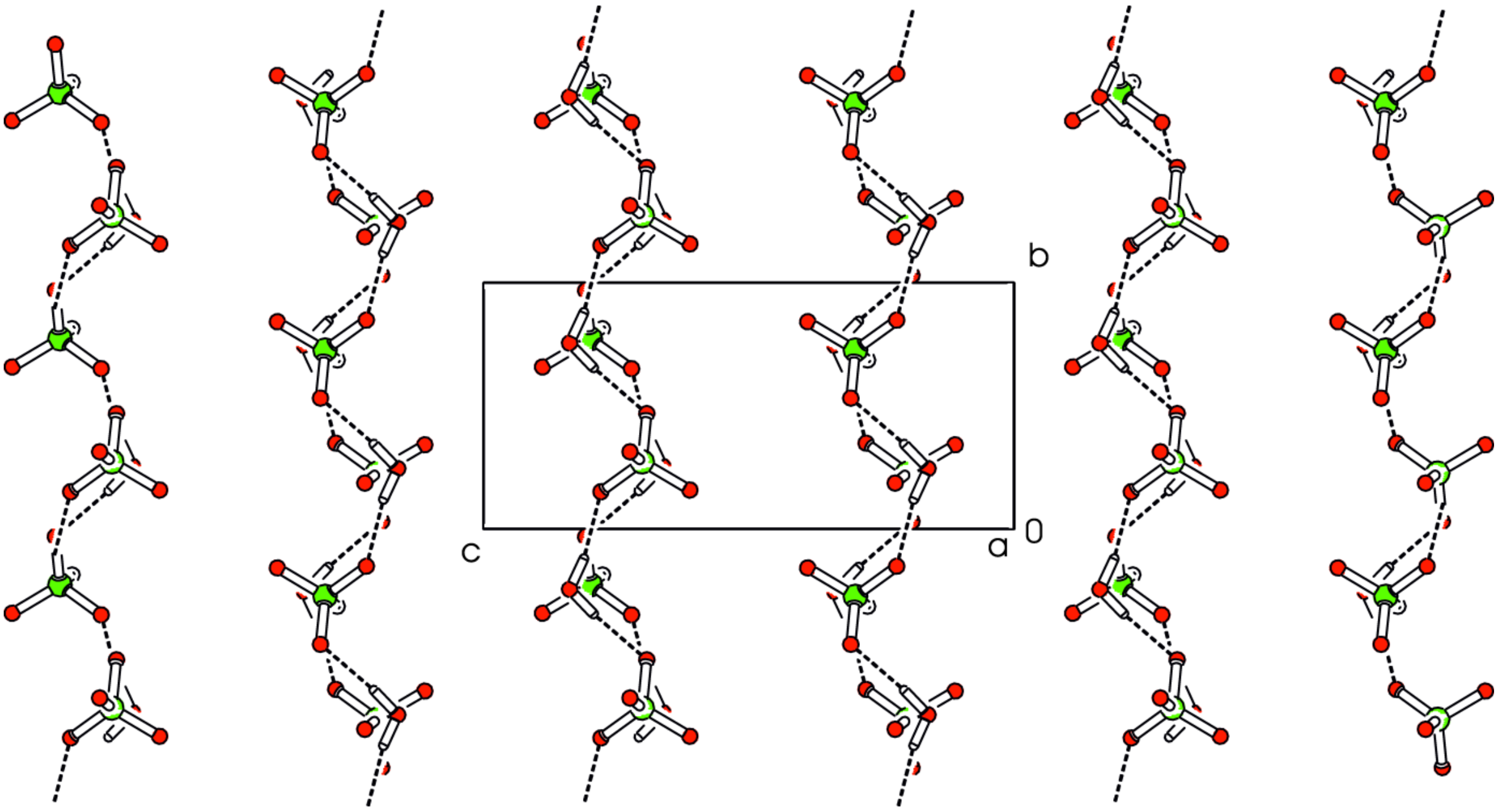

**Fig. S6.** The chains of alternating sulfate anions and water molecules in **(2-Amp)$_2$SO$_4$H$_2$O** crystal structure.

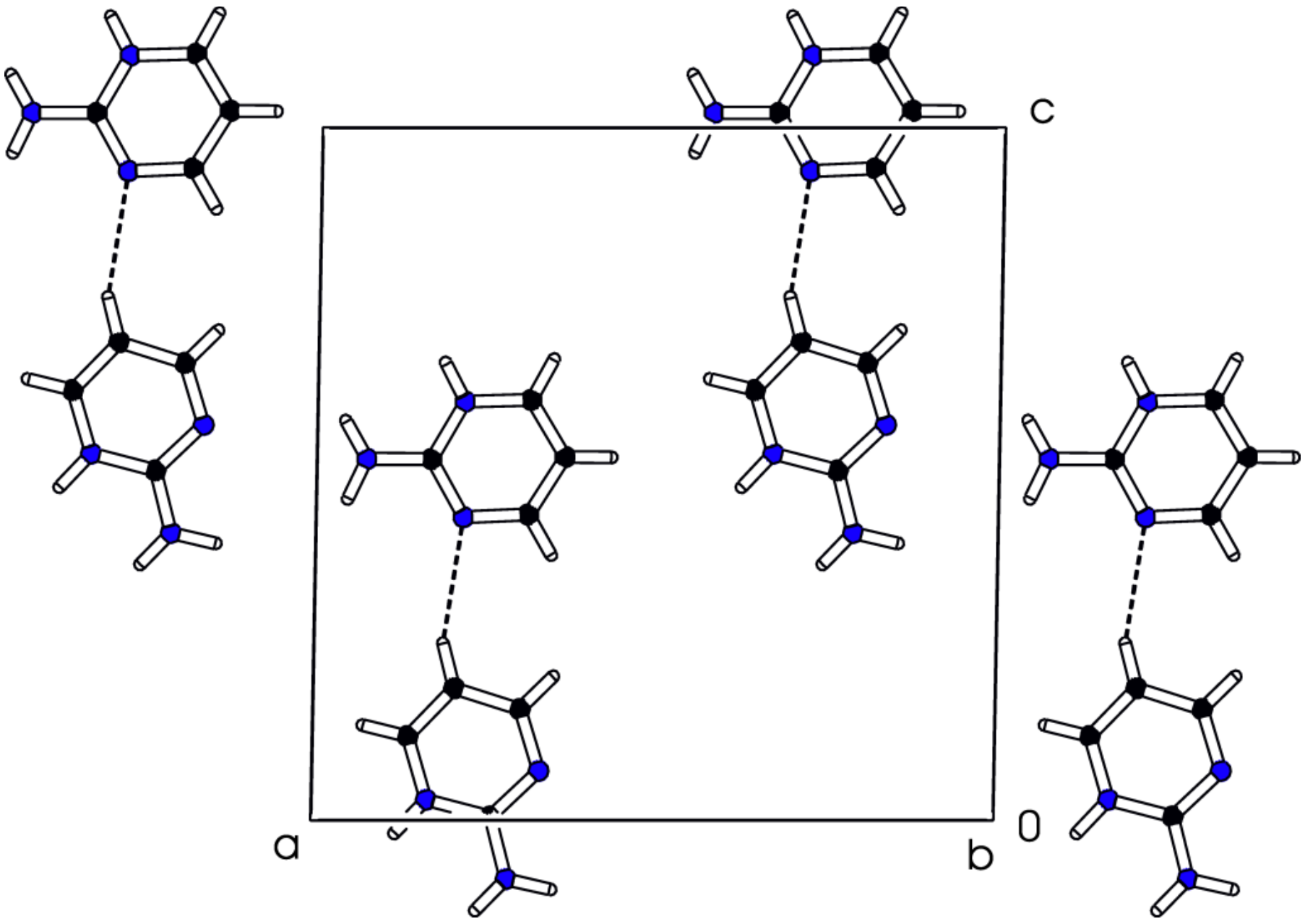

**Fig. S7.** The pairs of 2-aminopyrimidinium(1+) cations in **(2-Amp)$_2$SO$_4$H$_2$O** crystal structure.

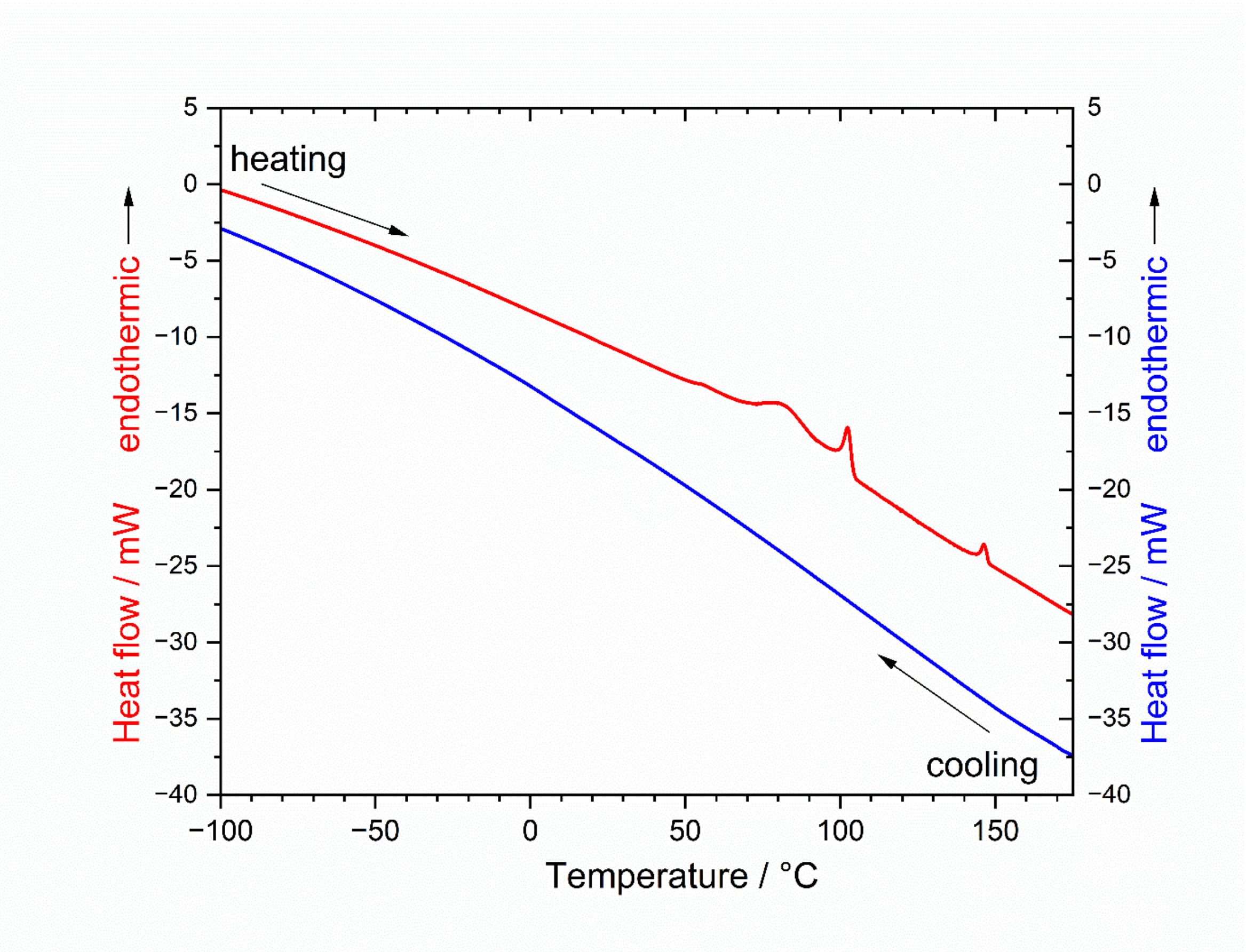

**Fig. S8**. DSC curves of **2-AmpSO$_4$H$_2$O**.

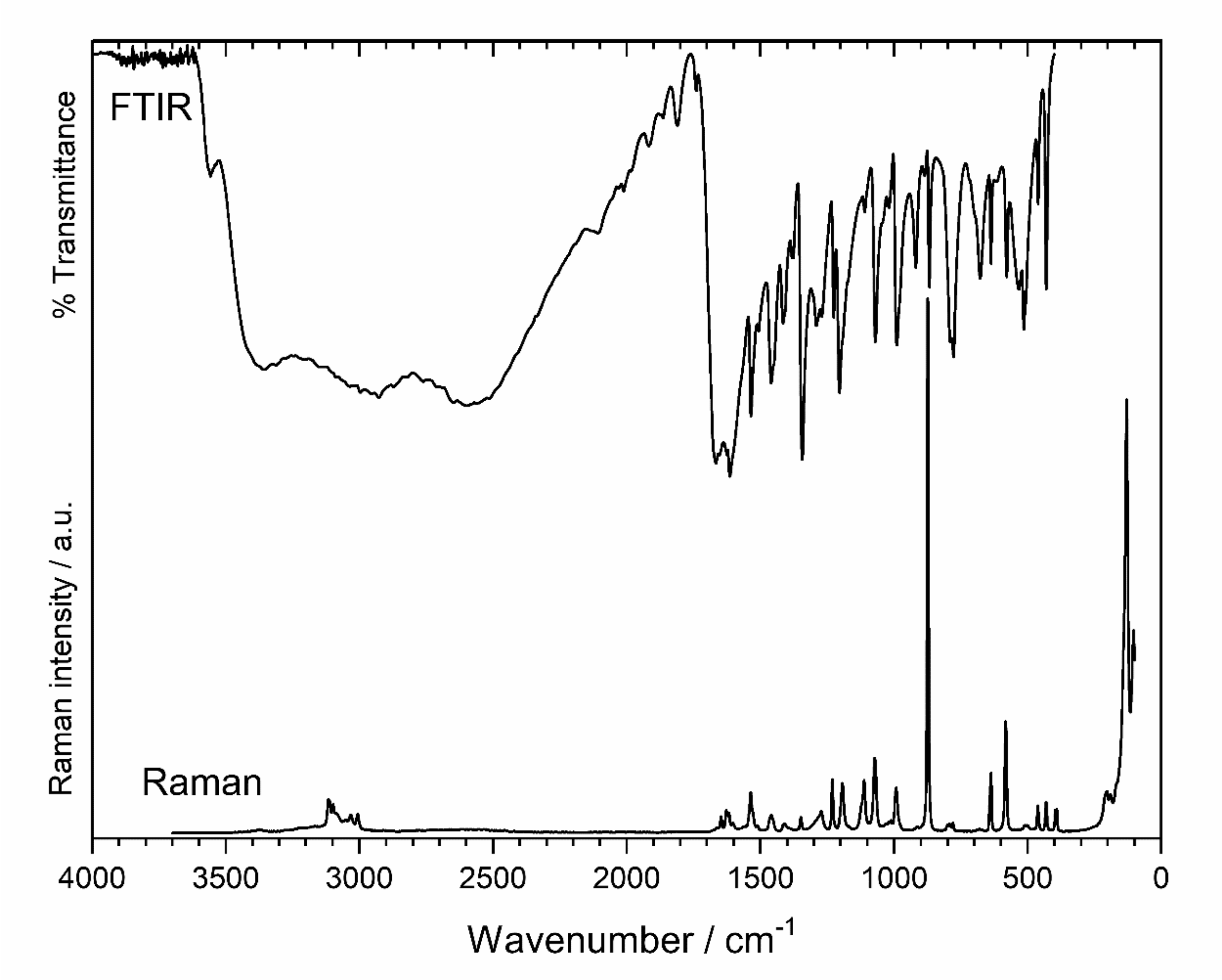

**Fig. S9.** FTIR (compiled from nujol and fluorolube mulls) and FT Raman spectra of 2-**AmpCl½H2O** crystals.

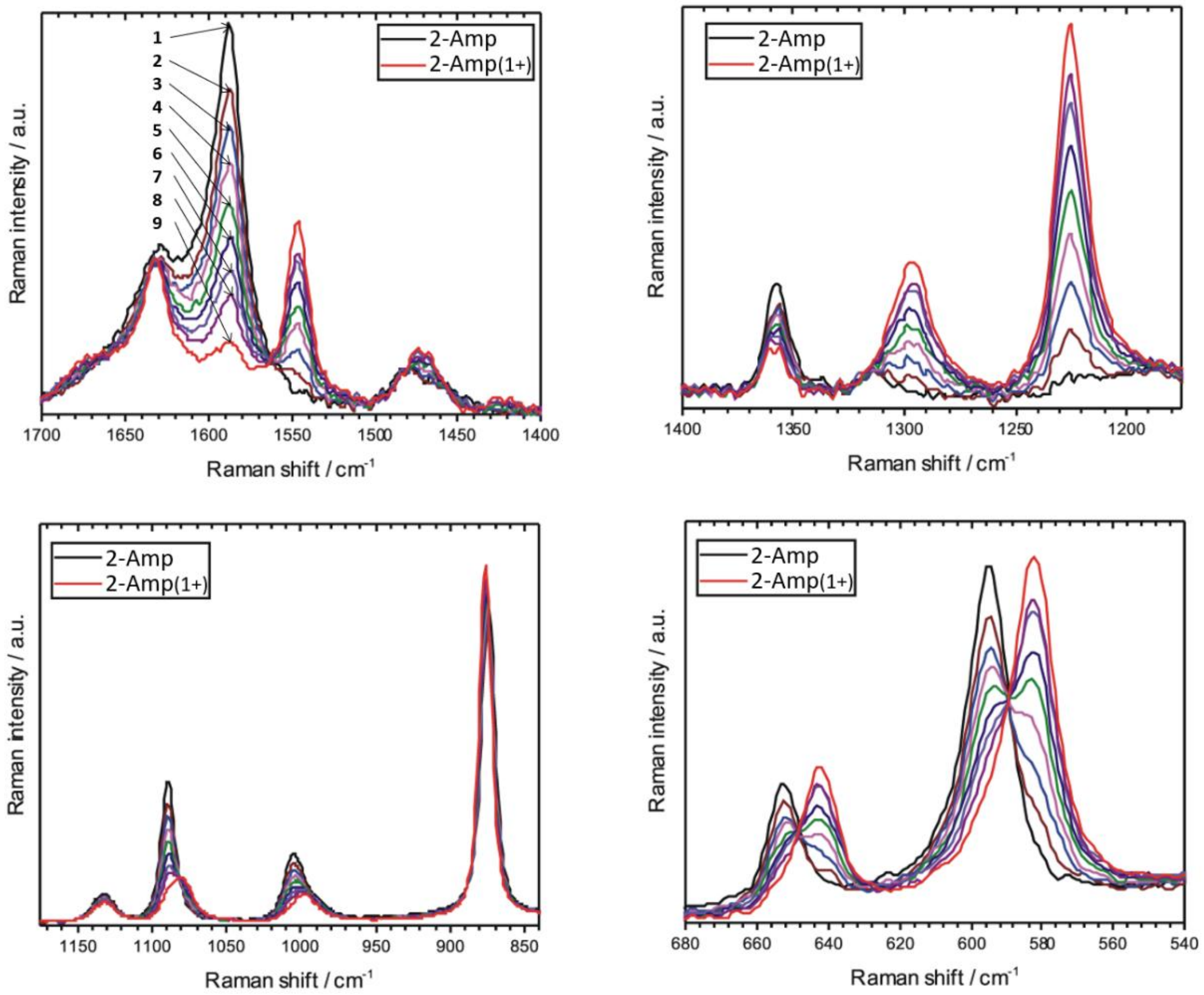

**Fig. S10.** The formation of **2-Amp**(1+) cation in aqueous solution studied by Raman spectroscopy (780 nm laser excitation). The particular spectra represent systems prepared by mixing of 2-aminopyrimidine solution with hydrochloric acid in the following molar ratios (base to acid) - i.e. 1:0 (spectrum **1**), 1:0.125 (spectrum **2**), 1:0.250 (spectrum **3**), 1:0.375 (spectrum **4**), 1:0.500 (spectrum **5**), 1:0.625 (spectrum **6**), 1:0.750 (spectrum **7**), 1:0.875 (spectrum **8**) and 1:1 (spectrum **9**).

The formation of **2-Amp**(1+) cation in aqueous solution was studied by Raman spectroscopy (Raman titration) using 780 nm laser excitation. The studied solutions were prepared by mixing of 1 mol/l solution of 2-aminopyrimidine with 4 mol/l solution of hydrochloric acid in the molar ratios (base to acid) ranging from 1:0 to 1:1. The resulting spectra are presented in Fig. S10. The formation of **2-Amp**(1+) cation is clearly demonstrated by the appearance of new Raman bands at ~1550 $cm^{-1}$, ~1300 $cm^{-1}$, ~640 $cm^{-1}$ and ~580 $cm^{-1}$.

The crystallization of equimolar 2-aminopyrimidine solution with hydrochloric acid led to the formation of the only product **2-AmpCl½H$_2$O** – see Table S23 and Fig. S9, Supplementary material.

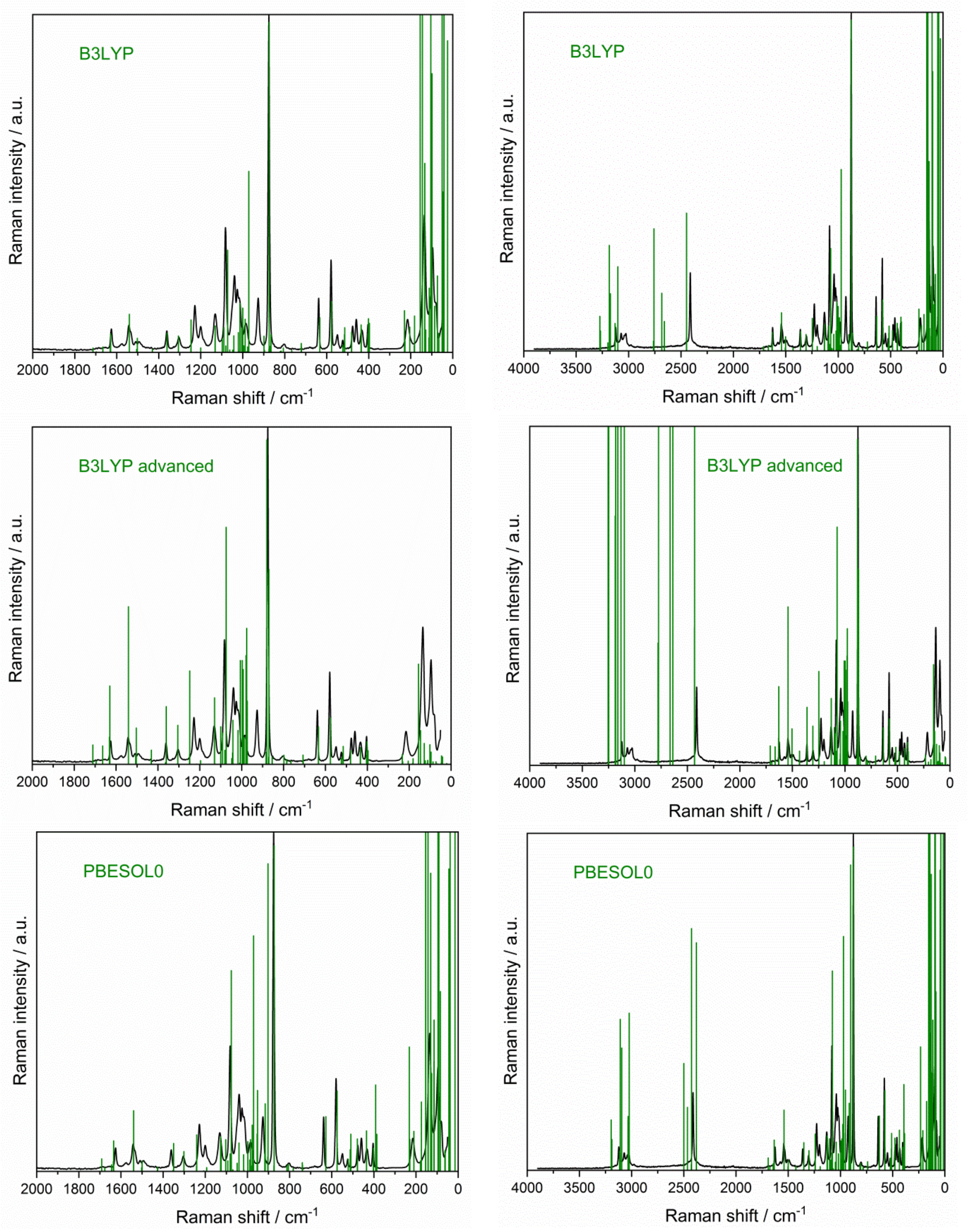

**Fig. S11**. The comparison of the recorded Raman spectrum of **2-AmpH$_2$PO$_3$** crystals and computed vibrational frequencies (green lines) using different computational approaches in 2000-0 cm$^{-1}$ (left column) and 4000-0 cm$^{-1}$ (right column) regions.

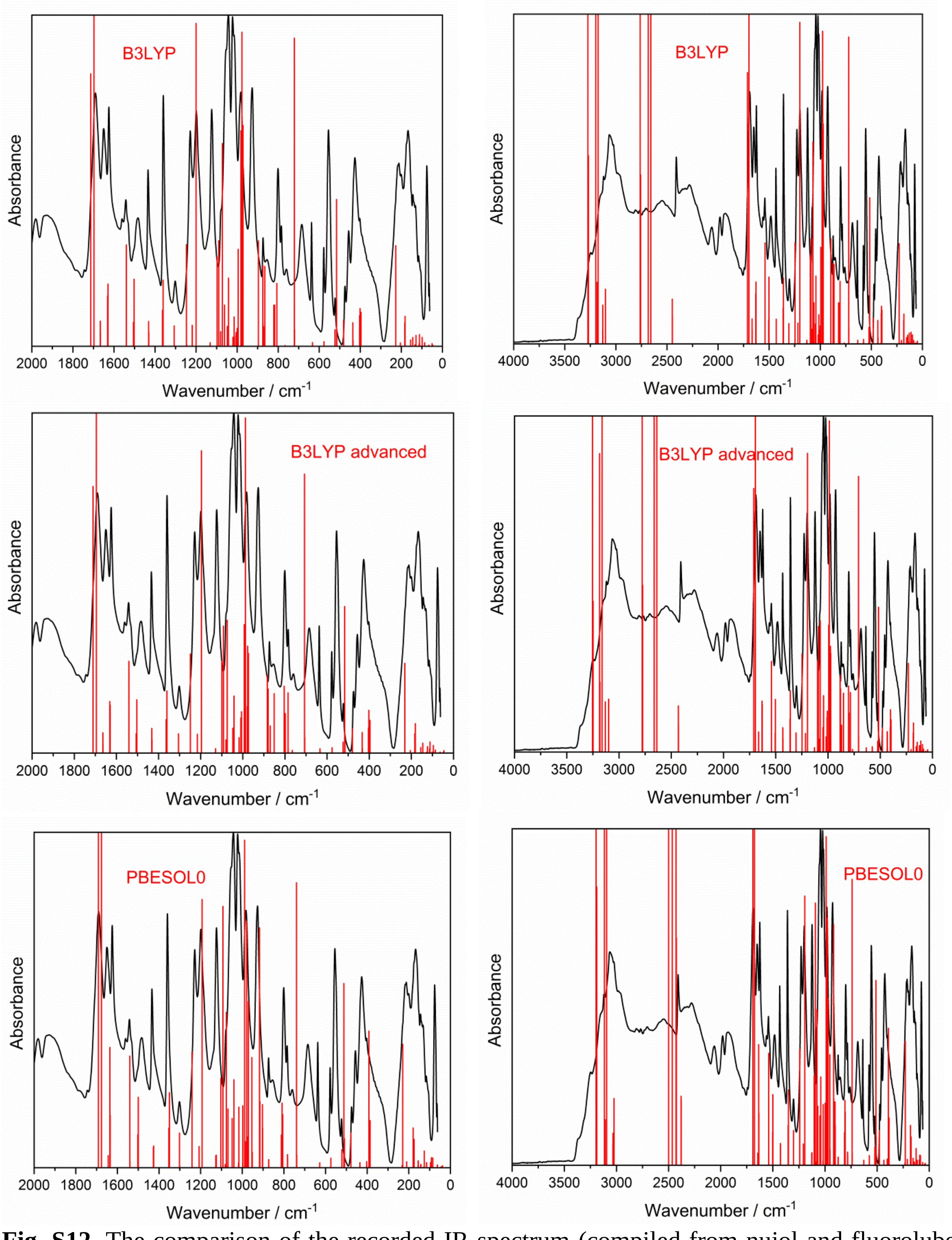

**Fig. S12.** The comparison of the recorded IR spectrum (compiled from nujol and fluorolube mulls) of **2-AmpH$_2$PO$_3$** crystals and computed vibrational frequencies (red lines) using different computational approaches in 2000-0 cm$^{-1}$ (left column) and 4000-0 cm$^{-1}$ (right column) regions.